\documentstyle[12pt,epsfig]{article}
\newlength{\dinwidth}
\newlength{\dinmargin}
\date{}
\newcommand{\half}{\mbox{\small$\frac{1}{2}$}}

\newcommand{\calH}{{\cal H}}
\newcommand{\be}{\begin{equation}}
\newcommand{\ee}{\end{equation}}
\newcommand{\bea}{\begin{eqnarray}}
\newcommand{\eea}{\end{eqnarray}}
\newcommand{\mye}{\mbox{e}}
\newcommand{\ph}{\phi}
\def\rellow#1#2{\mathrel{\mathop{\kern 0pt #1}\limits_{#2}}}

\newcommand{\str}{\rule{0ex}{2.7ex}}  % strut, to make a line higher
\newcommand{\strr}{\rule{0ex}{4.0ex}}  % strut, to make a line higher
\newcommand{\tabhline}{\\[0.3ex] \hline \str}

\newcommand{\Um}[1]{U_{#1,\mu}}
\newcommand{\Un}[1]{U_{#1,\nu}}
\newcommand{\Ut}[1]{U_{#1,\tau}}
\newcommand{\Tr}{\mbox{Tr}}
\newcommand{\foN}{ \mbox{ \small$\frac{1}{N}$ } }
\newcommand{\EW}[1]{ \langle #1 \rangle }
\newcommand{\vn}{\vec n}
\newcommand{\vs}{\vec \sigma}
\newcommand{\calP}{{\cal P}}
\newcommand{\Lamt}{ \Lambda_t^{\tau} }
\newcommand{\erfc}{\mbox{erfc}}
\newcommand{\delH}{\Delta\calH}
\def\bbZ{\mathchoice{\rm Z\kern-0.32 em{Z}}{\rm Z\kern-0.32 em{Z}}
                    {\rm Z\kern-0.28 em{Z}}{\rm Z\kern-0.28 em{Z}}}
\title{\vbox{\vspace{.1mm}}
Analysis and Development
of Stochastic Multigrid Methods in Lattice Field Theory}

\author{ \vbox{\vspace{7mm}}
   {\bf Martin Grabenstein\thanks{
supported by Deutsche Forschungsgemeinschaft}
                } \\[6mm]
     II. Institut f\"ur Theoretische Physik, Universit\"at Hamburg,
      \\[-1mm]
      Luruper Chaussee 149, 22761 Hamburg, Germany,
      \\[-1mm]
      {\tt \small  internet i$\!\, \not \! 0$2gra@dsyibm.desy.de }
      \\[20mm]
      preprint DESY 94-007, hep-lat/9401024}
\begin{document}
\maketitle \vfill
%%%%%%%%%%%%%%%%%%%%%%%%%%%%%%%%%%%%%%%%%%%%%%%%%%%%%%%%%%%%%%%%%%%%%%%%
% Abstract
%%%%%%%%%%%%%%%%%%%%%%%%%%%%%%%%%%%%%%%%%%%%%%%%%%%%%%%%%%%%%%%%%%%%%%%%
\mbox{}
\vfill

\renewcommand{\abstractname}{\protect Abstract}
\begin{abstract}
We study the relation between the dynamical critical behavior
and the kinematics of stochastic multigrid algorithms.
The scale dependence of acceptance rates for nonlocal Metropolis
updates is analyzed with the help of an approximation formula.
A quantitative study of the kinematics of 
multigrid algorithms in several interacting models is performed.
We find that for a critical
model with Hamiltonian $\calH(\ph)$
absence of critical slowing down
can only be expected if the expansion
of $\langle \calH (\ph + \psi) \rangle$ in terms of the shift $\psi$
contains no relevant term (mass term).
The predictions of this rule are verified in a multigrid Monte Carlo
simulation of the Sine Gordon model in two dimensions.

Our analysis can serve as a guideline for the development of new 
algorithms: We propose a new multigrid method for nonabelian
lattice gauge theory,
the time slice blocking. For $SU(2)$ gauge fields in two dimensions,
critical slowing down is almost completely eliminated by this method,
in accordance with the theoretical prediction.
The generalization of the time slice blocking to 
$SU(2)$ in four dimensions
is investigated analytically and by numerical simulations.
Compared to two dimensions, 
the local disorder in the four dimensional gauge field leads 
to kinematical problems.
\end{abstract} 

\vfill \mbox{}

\thispagestyle{empty}

\newpage
\pagenumbering{roman}
\tableofcontents

\newpage
\pagenumbering{arabic}

%%%%%%%%%%%%%%%%%%%%%%%%%%%%%%%%%%%%%%%%%%%%%%%%%%%%%%%%%%%%%%%%%%%%%%%%%
\clearpage

\section{Introduction}
\label{SECintro}

Monte Carlo simulations have become an important tool for the study of 
critical phenomena in statistical mechanics \cite{binder} and 
for nonperturbative calculations in 
Euclidian lattice field theory close to the continuum limit
\cite{rebbi}. 

 However, the method has limitations.
In the vicinity of a critical point the phenomenon of 
critical slowing down (CSD) is a serious problem:
For conventional local algorithms the autocorrelation time 
- that is, roughly speaking, the time needed to generate a new, 
``statistically independent'' configuration on a computer -
grows rapidly as the system approaches criticality.
More precisely, 
the autocorrelation time $\tau$ behaves like $\tau \sim \xi^z$, 
where $\xi$ denotes the spatial correlation
length, and $z$ is the dynamical critical exponent.  For conventional
local algorithms, $z \approx 2$. 
Thus, when the critical point is approached, there is a dramatic 
increase of computer time needed to calculate observables to a given accuracy.
It is therefore important to find Monte Carlo algorithms 
that have reduced CSD.
 
Accelerated algorithms that are still local, such
as overrelaxation or the optimized hybrid Monte Carlo algorithm, 
can sometimes reduce the dynamical critical exponent 
to $z \approx 1$ \cite{over,hybrid}.

For the complete elimination of CSD in the sense of $z \approx 0$,
nonlocal update algorithms are needed.
Heuristic arguments based on the picture that in
simulations with local algorithms ``information''
propagates though the lattice in the form of a random walk lead to 
a value of $z=2$.
Therefore it is natural to use nonlocal update schemes 
that perform changes of the configuration not only on the scale of
the lattice spacing but create fluctuations on all length scales,
in analogy to the physical fluctuations at a critical point.

Stochastic cluster algorithms are very successful in reducing 
CSD in spin models:
The first cluster algorithm 
was developed by Swendsen and Wang for 
discrete spin variables in Ising and Potts models
\cite{swendsenwang}.
This method could be successfully extended to continuous spin variables
in $O(N)$ nonlinear $\sigma$-models 
by Wolff \cite{wolff} and Hasenbusch \cite{mhbon}.
However, the generalization to $SU(N) \times SU(N)$
principal chiral models with $N > 2$, which are  
models with the same type of variables as QCD,
seems to be difficult~\cite{sunsun}.
There are arguments that 
cluster algorithms will not be efficient if
the spin variables take values in manifolds other than
spheres, products of spheres or the quotient of such a space 
by a discrete group~\cite{embedding}.

An alternative method to overcome CSD in the simulation of
models with continuous variables is multigrid Monte Carlo: 
Multigrid methods are well established tools for the solution
of discretized partial differential equations \cite{PDE,koelnporz}.
Their generalization to Monte Carlo simulations was proposed by Parisi
\cite{multigrid}.
Early ideas for nonlocal updating schemes
were formulated by H.\ Meyer-Ortmanns \cite{meyerortmanns}.
Goodman and Sokal generalized deterministic multigrid methods
to multigrid Monte Carlo in the
$\ph^4$ theory in two dimensions \cite{sokalprl}.
Mack presented a stochastic multigrid approach that was inspired by
constructive quantum field theory and renormalization group considerations
\cite{cargmack}.

In the present thesis, every algorithm that
performs stochastic updates of predesigned shape
on a hierarchy of length scales is called multigrid Monte Carlo algorithm.  

There are examples, e.g.\ the $SU(3) \times SU(3)$
principal chiral model, where no fast
cluster algorithms have been found whereas multigrid Monte Carlo
algorithms could reduce CSD considerably \cite{sunsun,hmsun}.

Presently, the only generally applicable method to study 
the dynamical critical behavior of Monte Carlo algorithms for
interacting models is numerical experiment.
For some models experiments show that the dynamical critical exponent
$z$ can be substantially reduced by a multigrid algorithm 
\cite{hmsun,hmm,laursen,sokalO4}.
For other models, still $z = 2$ is found \cite{sokalprl,linn}.
Theoretical insight into the critical dynamics of multigrid Monte Carlo
algorithms is therefore desirable.

We present a method that can help to judge which algorithm will have
a chance to overcome CSD in the simulation of a given model
{\em before} performing the simulation:
We study the kinematics of multigrid Monte Carlo algorithms.
By kinematics we mean the study of the scale (block size)
dependence of the Metropolis acceptance rates for nonlocal update
proposals.  We do not address the more difficult problem of
analytically calculating the dynamical critical behavior
from the stochastic evolution of the system.
However, we are able to find a heuristic relation  between the
kinematical and the critical dynamical behavior 
of multigrid Monte Carlo methods.
This relation is based on the fact that
sufficiently high acceptance rates
are necessary to overcome CSD.

We derive an approximation formula for the block size dependence of
acceptance rates for nonlocal Metropolis updates.  The influence of the
coarse-to-fine interpolation kernel (shape function) on the kinematics
in free field theory, where the formula is exact, is investigated in
detail.

The formula is then applied to several interacting
models and turns out to be a very
good approximation.  
By comparison with free field theory,
where CSD is eliminated by a multigrid algorithm,
we can formulate a necessary criterion 
for a given multigrid
algorithm to eliminate CSD:  For a critical
model with a fundamental Hamiltonian $\calH(\ph)$
absence of CSD can only be expected if the expansion
of $\langle \calH (\ph + \psi) \rangle$ in terms of the shift $\psi$
contains no relevant term (mass term).

Heuristically, the physical content of this criterion can be
summarized in a Leitmotiv:
A~piecewise constant update of a nonlocal domain should only have
energy costs proportional to the surface of the domain,
not energy costs proportional to the volume of the domain.

This seems to be a feature that multigrid algorithms share with
cluster algorithms~\cite{embedding}.

As a first test of the predictive power of the kinematical analysis
we perform a multigrid Monte Carlo simulation of the Sine Gordon model in two
dimensions.
There, the criterion tells that we have to expect CSD
caused by too small acceptance rates of the updates on large scales.
This prediction is confirmed by the numerical experiments.
In addition it is studied whether one can compensate for too small 
amplitudes of the updates on large blocks by accumulating
many of these updates randomly.

\bigskip 

After the introduction of the kinematical analysis of 
multigrid Monte Carlo algorithms in the context of spin models,
where the multigrid methods have already been developed,
we show that our method can also be helpful in the design of new
multigrid procedures.

One of the challenges in the development of fast Monte Carlo algorithms
is pure lattice gauge theory in four dimensions.
In particular the nonabelian gauge groups $SU(2)$ and $SU(3)$ are 
important for nonperturbative calculations
in the standard model of elementary particles.

For the dynamical critical behavior of the local
heat bath algorithm in $SU(3)$ only very crude estimates
are available up to now \cite{SU3z}, consistent with $z \approx 2$. 
The present state-of-the-art algorithm for nonabelian
gauge fields in four dimensions 
is overrelaxation \cite{overgauge}.
For this algorithm, first estimates for $z$ in $SU(2)$
lattice gauge theory gave
$z = 1.0(1)$ in physically small volumes 
\cite{luescherwolff}.
Effort was also spent in developing nonlocal algorithms
for gauge theories. 
A fast cluster algorithm
was found for $3 +1$-dimensional $SU(2)$ 
gauge theory on a $L^3\times T$ lattice at finite temperature,
but only in the special case $T=1$ \cite{su2fint}.
A cluster algorithm for $U(1)$ gauge theory
in two dimensions is based on the 
reduction of the gauge theory
to a one dimensional $XY$ model \cite{sinclair}.
However, apart from these special cases, no
efficient cluster algorithm for
continuous gauge groups has been found up to now.

Therefore possible
developments of stochastic multigrid methods
for pure gauge fields are of particular interest.
Multigrid algorithms for $U(1)$ gauge models were
introduced and studied in two and four dimensions
\cite{laursen,newgauge}.
A different but related nonlocal updating scheme for
abelian lattice gauge theory is the multiscale 
method \cite{adler}.

We propose a new multigrid algorithm for nonabelian gauge
theory and analyze its kinematics. 

To gain experience, we first study the case of gauge group $SU(2)$ in
two dimensions. We introduce a multigrid method
for nonabelian gauge theory
that treats different time slices independently:
the time slice blocking algorithm.
The theoretical analysis predicts
that CSD can be eliminated by the time slice blocking.
By numerical experiments on systems with
lattice sizes up to $256^2$ we check
whether this is indeed the case.

In a second step we generalize 
the time slice blocking  
to $SU(2)$ in four dimensions.  
Compared to the two dimensional case we have to face additional 
difficulties that are caused by the local disorder of the gauge fields.
Our approximation formula turns out
to be very reliable also in this case and allows for a prediction of
acceptance rates for a large class of nonlocal updates.
We attempt to estimate the kinematical behavior of the
proposed algorithm in the weak coupling limit
and study whether a reduction of CSD can be expected.
In numerical simulations, the time slice blocking algorithm
in $SU(2)$ in four dimensions is compared with
a local heat bath algorithm.

When Parisi proposed the use of multigrid methods
for Monte Carlo simulations in 1983,
he stated \cite{multigrid}:
``On the contrary [{\it to quadratic actions}] the application 
of the multigrid method to 
the gauge field sector seems to be particularly painful
but it may be rewarding.'' 

We attempt to describe where we stand 
ten years after Parisi's speculation.

\bigskip 

This thesis is organized as follows:
In the first part (sections \ref{SECunigrid}-\ref{SECsg}) 
the kinematical analysis of multigrid algorithms is 
introduced and discussed.
In the second part (sections \ref{SECSU22acc}-\ref{SECSU24sim}) 
the  method of the kinematical analysis 
is applied to the development of new multigrid algorithms
for nonabelian gauge fields.

In section \ref{SECunigrid} we
introduce multigrid Monte Carlo algorithms.
Section \ref{SECapprox} contains the derivation of our
approximation formula for acceptance rates.
Several coarse-to-fine-interpolation
kernels are discussed in section \ref{SECkernels}.
In section \ref{SECfree} the
acceptance rates in free field theory are examined in detail.
The main idea of the thesis is introduced and 
discussed in section \ref{SECappl}:
The kinematical analysis for the Sine Gordon, XY, $\phi^4$,
$O(N)$ and $CP^{N-1}$ models is
presented.
In section \ref{SECsg} the prediction of this analysis
for the Sine Gordon model is verified by a multigrid Monte Carlo simulation.

Section \ref{SECSU22acc} starts with the discussion of abelian gauge
fields in two dimensions.
There are two directions of increasing the difficulty of the problem:
from abelian gauge group to nonabelian gauge group and from 
two dimensions to four dimensions.
Most of the concepts that are needed for the treatment of nonabelian gauge
fields are introduced in the context of $SU(2)$ lattice gauge theory
in two dimensions. 
A new multigrid procedure,
the time slice blocking, is introduced.
In section \ref{SECSU22sim}, $SU(2)$ lattice gauge theory
in two dimensions is simulated
with the time slice blocking. 

The new procedure is generalized from $SU(2)$ in two dimensions 
to $SU(2)$ in four dimensions in
section \ref{SECSU24acc}. 
Section \ref{SECSU22sim} contains the results of a
multigrid Monte Carlo simulation of $SU(2)$ lattice gauge theory
in four dimensions. 
A summary is given in section \ref{SECsummary}.

Parts of this thesis have been published before in refs.\ 
\cite{acceptance,kinematics,theoretical,sgpaper,sgproc}.

\clearpage

\section{Multigrid Monte Carlo algorithms}

\label{SECunigrid}

\setcounter{equation}{0}

We consider lattice models with partition functions
\be\label{PARTI}
Z= \int \prod_{x \in \Lambda_0} d\ph_x \, \exp(-\calH(\ph)) \,
\ee
on hypercubic $d$-dimensional lattices $\Lambda_0$ with periodic
boundary conditions. The lattice spacing is set to one. We
use dimensionless spin variables $\ph_x$.
An example is single-component $\phi^4$ theory, defined by the
Hamiltonian
\be
\calH(\ph) = \half(\ph, -\Delta \ph)
           + \frac{m_o^2}2 \sum_x \ph_x^2
           + \frac{\lambda_o}{4!} \sum_x \ph_x^4 \, ,
\ee
where
\be \label{DELTA}
(\ph, -\Delta \ph) = \sum_{<x,y>} (\ph_x - \ph_y)^2 \, .
\ee
The sum in eq.\ (\ref{DELTA}) is over all nearest neighbor pairs
in the 
lattice\footnote{The definitions for lattice gauge theory 
will be introduced in  section \ref{SECSU22acc}}.

\subsection{Local Metropolis updating}

A standard algorithm to perform Monte Carlo simulations in a model
of the type defined above is the local Metropolis algorithm:
One visits in a regular or random order the sites of the lattice
and performs the following steps: At site $x_o$, one
proposes a shift
\be
\ph_{x_o} \rightarrow  \ph_{x_o}' = \ph_{x_o} + s \, .
\ee
The configuration $\{ \ph_x \}$ remains unchanged for $x \neq x_o$.
$s$ is a random number selected according to an a priori
distribution $\rho(s)$ which is symmetric with respect to
$s \rightarrow - s$. E.g., one selects $s$ with uniform
probability from an interval $[-\varepsilon,\varepsilon]$.
Then one computes the change of the Hamiltonian
\be
\Delta \calH = \calH(\ph')-\calH(\ph) \, .
\ee
Finally the proposed shift is accepted with
probability $\min [1,\exp(-\Delta \calH)]$. Then one proceeds
to the next site.

The local Metropolis algorithm suffers from CSD when
the correlation length in the system becomes large: long
wavelength fluctuations cannot efficiently be generated
by a sequence of local operations.
It is therefore natural to study nonlocal generalizations
of the update procedure defined above.

\subsection{Nonlocal multigrid updating}

%%%%%%%%%%%%%%%%%%%%%%%%%%%%%%%%%%%%%%%%%%%%%%%%%%%%%%%%%%%%%%%%%%%
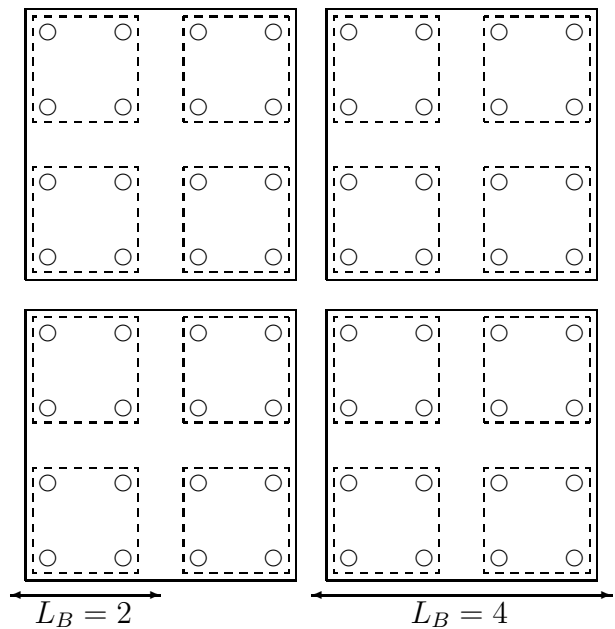
\begin{figure}
\begin{center}
\begin{picture}(80,80)(0,-7)
\multiput(0,0)(0,10){8}{\multiput(0,0)(10,0){8}{\circle{2.0}}}
\multiput(-2,-2)(20,0){4}{\multiput(0,0)(0,20){4}{\dashbox{1}(14,14){}}}
\multiput(-3,-3)(40,0){2}{\multiput(0,0)(0,40){2}{\dashbox{36}(36,36){}}}
\put(0, -5){\vector(1,0){15}}
\put(10, -5){\vector(-1,0){15}}
\put(60, -5){\vector(1,0){15}}
\put(60, -5){\vector(-1,0){25}}
\put(-1,-10){\makebox(10,5)[c]{ $L_B = 2$}}
\put(49,-10){\makebox(10,5)[c]{ $L_B = 4$}}
\end{picture}
\caption[dummy]{\label{blocking} 
                \sl Division of the finest lattice 
                in block lattices of  
                $L_B \times L_B$-blocks in two dimensions}
\end{center}
\end{figure}
%%%%%%%%%%%%%%%%%%%%%%%%%%%%%%%%%%%%%%%%%%%%%%%%%%%%%%%%%%%%%%%%%%%

Consider the fundamental lattice $\Lambda_0$ as divided in
hypercubic blocks of size $l^d$, 
where $l$ denotes the coarsening factor.
Typical coarsening factors are $l = 2$ or $3$.
This defines a block lattice
$\Lambda_1$. 
By iterating this procedure one obtains a whole
hierarchy of block lattices $\Lambda_0, \Lambda_1, \dots, \Lambda_K$
with increasing lattice spacing (see figure \ref{blocking}).
This hierarchy of lattices is called multigrid.

Let us denote block lattice points in $\Lambda_k$ by $x'$.
Block spins $\Phi_{x'}$ are defined on block lattices
$\Lambda_k$. They are averages of the fundamental field $\ph_x$
over blocks of side length $L_B=l^k$:
\be\label{average}
\Phi_{x'} = L^{(d-2)/2}_B \, L^{-d}_B  \sum_{x \in x'} \ph_x \, .
\ee
The sum is over all points $x$ in the block $x'$.
The $L_B$-dependent factor in front of the average
comes from the fact that the corresponding dimensionful
block spins are measured in units of the
block lattice spacing:
A scalar field $\ph(x)$ in $d$ dimensions has canonical
dimension $(2-d)/2$. Thus $\ph(x) = a^{(2-d)/2}\ph_x$, where
$a$ denotes the fundamental lattice spacing. Now measure the
dimensionful block spin $\Phi(x')$ in units of the
block lattice spacing $a'$:
$\Phi(x') = a'^{(2-d)/2}\Phi_{x'}$ , with $a' = a L_B$.
If we average in a natural way
$\Phi(x')= L_B^{-d}\sum_{x \in x'}\phi(x)$ and return to
dimensionless variables, we obtain eq.\ (\ref{average}).

%%%%%%%%%%%%%%%%%%%%%%%%%%%%%%%%%%%%%%%%%%%%%%%%%%%%%%%%%%%%%%%%%%%
\begin{figure}
\begin{center}
%%%%%%%%%%%%%%%%%%%%%%%%%%%%%%%%%%%%%%%%%%%%%%%%%%%%%%%%%%%%%%%%%%%
\begin{picture}(120,40)(0,0)
\put(10, 10){\vector(1,0){80} }
\multiput(20,9)(10,0){7}{\line(0,1){2}}
\put(10, 10){\vector(0,1){25} }
\put(2, 25){\makebox(10,10){$\psi_x$}}
\put(80, 2){\makebox(10,10){$x$}}
\put(40,5){\vector(-1,0){15} }
\put(50,5){\vector(1,0){15} }
\put(40, 0){\makebox(10,10){$L_B$}}
\thicklines
\put(25,10){\line(0,1){7.5} }
\put(25,17.5){\line(1,0){40} }
\put(65,10){\line(0,1){7.5} }
\put(25,10){\line(-1,0){15} }
\put(65,10){\line(1,0){24} }
\end{picture}

\begin{picture}(120,40)(0,0)
\put(10, 10){\vector(1,0){80} }
\multiput(20,9)(10,0){7}{\line(0,1){2}}
\put(10, 10){\vector(0,1){25} }
\put(2, 25){\makebox(10,10){$\psi_x$}}
\put(80, 2){\makebox(10,10){$x$}}
\put(40,5){\vector(-1,0){15} }
\put(50,5){\vector(1,0){15} }
\put(40, 0){\makebox(10,10){$L_B$}}
\thicklines
\put(20,10){\line(2,1){25} }
\put(70,10){\line(-2,1){25} }
\put(20,10){\line(-1,0){10} }
\put(70,10){\line(1,0){19} }
\end{picture}
\caption[dummy]{\label{kernels} \sl
                Coarse-to-fine interpolation kernels $\psi$
                in one dimension:
                Top: piecewise constant kernel,
                bottom: piecewise linear kernel} 
\end{center}
\end{figure}
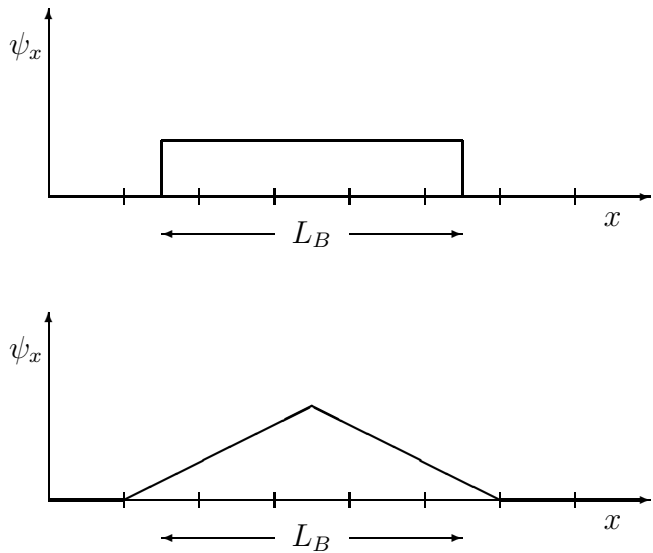
%%%%%%%%%%%%%%%%%%%%%%%%%%%%%%%%%%%%%%%%%%%%%%%%%%%%%%%%%%%%%%%%%%%

A nonlocal change of the configuration $\ph$ consists
of a shift
\be \label{nonlocal}
\ph_x \rightarrow \ph_x + s \, \psi_x \, .
\ee
$s$ is a real parameter, and the
``coarse-to-fine interpolation kernel'' (or shape function)
$\psi_x$ determines the shape of the nonlocal change. 
$\psi$ is
normalized according to
\be\label{normpsi}
L^{-d}_B \sum_{x \in x'} \psi_x = L^{(2-d)/2}_B \delta_{x',x_o'} \, .
\ee
Note that by the nonlocal change (\ref{nonlocal}), the block spin
is moved as $\Phi_{x'} \rightarrow \Phi_{x'} + s $
for $x'=x_o'$, and remains unchanged on the other blocks.
The simplest choice of the kernel
$\psi$ that obeys the constraint
(\ref{normpsi}) is a piecewise constant
kernel: $\psi_x = L^{(2-d)/2}_B$, if $x \in x_o'$, and $0$ otherwise.
Other kernels are smooth and thus avoid
large energy costs from the block boundaries. 
A systematic study of different kernels will be given in section
\ref{SECkernels}.
Two one dimensional examples for $\psi$ are given in figure \ref{kernels}. 

\subsection{The average Metropolis acceptance rate $\Omega(s)$}

The $s$-dependent Metropolis acceptance rate for such
proposals is given by
\be\label{omega}
\Omega(s) = \bigl<
\min \lbrack 1 , \exp( - \Delta \calH) \rbrack \bigr> \, .
\ee
Here, $\bigl< (.)\bigr>$ denotes the expectation value in the
system defined by eq.\ (\ref{PARTI}). Furthermore,
\be
\Delta \calH = \calH(\ph + s\psi)-\calH(\ph) \, .
\ee
$\Omega(s)$ can be interpreted as the acceptance rate for
shifting block spins by an amount of $s$, averaged over
a sequence of configurations generated by a Monte Carlo simulation.
As a static observable, 
$\Omega(s)$ is independent of the algorithm that we 
use to compute it.
$\Omega(s)$ is a useful quantity when one wants to know 
whether updates with increasing nonlocality (i.e.\ increasing
block size $L_B$) can be performed in an efficient way. Of course,
different choices of the kernel $\psi$ result in different acceptance
rates.

In actual Monte Carlo simulations, $s$ is not fixed.  In the same way as
in the local Metropolis algorithm, $s$ is a random number distributed
according to some a priori probability density.
If we choose $s$ to be
uniformly distributed on the interval $[-\varepsilon,\varepsilon]$, the
integrated acceptance rate $P_{acc}$ (as customarily measured in Monte
Carlo simulations) is obtained by averaging $\Omega(s)$ as follows:
\be \label{integrated_acc}
P_{acc}(\varepsilon)=\frac{1}{2\varepsilon}
                 \int_{-\varepsilon}^{\varepsilon} ds\,\Omega(s)  \ \ .
\ee
It turns out to be a good rule to adjust
the maximum Metropolis step size $\varepsilon$ such that
\mbox{$P_{acc}(\varepsilon) \approx 0.5$.}

\subsection{Unigrid versus recursive multigrid}

We consider every algorithm that updates stochastic
variables on a hierarchy of length scales as multigrid Monte Carlo
algorithm.  However, there are two different classes of multigrid
algorithms:  multigrid algorithms in a unigrid implementation
and ``recursive'' multigrid algorithms.

In the unigrid formulation one considers nonlocal updates of the
form (\ref{nonlocal}). Updates on the various layers
of the multigrid are formulated on the level of the finest
lattice $\Lambda_0$.
There is no explicit reference to block spin variables $\Phi$ defined on
coarser layers $\Lambda_k$ with $k > 0$.
In addition, unigrid also refers to a
computational scheme: Nonlocal updates are performed directly 
in terms of the variables on the 
finest grid $\Lambda_0$ in practical simulations.
An example for a unigrid implementation is explained for 
two dimensional nonabelian gauge
theory in section~\ref{SECSU22sim}.

In contrast, the recursive multigrid formulation consists of
recursively calculating conditional Hamiltonians that depend on the block
spin variables $\Phi$ on coarser layers $\Lambda_k$.  This formulation
is possible if the conditional Hamiltonians are of the same type or
similar to the Hamiltonian on the finest lattice.  Then, the conditional
probabilities used for the updating on the $k$-th layer can be computed
without always going back to the finest level $\Lambda_0$. 
Therefore, an
recursive multigrid implementation reduces the computational work on
the coarser layers (see the work estimates below).
At least in free field theory, a recursive multigrid implementation
with smooth interpolation
is possible using $9$-point prolongation kernels 
(a special case of piecewise linear kernels as shown in figure \ref{kernels})
in two dimensions
and generalizations thereof in higher dimensions \cite{brandt,sokalrev}.
Generally, a recursive multigrid implementation for interacting
models is only feasible in special cases with piecewise constant
kernels.
Examples for such implementations are given for the Sine Gordon model
in section \ref{SECsg} and in the four dimensional nonabelian gauge
theory in section \ref{SECSU24sim} below.

An algorithm formulated in the recursive multigrid style can
always be translated to the unigrid language (that is the way we are going
to use the unigrid formulation for the analysis of
multigrid algorithms). The reverse is not true, since not all
nonlocal changes of the field
on the finest lattice can be
interpreted as updates of a single block spin variable of a recursive
multigrid.
As an example, one can use stochastically overlapping
blocks in the unigrid style
by translating the fields by a randomly chosen distance
\cite{hmsun}.

If we formulate our kinematical analysis in the unigrid
language we nevertheless can include all algorithms formulated in the
recursive multigrid style.

\subsection{Multigrid cycles}

%%%%%%%%%%%%%%%%%%%%%%%%%%%%%%%%%%%%%%%%%%%%%%%%%%%%%%%%%%%%%%%%%%%%%%%%%%%%%%
\begin{figure}[htb]
\addtolength{\unitlength}{-0.5mm}
\begin{center} \begin{picture}(270,100)(10,0)
%
% Lattice-Names
%
\thicklines
\put(10,65){\makebox(10,10){$\Lambda^0$}}
\put(10,45){\makebox(10,10){$\Lambda^1$}}
\put(10,25){\makebox(10,10){$\Lambda^2$}}
\put(10, 5){\makebox(10,10){$\Lambda^3$}}
\put(270,65){\makebox(10,10){$L_B = 1$}}
\put(270,45){\makebox(10,10){$L_B = 2$}}
\put(270,25){\makebox(10,10){$L_B = 4$}}
\put(270, 5){\makebox(10,10){$L_B = 8$}}
%
% V-cycle
%
\put(30,70){\line(1,-2){30}}
\put(90,70){\line(-1,-2){30}}
\multiput(30,70)(10,-20){4}{\circle*{3}}
\multiput(90,70)(-10,-20){4}{\circle*{3}}
%
% W-cycle
%
\put(140,10){\line(-1,2){30}}
\put(140,10){\line( 1,2){10}}
\put(160,10){\line( 1,2){20}}
\put(160,10){\line(-1,2){10}}
\put(200,10){\line(-1,2){20}}
\put(200,10){\line( 1,2){10}}
\put(220,10){\line( 1,2){30}}
\put(220,10){\line(-1,2){10}}
\multiput(140,10)(-10,20){4}{\circle*{3}}
\put(150,30){\circle*{3}}
\multiput(160,10)( 10,20){3}{\circle*{3}}
\multiput(200,10)(-10,20){2}{\circle*{3}}
\put(210,30){\circle*{3}}
\multiput(220,10)( 10,20){4}{\circle*{3}}
\end{picture} \end{center}
\caption{\sl V-cycle ($\gamma=1$) and W-cycle ($\gamma=2$)\label{cycles}}
\end{figure}
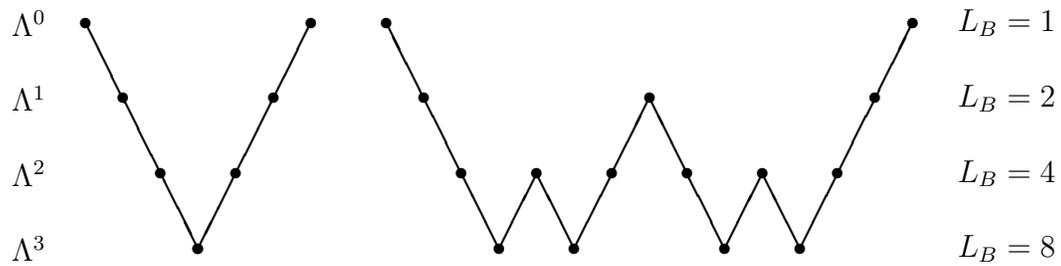
\addtolength{\unitlength}{0.5mm}
%%%%%%%%%%%%%%%%%%%%%%%%%%%%%%%%%%%%%%%%%%%%%%%%%%%%%%%%%%%%%%%%%%%%%%%%%%%%%%

The sequence of sweeps through the different layers $\Lambda_k$ of the
multigrid is organized in a periodic scheme called cycle
\cite{koelnporz}.
The most important cycles are illustrated in figure \ref{cycles}.  
The simplest scheme is the V-cycle:  The sequence of
layers visited in turn is $\Lambda_0,\Lambda_1,\ldots \Lambda_K,
\Lambda_{K-1}\ldots \Lambda_1$.  More general cycles are characterized
by the cycle control parameter $\gamma$.  The rule is that from an
intermediate layer $\Lambda_k$ one proceeds $\gamma$ times to the next
coarser layer $\Lambda_{k+1}$ before going back to the next finer layer
$\Lambda_{k-1}$.  A cycle control parameter $\gamma > 1$ samples coarser
layers more often than finer layers.  With $\gamma = 1$ we obtain the
V-cycle.  $\gamma = 2$ yields the W-cycle that is frequently used with
piecewise constant kernels.

The computational work estimates for the different cycles are as follows
\cite{sokalrev}: The work for a recursive multigrid cycle is
$\sim L^d$
if $\gamma < l^d$, where $L$ denotes the lattice size
and $l$ is the coarsening factor.
The work for a unigrid cycle is $\sim L^d \log L$ if $\gamma=1$,
and $\sim L^{d +\log_l \gamma}$ if $\gamma > 1$. 

If one wants the computational work in the unigrid style
not to exceed (up to a logarithm) an amount
proportional to the volume $L^d$ of the lattice,
one is restricted to use a V-cycle.
Simulations with $\gamma > 1$ (e.g.\ a W-cycle)
can only be performed in the recursive multigrid style.

\clearpage

\section{An approximation formula for $\Omega(s)$}

\label{SECapprox}

\setcounter{equation}{0}

In this section we shall derive an approximate formula
for the quantity $\Omega(s)$ defined in (\ref{omega}).
We can write $\Omega(s)$ as
\be\label{intrep}
\Omega(s)= \int du \, \min(1, \mye^{-u}) \int \frac{dp}{2\pi}
\, \mye^{-ipu}
\, \bigl< \mye^{ip \Delta \calH} \bigr> \  .
\ee
Let us assume that the probability distribution of $\Delta \calH$
is approximately Gaussian. We parameterize this
distribution as follows:
\be\label{itsgaussian}
\mbox{dprob}(\Delta\calH) \propto
d \Delta\calH \,
\exp \bigl(- \frac{1}{4h_2} (\Delta\calH - h_1)^2 \bigr) \  ,
\ee
with $h_1 = \EW{\Delta\calH}$ and $h_2 = \half \bigl(
\EW{\Delta\calH^2} - \EW{\Delta\calH}^2 \bigr)$.
Then
\be \label{gaussew}
\EW{  \exp(ip \delH)} \approx
\exp(ih_1 p - h_2 p^2 )\  .
\ee
We now show that from the assumption that the probability 
distribution of $\Delta \calH$
is Gaussian we can derive that $h_1 = h_2$. 
The starting point is the identity
\be
1 = 
\frac{1}{Z}\int \prod_{x \in \Lambda_0} d\ph_x 
\, \exp(-\calH(\ph)) \ .
\ee
If we use the translational invariance of the
measure $\prod_{x \in \Lambda_0} d\ph_x =
\prod_{x \in \Lambda_0} d(\ph_x + s\psi_x)$ we get
\be
1 = \frac{1}{Z}\int \prod_{x \in \Lambda_0} d\ph_x \,  
\exp(-\calH(\ph+s\psi))
= \frac{1}{Z}\int \prod_{x \in \Lambda_0} d\ph_x \, 
\exp(-\calH(\ph+s\psi)+\calH(\ph)-\calH(\ph)) \ .
\ee
With the definition $\Delta\calH =
\calH(\ph+s\psi)-\calH(\ph)$ we obtain
\be\label{emdeltah}
1 
= \frac{1}{Z}\int \prod_{x \in \Lambda_0} d\ph_x \  
\exp(-\Delta\calH)\exp( -\calH(\ph))
= \left\langle
\exp(-\Delta\calH)
\right\rangle \ .
\ee
This identity is independent of the form of 
the Hamiltonian.
If we now set $p = i$ in eq.\ (\ref{gaussew}) and compare this 
with eq.\ (\ref{emdeltah}), we are led to
$h_1 = h_2$. Therefore eq.\ (\ref{gaussew}) reduces to
\be 
\EW{  \exp(ip \delH)} \approx
\exp[(ip - p^2 ) h_1]\  ,
\ee
and the integrations in eq.\ (\ref{intrep})
can be performed exactly since there are only
Gaussian integrals involved.
The result is
\be\label{formula}
\Omega(s) \approx
\erfc ( \half \sqrt{h_1} ) \  .
\ee
with
$
\erfc(x) = {2/\sqrt{\pi}} \int_x^{\infty} \!
dt \, \exp(-t^2)
$.
(For an analogous result in the context of
hybrid Monte Carlo see \cite{hybrid_cumul}.)

For free massless field theory with Hamiltonian
$\calH(\ph)= \half(\ph,-\Delta \ph)$, we get
$h_1 = \half\alpha\, s^2$ with $\alpha =(\psi,-\Delta\psi)$,
and our approximation formula becomes {\em exact}:
\be\label{omega_free}
\Omega(s) = \erfc \left(
\sqrt{\frac{\alpha}{8}}
 \, \vert s \vert \right) \, .
\ee
Eq. (\ref{omega_free}) can be checked directly by using
$\EW{\exp(ip \delH)} = \exp[(ip - p^2)h_1]$ in eq.\ (\ref{intrep}).
This relation is exact in free field theory.

\clearpage

\section{Coarse-to-fine interpolation}

\label{SECkernels}

\setcounter{equation}{0}

In this section we shall discuss several choices of the coarse-to-fine
interpolation kernels. In order to have a ``fair'' comparison,
all kernels $\psi$ will be normalized according to eq.\ (\ref{normpsi}).

\subsection{Optimal interpolation kernels}

In free massless field theory, the quantity $\alpha = (\psi,-\Delta\psi)$
characterizes the decrease of the acceptance rate $\Omega(s)$ as given
in (\ref{omega_free}) with increasing shift $s$.  Therefore it is natural
to minimize $\alpha$ in order to maximize $\Omega(s)$ for fixed $s$.

The optimal kernel $\psi^{exact}$ from the point of view of acceptance
rates can be defined as follows: minimize the quadratic form
\be\label{qform}
\alpha\,=\,(\psi,-\Delta \psi)
\ee
under the constraints that the average of $\psi$ over the
``central block'' $x_o'$ is given by $L_B^{(2-d)/2}$,
and its average over blocks $x' \neq x_o'$ vanishes:
\be
L^{-d}_B \sum_{x \in x'} \psi_x = L^{(2-d)/2}_B \delta_{x',x_o'}
\, \mbox{\ for all \ } x' \in \Lambda_k    \ \ .
\ee
This variational problem can be solved with the help of
Fourier methods. The result is
\be
\psi^{exact}_x = L_B^{{(2+d)}/{2}} {\cal A}_{x,x_o'}\; ,
\ee
where ${\cal A}_{x,x_o'}$ denotes the Gaw\c{e}dzki-Kupiainen kernel
(see, e.g.\, \cite{cargmack}).  The use of this kernel leads to a
complete decoupling of the different layers of the multigrid.  This way
of interpolating from a coarser block lattice $\Lambda_k$ to the fine
lattice $\Lambda_0$ is well known in rigorous renormalization group
theory \cite{rigor}.  It is interesting that considerations about
optimizing acceptance rates in a stochastic multigrid procedure lead to
the same choice of the interpolation kernel.

Because $\psi^{exact}$ is nonvanishing on the whole lattice, it
is not convenient for numerical purposes.  For an attempt to change the
block spin $\Phi_{x_o'}$ on block $x_o'$ one has to calculate 
contributions to the
change of the Hamiltonian from all lattice points.  Therefore the
computational work for a single update is proportional to the volume.

\subsection{Truncation of the support of the optimal kernel}

\subsubsection{Kernels with support on a block and its nearest neighbors}

We define a ``truncated kernel'' $\psi^{trunc}$ by restricting the
support of $\psi$ on the block $x_o'$ and its nearest neighbor blocks
$y_o'$
\be
\psi_x^{trunc} = 0 \mbox{\ if \ } x \not \in x_o' \mbox{\ or \ } x
\not \in y_o',
\mbox{\ where \ } y_o'\, n.n. \, x_o'\;.
\ee
In other words, the Laplacian in eq.\ (\ref{qform}) is replaced by a
Laplacian $\Delta_D$ with Dirichlet boundary conditions on the boundary
of the support of $\psi$.  We again minimize $\alpha=(\psi,-\Delta_D
\psi)$ under the $2d+1$ constraints that the average of $\psi$ over the
blocks $x_o'$ and its nearest neighbor blocks is given.  This
minimization can be performed numerically by a relaxation procedure.  In
order to maintain the normalization condition, one always updates
simultaneously two spins residing in the same block, keeping their sum
fixed.  The $\psi^{trunc}$-kernels were used in a multigrid simulation
of the $\phi^4$ model in four dimensions \cite{phifour}.

\subsubsection{Kernels with support on a single block}

From a practical point of view, it is convenient to use kernels that
have support on a single block $x_o'$, i.e.\
\be
\psi_x = 0 \,\,\,\mbox{if}\;x \not \in x_o'\;.
\ee
We define a kernel $\psi^{min}$ with this property by minimizing
$\alpha=(\psi,-\Delta_{D,x_o'} \psi)$ under the constraint that the
average of $\psi$ over the block $x_o'$ is given.  The Laplacian with
Dirichlet boundary conditions on the boundary of $x_o'$ is defined as
follows:
 \be
  (\Delta_{D,x_o'} \phi )_x =
   \left[ -2d\,\phi_x + \strr \right.
 \sum_{\stackrel{{\scriptstyle y \, n.n. x}}{y \in
x_o'}}
   \left. \strr \phi_{y} \, \right]
    \mbox{\ \ \ for} \;x \in x_o'\ .
 \ee
$\psi^{min}$ can be calculated using an orthonormal set of
eigenfunctions of $\Delta_{D,x_o^{\prime}}$.

\subsection{Other kernels with support on the block}

We shall now discuss other kernels with support on the block
that are frequently used in the literature.

\subsubsection{Piecewise constant interpolation} 

Piecewise constant interpolation kernels are defined by
\be
 \psi^{const}_x=\left\{
\begin{array}{ll}
{L_B}^{(2-d)/2} &\mbox{for}\; x \in x_o' \\
0&\mbox{for}\; x \not \in x_o' \; .\\
\end{array}
\right.
\ee
This kernel has the advantage that for many models the conditional
Hamiltonians used for updating on coarse lattices are of the same type or
similar to the Hamiltonian on the finest lattice.  This means that the
conditional probabilities used for the updating on the $k$-th layer can
be computed without always going back to the finest level $\Lambda_o$.
Therefore, an recursive multigrid implementation with  
cycle control parameters $\gamma > 1$, e.g.\ a W-cycle, can be
used.

Piecewise constant kernels in a recursive multigrid implementation 
with a W-cycle will be used 
in the Sine Gordon model
in section \ref{SECsg} and in the four dimensional nonabelian gauge
theory in section \ref{SECSU24sim}.

\subsubsection{Piecewise linear interpolation} 

We consider the block
\be
 x_o' = \left\{ x \, \vert \,
 x^{\mu}\ \in \bigl\{1,2,3,\ldots,L_B \bigr\},\,\mu=1,\dots, d \right\}
\, .
\ee
The kernels for other blocks are simply obtained by translation.
For $L_B$ even, $\psi^{linear}$ is given by
\be
 \psi^{linear}_x={\cal N}\prod_{\mu = 1}^{d}
 \left\{\frac{L_B+1}{2}-\left|x^{\mu}-\frac{L_B+1}{2}\right|\right\}
\;\;\;\mbox{for}\; x \in x_o'  \ \ .
\ee
${\cal N}$ is a normalization constant.
Piecewise linear kernels are going to be used in the 
simulation of two dimensional nonabelian gauge
theory in section \ref{SECSU24sim}. There, smooth nonlocal
heat bath updates can be performed with linear interpolation.

\subsubsection{Ground state projection} 

Ground state projection kernels are defined as follows:
$\psi^{sine}$ is the eigenfunction corresponding to the
lowest eigenvalue of the negative Laplacian with Dirichlet
boundary conditions
$-\Delta_{D,x_o'}$ :
\be
 \psi^{sine}_x = \left\{
\begin{array}{ll}
{\displaystyle {\cal N}\prod_{\mu=1}^d
\sin(\frac{\pi}{ L_B+1 } x^{\mu})}
&\mbox{for}\; x \in x_o' \\
&\\
0&\mbox{for}\; x \not \in x_o' \ \ .\\
\end{array}\right.
\ee
Again, ${\cal N}$ denotes a normalization constant.
Note that this kernel is different from $\psi^{min}$.
A generalization of this kernel was introduced for scalar fields in the
background of nonabelian gauge fields in ref.\  \cite{thomask}.

\subsection{Results for $\alpha = (\psi, -\Delta \psi)$
in two and four dimensions}

%%%%%%%%%%%%%%%%%%%%%%%%%%%%%%%%%%%%%%%%%%%%%%%%%%%%%%%%%%%%%%%%%%%%%%%%%
\begin{table}
 \centering
 \caption[dummy]{\label{tab2} Results for $\alpha=(\psi,-\Delta\psi)$
   in 2 dimensions, $512^2$ lattice}
 \vspace{2ex}
\begin{tabular}{|c||c|c|c|c|c|c|c|c|}
\hline\str
 kernel & $L_B$=2   & $L_B$=4   & $L_B$=8  & $L_B$=16
        & $L_B$=32  & $L_B$=64  & $L_B$=128 & $L_B$=256
\\[.5ex] \hline \hline \str
 exact  & 6.899 & 8.902 & 9.705 & 9.941 & 10.00 & 10.02 & 10.18 & 13.11
\tabhline
 trunc  & 7.000 & 9.405 & 10.73 & 11.38 & 11.69 & 11.84 & 11.92 & --
\tabhline
 min    & 8.000 & 13.24 & 18.48 & 22.58 & 25.23 & 26.76 & 27.59 & 28.02
\tabhline
 sine   & 8.000 & 13.62 & 19.34 & 23.78 & 26.62 & 28.25 & 29.13 & 29.58
\tabhline
 linear & 8.000 & 15.80 & 24.58 & 31.84 & 36.68 & 39.51 & 41.05 & 41.84
\tabhline
 const  & 8.000 & 16.00 & 32.00 & 64.00 & 128.0 & 256.0 & 512.0 & 1024
\\[.3ex] \hline
\end{tabular} \end{table}
%%%%%%%%%%%%%%%%%%%%%%%%%%%%%%%%%%%%%%%%%%%%%%%%%%%%%%%%%%%%%%%%%%%%%%%%%

The results for the quantities $\alpha = (\psi,-\Delta\psi)$ for
different kernels in two dimensions are presented in table \ref{tab2}.
We used a $512^2$ lattice ($\psi^{exact}$ depends on the lattice
size).  The different kernels are ordered according to increasing
value of $\alpha$.

The values of $\alpha^{exact}$ and $\alpha^{trunc}$ are close together.
This shows that the truncation of the support of $\psi$ to the block and
its nearest neighbor blocks is a good approximation
to $\psi^{exact}$ (in the sense of acceptance rates).
The
value of $\alpha^{exact}$ for $L_B=256$ is remarkably higher than on
smaller blocks.  This is a finite size effect because the block lattice
consists only of $2^2$ points.  Since the nearest neighbors overlap on a
$2^2$ lattice, no result for $\alpha^{trunc}$ is quoted for $L_B=256$.
The values of $\alpha$ for the smooth kernels with support on the block
$\psi^{min}, \psi^{sine}$ and $\psi^{linear}$ are of the same magnitude.
We can see that $\psi^{sine}$ is almost as good as the optimal $\psi =
\psi^{min}$.
In contrast to all smooth kernels, $\alpha^{const}$ grows linear
in $L_B$.

%%%%%%%%%%%%%%%%%%%%%%%%%%%%%%%%%%%%%%%%%%%%%%%%%%%%%%%%%%%%%%%%%%%%%%%%%
\begin{table}
 \centering
 \caption[dummy]{\label{tab1} Results for $\alpha=(\psi,-\Delta\psi)$
   in 4 dimensions, $64^4$ lattice}
 \vspace{2ex}
\begin{tabular}{|c||c|c|c|c|c|}
\hline\str
 kernel & $L_B$=2 & $L_B$=4 & $L_B$=8 & $L_B$=16 & $L_B$=32
\\[.5ex] \hline \hline \str
 exact  & 14.48 & 20.38 & 23.48 & 24.71 & 30.62
\tabhline
 trunc  & 14.67 & 21.61 & 26.54 & 29.26 & --
\tabhline
 min    & 16.00 & 27.72 & 41.56 & 54.18 & 63.33
\tabhline
 sine   & 16.00 & 30.37 & 48.46 & 64.85 & 76.44
\tabhline
 linear & 16.00 & 39.02 & 70.78 & 101.0 & 122.9
\tabhline
 const  & 16.00 & 32.00 & 64.00 & 128.0 & 256.0
\\[.3ex] \hline
\end{tabular} \end{table}
%%%%%%%%%%%%%%%%%%%%%%%%%%%%%%%%%%%%%%%%%%%%%%%%%%%%%%%%%%%%%%%%%%%%%%%%%

The results for different kernels in four dimensions are presented in
table \ref{tab1}.  Here we used a $64^4$ lattice.  In principle, the
$\alpha$'s behave as in two dimensions.  The values of $\alpha^{linear}$
for small blocks are higher than $\alpha^{const}$.  The pyramids of the
piecewise linear kernels have a lot of edges in four dimensions which
lead to high costs in the kinetic energy.

The $L_B$-dependence of the $\alpha$'s in $d$ dimensions is
\be
\begin{array}{lcll}
\alpha &=& 2 d L_B \quad &\mbox{for piecewise constant kernels} \, ,
\nonumber \\
\alpha &\rellow{\longrightarrow}{L_B >\!\!> 1}&
const \quad &\mbox{for smooth kernels} \, .
\end{array}
\ee
As an example, the expression for $\alpha^{sine}$ in $d$ dimensions
is
\be
\alpha^{sine}=L_B^{2+d}(L_B+1)^d2^{d+2}d
\sin^{4d+2}\left[\frac{\pi}{2(L_B+1)}\right]
\sin^{-2d}\left(\frac{\pi}{L_B+1}\right) \ \ .
\ee
For large block sizes we find
\be
\alpha^{sine}\rellow{\longrightarrow}{L_B >\!\!> 1}
d\,\frac{\pi^{2d+2}}{2^{3d}}\,=\,const  \ \ .
\ee
From table \ref{tab2} we observe that in two dimensions $\alpha$
becomes
almost independent of $L_B$ for the smooth kernels if the block size is
larger than 16.  In four dimensions (table \ref{tab1}), we find
$\alpha(L_B) \sim const$ only for $\alpha^{exact}$.  The other
$\alpha$'s for the smooth kernels have not become independent of $L_B$
for the block sizes studied.

\clearpage

\section{Acceptance rates in free field theory}

\label{SECfree}

\setcounter{equation}{0}

In this section we discuss the scale dependence of Metropolis
acceptance rates 
of multigrid Monte Carlo algorithms in free field theory.
The kinematical behavior can be related to the dynamical 
critical behavior of the algorithms.
This will be the starting point for the theoretical analysis
of interacting models in section 
\ref{SECappl}. In free field theory CSD is eliminated 
by a multigrid algorithm.

\subsection{Massless free field theory}

Recall the exact result (\ref{omega_free})
$$
\Omega(s) = \erfc \left(
\sqrt{\frac{\alpha}{8}}
 \, \vert s \vert \right) \, 
$$
in massless free field theory.
$\Omega(s)$ is only a function of the product $\alpha s^2$.
If we increase the block size $L_B$,
the quantity $\alpha$ increases as a function $\alpha(L_B)$ of 
the block size.
In order to keep $\Omega(s)$ fixed when $L_B$ is increased,
we have to rescale the changes $s$ like 
$1/\sqrt{\alpha(L_B)}$.
As a consequence, to maintain a constant
acceptance rate in massless free field theory,
$s$ has to be scaled
down like $1/\sqrt{L_B}$ for piecewise constant kernels, whereas for
smooth kernels the acceptance rates for large $L_B$
do not depend on the block size.

Note that this behavior of the acceptance rates for large $L_B$ is not
yet reached in four dimensions for the block sizes studied (except for
$\psi^{exact}$).  See also the discussion of the
Metropolis step size below. 

Let us relate the scale dependence of the Metropolis amplitude $s$
with the dynamical critical behavior.
For smooth kernels we have 
$s \sim const$ for large $L_B$. This indicates that the fluctuations 
generated by the algorithm are of the same size on all length scales.
Multigrid simulations of free massless field theory with smooth interpolation
and a V-cycle yield $z = 0$ \cite{sokalothers,linn}.

For piecewise constant kernels we have to scale down the 
Metropolis amplitude $s$
like $s~\sim~1/\sqrt{L_B}$. 
In simulations with piecewise constant
interpolation and a V-cycle, $z = 1$ was found \cite{sokalunp,linn,mikeska}.
At least for free field theory, this disadvantage of the piecewise
constant kernels can be compensated for by using a W-cycle instead of a
V-cycle, resulting in $z = 0$ \cite{sokalunp,mikeska,mikeskapriv}. 

Smooth kernels in a unigrid implementation 
can be used only in V-cycle algorithms.
An exception are $9$-point prolongation kernels in
two dimensions and generalizations thereof in higher dimensions.
They can also be used in a recursive multigrid implementation
with a W-cycle, at least in free field theory
(cf.\ section \ref{SECunigrid}).

\begin{figure}[htbp]
\setlength{\unitlength}{1mm}
 \begin{center}
   \begin{minipage}[t]{145mm}
      % GEP-File:  GEP.EPS2
      \begin{picture}(145,95)(0,-2)
         \hspace*{5mm} 
         \epsfig{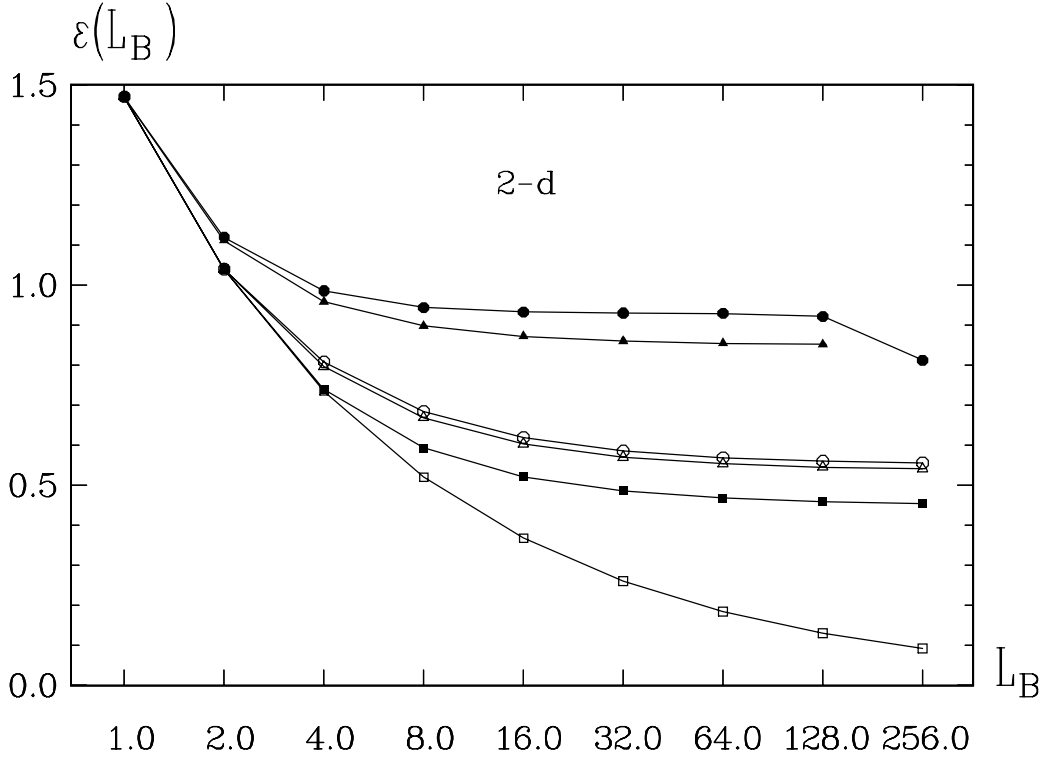}
      \end{picture}
      % GEP-File:  GEP.EPS4
      \begin{picture}(145,105)(0,-2)
         \hspace*{5mm} 
         \epsfig{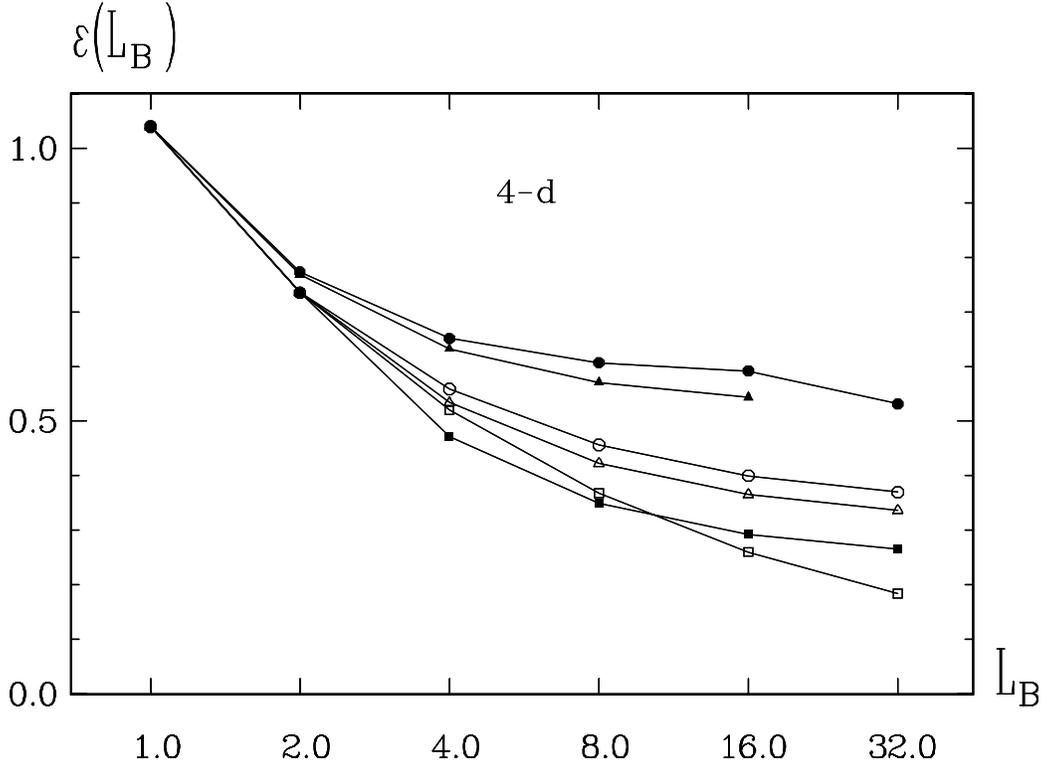}
         \hspace*{-5mm}
      \end{picture}
      \vspace{-7mm} 
      \caption[ Metropolis step sizes in two and four Dimensions]
              {\label{EPS}
               \sl Top: Metropolis step sizes $\varepsilon(L_B)$
                   for massless free field theory
                         in two dimensions, $512^2$-lattice.
                   Bottom: $\varepsilon(L_B)$
                   for massless free field theory
                   in four dimensions, $64^4$-lattice.
                   $\varepsilon(L_B)$ is choosen in such a way that
                   always $P_{acc} = 0.5$ holds.
                   Symbols:
                   full circles: $\psi^{exact}$,
                   full triangles: $\psi^{trunc}$,
                   empty circles: $\psi^{min}$,
                   empty triangles: $\psi^{sine}$,
                   full squares: $\psi^{linear}$,
                   empty squares: $\psi^{const}$.
                   Lines are only drawn to guide the eye.}
   \end{minipage}
 \end{center}
\end{figure}

\subsection{Scale dependence of the Metropolis step size}

We now illustrate what this rescaling of $s$ means for the
Metropolis step size $\varepsilon$ in an actual multigrid Monte Carlo
simulation.  Let us inspect the integrated acceptance 
probability defined in
eq.\ (\ref{integrated_acc}).  If we insert the exact result
(\ref{omega_free}) for massless free field theory
\be
P_{acc}(\varepsilon)=
\frac{1}{2\varepsilon}\int_{-\varepsilon}^{\varepsilon} ds\,
\Omega(s) =
\frac{1}{2\varepsilon}\int_{-\varepsilon}^{\varepsilon} ds\,
\erfc \left( \sqrt{\frac{\alpha}{8}}
 \, \vert s \vert \right) \, ,
\ee
we get
\be
P_{acc}(\varepsilon)
=\mbox{erfc}\left(\sqrt{\frac{\alpha}{8}}\varepsilon\right)
+\frac{1}{\sqrt{\frac{\pi\alpha}{8}}\varepsilon}
\left[\str 1-e^{-\frac{\alpha}{8}\varepsilon^2 }\right]\ \ .
\ee
$P_{acc}$ is only a function of the product $\alpha
\varepsilon^2$.  In order to keep $P_{acc}$ fixed (to, e.g.\ 50
percent) we
have to rescale $\varepsilon(L_B)$ like $1/\sqrt{\alpha(L_B)}$, exactly in
the same way as we had to rescale $s$ to keep $\Omega(s)$ fixed.
This $L_B$-dependence is plotted in figure \ref{EPS}.

\subsection{Intuitive random walk picture of the W-cycle}

In free field theory it is possible to overcome the 
the disadvantage $\varepsilon(L_B) \sim 1/\sqrt{L_B}$ of the piecewise
constant kernels compared to $\varepsilon(L_B) \sim const$ for smooth kernels
by using a W-cycle instead of a
V-cycle. This can be made plausible by a
simple random walk argument.

A constant accumulated
step size on all length scales can be achieved in the following way:
For step sizes scaling down like
$\varepsilon(L_B) \sim 1/\sqrt{L_B}$ and a coarsening by a factor of two,
the Metropolis step size on a next coarser lattice
is too small by a factor of $1/\sqrt{2}$.
If we assume that subsequent update steps within a 
multigrid cycle are statistically independent and that these 
steps accumulate in a random-walk like way,
we can expect to compensate for this decrease by increasing
the number of updates on the next coarser grid by a factor of two.
This can be achieved by a higher cycle with  
cycle control parameter
$\gamma = 2$.
Since higher cycles are defined recursively by going 
from an intermediate block lattice
$\gamma$ times to the next coarser lattice before going back
to the next finer lattice, 
$\gamma$ times more updates on each coarser
lattice are performed.

In free field theory, the assumptions of this argument seem to be correct,
and the random walk argumentation
can explain that for piecewise constant kernels and a W-cycle
CSD is eliminated.

\subsection{Massive free field theory}

\label{SUBSECmassive}

We now discuss massive free field theory with Hamiltonian
$\calH(\ph)= \half(\ph,[-\Delta + m^2] \ph)$. We find
$h_1 =\half\alpha_m\, s^2 $, with $\alpha_m$ given by
\be
\alpha_m = (\psi,[-\Delta + m^2] \psi)
         = \alpha
+  m^2 \sum_{x \in \Lambda_0} \psi_x^2 \, .
\ee
Therefore the exact result is
$\Omega(s) = \erfc \left( \sqrt{\alpha_m/8}\, \vert s \vert \right)$.
A term $\sum_x \psi_x^2$ scales $\sim L_B^2 $ in arbitrary
dimensions.
For piecewise constant kernels
\be
\sum_{x \in \Lambda_0}
  \left(\psi^{const}_x \right)^2 \,
                                    = \, L_B^2 \ .
\ee
For $\psi^{sine}$-kernels  we find
\be
\sum_{x \in \Lambda_0}
                       \left(\psi^{sine}_x \right)^2
= 2^{d} L_B^{2+d}(L_B+1)^d
\sin^{4d}\left[\frac{\pi}{2(L_B+1)}\right]
\sin^{-2d}\left(\frac{\pi}{L_B+1}\right) \ \ ,
\ee
and for large block sizes
\be
\sum_{x \in \Lambda_0}\left(\psi^{sine}_x \right)^2
\rellow{\longrightarrow}{L_B >\!\!> 1}
\frac{\pi^{2d}}{2^{3d}} L_B^2 \ \ .
\ee
If the block size $L_B$ is smaller than the correlation
length $\xi = 1/m$, $h_1$ is still dominated by the kinetic term
$s^2 (\psi,-\Delta\psi)$,
and the discussion is the same as in the massless case.

As soon as the block size $L_B$ becomes larger than $\xi$, $h_1$ is
dominated by the ``mass'' term $s^2 m^2 \sum_x \psi^2_x \sim s^2 L_B^2$, and
$s$ has to be rescaled like $s \sim 1/L_B$ in order to maintain
constant acceptance rates.  Of course this is a dramatic decrease for
large block sizes compared to $s \sim const$ 
in the massless case (using smooth kernels).
Block spins on large blocks are essentially
``frozen''.  However this is not a problem for the performance of the
algorithm in massive free field theory: The effective probability
distribution for the block spins $\Phi$ is given by
$\exp(-\calH_{\mbox{\scriptsize eff}}(\Phi))$, where
$\calH_{\mbox{\scriptsize eff}}(\Phi)$ denotes the
effective Hamiltonian in the sense of the block spin renormalization
group \cite{schladming}.  The physical fluctuations of the block spins
are dictated by an effective mass term
\be
 m_{\mbox{\scriptsize eff}}^2 \sum_{x' \in \Lambda_k} \Phi_{x'}^2\,
\; \;\; \mbox{with}\;\;\; m_{\mbox{\scriptsize eff}}^2 \sim m^2
                             L_B^2  \, .
\ee
Thus, the algorithmic fluctuations (described by the mass term $m^2
\sum_x \psi_x^2 \sim m^2 L_B^2$) and the physical fluctuations
(described by the effective mass $\sim m^2 L_B^2$) behave similarly, and
the multigrid algorithm is able to create fluctuations just of the size
that is needed by the physics of the model.
Moreover there is no need to do updates at length scales
larger than $\xi$ in order to beat CSD.

In this sense, the discussed algorithmic mass term $m^2 \sum_x \psi^2_x$
is well behaved for free field theory, since it decreases with the
physical mass in the vicinity of the critical point. 
Stated differently, in massive free field theory the algorithmic
mass term scales with the physical mass.
As we shall see in
section \ref{SECappl}, for interacting models close to criticality, a
different scenario is possible.  There, it can happen that an
algorithmic mass term $\sim \sum_x \psi_x^2$ persists, whereas the
physical mass vanishes.  If this happens, the multigrid algorithm is
not able to produce the large critical fluctuations required by the
physics, and we can {\em not} expect that CSD will be eliminated.

\subsection{Behavior of a term $\sum_x \psi_x^4$}

The $L_B$-dependence of a term $\sum_x \psi_x^4$ will also be
needed in the study of the $\phi^4$ theory in section~\ref{SECappl}
below.
In $d$ dimensions such a term scales $\sim L_B^{4-d}$:
For piecewise constant kernels
\be
\sum_{x \in \Lambda_0}
                       \left(\psi^{const}_x \right)^4
                       \, = \, L_B^{4-d} \ ,
\ee
whereas using $\psi^{sine}$-kernels  we find
\be
\sum_{x \in \Lambda_0}
                       \left(\psi^{const}_x \right)^4
=6^d L_B^{4+2d}(L_B+1)^{d}
\sin^{8d}\left[\frac{\pi}{2(L_B+1)}\right]
\sin^{-4d}\left(\frac{\pi}{L_B+1}\right) \ \ .
\ee
In the limit of large block sizes this term behaves like
\be
\sum_{x \in \Lambda_0}
                       \left(\psi^{sine}_x \right)^4
\rellow{\longrightarrow}{L_B >\!\!> 1}
\left(\frac{3\pi^4}{128}\right)^d L_B^{4-d} \ \ .
\ee

\subsection{Degree of relevance}

In order to summarize the different large-$L_B$-behavior of local
operators in the kernel $\psi$, let us introduce the
{\em degree of relevance} in the sense of the perturbative
renormalization group:
The (superficial) degree $r$ of relevance of a local operator in
$\psi$ which is a polynomial of $m$ scalar fields
with $n$ derivatives is defined by $r=d+m(2-d)/2-n$.
This definition is valid for smooth kernels.
For large $L_B$, an operator with degree of relevance
$r$ behaves like $L_B^r$.
An operator is called relevant if $r > 0$.
As we have seen in the examples above, a mass term has $r=2$, and a
$\psi^4$-term has $r=4-d$.  A kinetic term $\alpha=(\psi,-\Delta \psi)$
has $r=0$ for smooth kernels.

The only difference for piecewise constant kernels is that a kinetic
term behaves like $\alpha=(\psi,-\Delta \psi) \propto L_B$.
  
\clearpage

\section{Acceptance rates for interacting models}
\label{SECappl}

\setcounter{equation}{0}

In this section, we shall apply formula (\ref{formula}) in the discussion
of multigrid procedures for various spin models in two
dimensions:  the Sine Gordon model, the XY model, the
single-component $\ph^4$ theory,
the $O(N)$ nonlinear $\sigma$-model and the $CP^{N-1}$ model.
The scale dependence of acceptance
rates for interacting models will be compared with the
kinematical behavior in
free field theory, where CSD is known to be eliminated
by a multigrid algorithm. 
A necessary criterion for the successful acceleration
of a simulation by a multigrid algorithm is formulated.

\subsection{The two dimensional Sine Gordon model}

\label{SUBSECsinegordon}

The two dimensional Sine Gordon model is defined by the
Hamiltonian
\be
\calH(\ph) = \frac1{2\beta} ( \ph, -\Delta \ph)
- \zeta \sum_x \cos 2\pi \ph_x  \, .
\ee

From the point of view of statistical mechanics, this system
can be considered
as a two dimensional surface in a periodic potential.
 The model exhibits a Kosterlitz-Thouless phase transition
at $\beta_c(\zeta)$.
In the limit of
vanishing fugacity $\zeta$, $\beta_c$ takes the value $2/\pi=0.6366\ldots$~.
For $\beta > \beta_c$ the model is in the massless (rough) phase. 
The fluctuations of the surface are given by the surface thickness
\begin{equation} \label{thickness}
  \sigma^2  = \left \langle (\phi_x - \overline{\phi})^2
              \right \rangle \ ,
\end{equation}
where $\overline{\phi}$ denotes the average of the field over the lattice.
In the massless (rough) phase, the surface thickness 
$\sigma^2$ scales with $\log L$ \cite{sos}.
In this phase,
the cosine-term of the Hamiltonian is irrelevant, 
and the flow of the effective Hamiltonian (in the sense of the
block spin renormalization group) converges to that of a massless free
field theory:  the long distance behavior of the theory is that of a
Gaussian model.  Since multigrid algorithms have proven to be efficient
in generating long wavelength Gaussian modes, one might naively conclude
that multigrid should be the right method to fight CSD
in the simulation of the Sine Gordon model in the massless phase.  But
this is not so:
  
The energy change of a nonlocal update 
$\phi_x \rightarrow \phi_x + s\psi_x$ is
$$
\Delta \calH = 
\calH(\ph+s\psi) - \calH(\ph) 
= 
\frac{s^2}{2\beta} ( \psi, -\Delta \psi)
+ \frac{s}{\beta} ( \psi, -\Delta \ph)
$$
\be
- \zeta \sum_x \left\{
\cos(2\pi \ph_x)\left[\cos(2\pi s\psi_x) -1\right]
-\sin(2\pi \ph_x)\sin(2\pi s\psi_x) 
\right\} \ .
\ee
The Hamiltonian is globally invariant under $\ph_x \rightarrow -\ph_x$.
Therefore, $\langle \ph_x \rangle = 0$ and 
$\langle \sin 2\pi\ph_x \rangle = 0$ on finite lattices.
Using this, we find for the average
energy change  $h_1 = \langle \Delta \calH \rangle$ the expression
\be\label{h1h2}
h_1
= \frac{\alpha}{2\beta} s^2 + \zeta C \sum_x \lbrack 1 - \cos(2\pi s\psi_x)
\rbrack \, , \ee
with $C = \langle \cos (2\pi\ph_x) \rangle$.
Recall that $h_1$ is the quantity that determines the
acceptance rates $\Omega(s)$ by eq.\ (\ref{formula}):
$$
\Omega(s) \approx \erfc ( \half \sqrt{h_1} ) \, .
$$

%------------------------------------------------------------------------
\begin{figure}[htbp]
 \begin{center}
\begin{minipage}[t]{145mm} 
% GEP-File:  GEP.OMSG
  \begin{picture}(145,90)(0,-2)
    \hspace*{5mm} 
    \epsfig{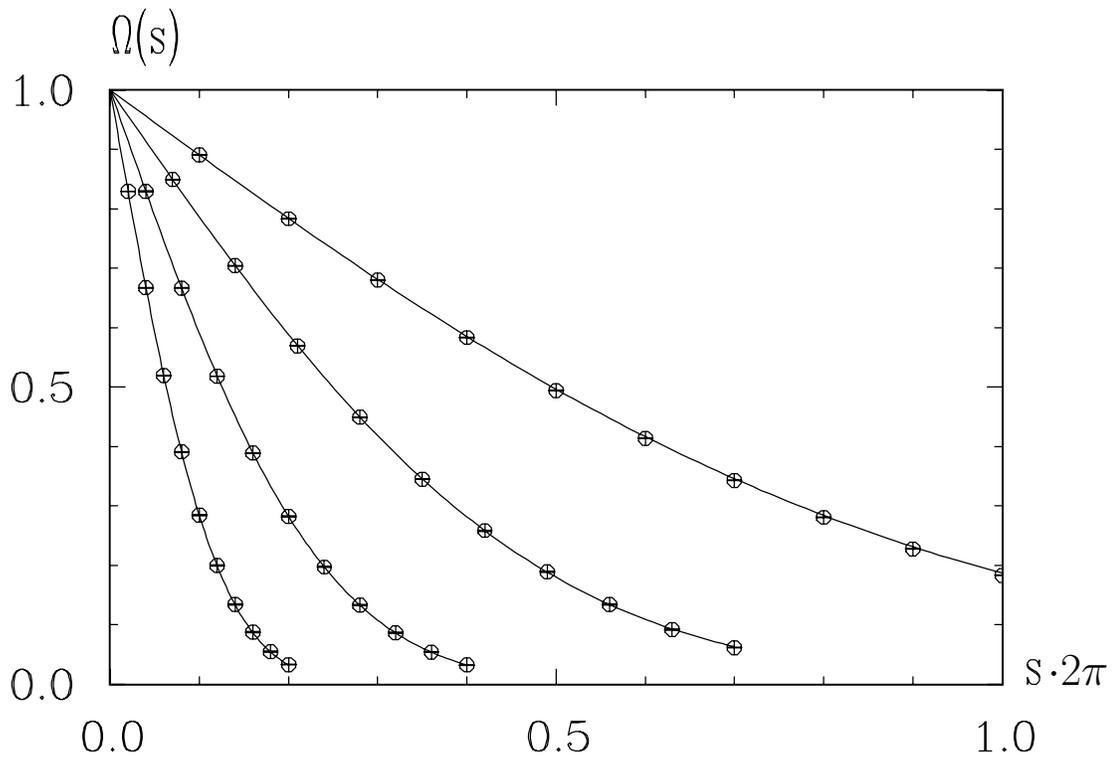}
    \put(-2,2.4){\makebox(10,10){ \Large $\cdot  2\pi$}}
  \end{picture}
  \caption[\Omega(s) in the 2d Sine Gordon model]
 {\label{OMSG}
\sl $\Omega(s)$ for piecewise constant kernels
in the two dimensional Sine
Gordon model, $\beta=1.0$, $\zeta=1.0$.  From top to bottom:
$L_B=4,8,16,32$ on a $16^2,32^2,64^2,128^2$ lattice, respectively.
Points with error bars:  Monte Carlo results, lines:  analytical
results  }
 \end{minipage}
  \end{center}
   \end{figure}
%************************************************************************

%
The essential point is that the second term in (\ref{h1h2}) is
proportional to the block volume $L^2_B$ for piecewise constant
{\em and} for smooth kernels (cf.\ the discussion in section
\ref{SECfree}).
This can be checked for small $s$ by expanding the cosine term in $s$.
One therefore has to face a dramatic decrease of acceptance when the
blocks become large, even for small fugacity $\zeta$.  A constant
acceptance rate is achieved only when the proposed update 
steps are scaled down
like $1/L_B$.  It is therefore unlikely that any multigrid algorithm
- based on nonlocal updates of the type discussed here - will
be successful for this model. This prediction will be verified by a 
multigrid Monte Carlo simulation of the two dimensional
Sine Gordon model in section \ref{SECsg}.

We demonstrate the validity of formula (\ref{formula}) (using a Monte
Carlo estimate for $C$) by comparing with Monte Carlo results at $\beta
= 1.0$, $\zeta=1.0$.  This point is in the massless phase.
There, the physical scale of the system is set by the linear
size $L$ of the lattice.
In figure~\ref{OMSG} we show both the numerical and analytical results for
$\Omega(s)$ for $L_B=4,8,16,32$ on lattices of size
$16^2,32^2,64^2 \mbox{and} 128^2$, respectively.

We tested the precision of our approximation formula for piecewise
constant kernels only.  However, we have no doubts that the quality of
the approximation is also very good for other interpolation kernels $\psi$.

\subsection{The two dimensional XY model}
\label{SUBSECxy}

We now discuss the
two dimensional XY model, defined by the partition function
\be \label{xy_part}
Z = \int \prod_x d\theta_x \, \exp \bigl(
\beta \sum_{<x,y>} \cos(\theta_x-\theta_y) \bigr) \, .
\ee
The sum in the exponent
is over all unordered pairs of nearest neighbors in the
lattice.
As the Sine Gordon model, the XY model has a massless (spin wave) phase
for $\beta > \beta_c$, and a massive (vortex) phase for $\beta < \beta_c$.  The
best available estimate for the critical coupling is $\beta_c =
1.1197(5)$~\cite{rough}.

Nonlocal updates are defined by
\be \label{xy_update}
\theta_x \rightarrow
\theta_x + s \psi_x \, ,
\ee
with $\psi$ obeying again the normalization
condition (\ref{normpsi}). To define a (linear) block spin, we
rewrite the partition function (\ref{xy_part}) in terms of
2-component unit vector spin variables $s_x$:
\be
Z = \int \prod_x \left( d^2 s_x \,  \delta(s_x^2-1) \right)
\exp \bigl( \beta \sum_{<x,y>} s_x \cdot s_y \bigr)
\ee
The block spins $S_{x'}$ are then defined as block averages
of the unit vectors $s_x$.

Note that the proposal (\ref{xy_update}) changes the block
spin by an amount $\approx s$ only when the spins inside the
block are sufficiently aligned. This will be the case in the spin wave
phase for large enough $\beta$. For smaller $\beta$, the
correct (or ``fair'') normalization of the kernels $\psi$ is
a subtle point. We believe, however, that our argument is
not affected by this in a qualitative way.

Let us inspect the energy change of the nonlocal update 
(\ref{xy_update}):
$$
\Delta \calH 
= 
-\beta \sum_{<x,y>} 
\left\{
\cos(\theta_x-\theta_y +s(\psi_x-\psi_y)) 
-\cos(\theta_x-\theta_y) 
\right\}
$$
\be
= -\beta \sum_{<x,y>}
\left\{
 \cos(\theta_x-\theta_y ) \left[\cos(s(\psi_x-\psi_y)) -1
\right]
 -\sin(\theta_x-\theta_y ) \sin(s(\psi_x-\psi_y)) 
\right\} \ .
\ee
Since the Hamiltonian is globally invariant under 
$\theta_x \rightarrow -\theta_x$, the expectation value
$\langle \sin(\theta_x-\theta_y ) \rangle$ vanishes on finite lattices.
Therefore,
the relevant quantity $h_1$ is 
\be
h_1 = \beta E \sum_{<x,y>}
\lbrack 1 - \cos( s( \psi_x - \psi_y) )\rbrack \, ,
\ee
with
$E=\langle \cos(\theta_x - \theta_y) \rangle $,
$x$ and $y$ nearest neighbors.
For piecewise constant kernels,
$h_1$ is proportional to $L_B$.
For smooth kernels $h_1$ will become independent of $L_B$ for large
enough blocks. For small $s$,
\be
h_1 \approx \half s^2 \beta E \sum_{<x,y>} (\psi_x -\psi_y)^2
= \half s^2 \beta E \alpha \, .
\ee
As above, $\alpha=(\psi,-\Delta \psi)$. This quantity becomes nearly
independent of $L_B$ already for $L_B$ larger than $16$
(cf.\ section \ref{SECkernels}).

From the point of view of acceptance rates the XY model therefore
behaves like massless free field theory.  A dynamical critical
exponent $z$ consistent with zero was observed in the massless phase
\cite{xy}. In this phase, the physically dominant excitations
are spin waves. 
From this comparison we conclude that the
analysis of the kinematics of multigrid algorithms
can tell us whether spin wave excitations can be accelerated by the 
algorithm.

In the massive phase the physically important excitations are vortices,
which are topologically nontrivial objects. 
In this phase multigrid Monte Carlo simulations
with piecewise constant kernels yielded $z \approx 1.4$ \cite{xy}.
This is an example for the fact that good acceptance rates are not 
sufficient to overcome CSD. 
The kinematical analysis is not sensitive to
topological problems.

\begin{figure}[htbp]
 \begin{center}
  \begin{minipage}[t]{145mm}
  % GEP-File:  GEP.OMXY
  \begin{picture}(145,90)(0,-2)
   \epsfig{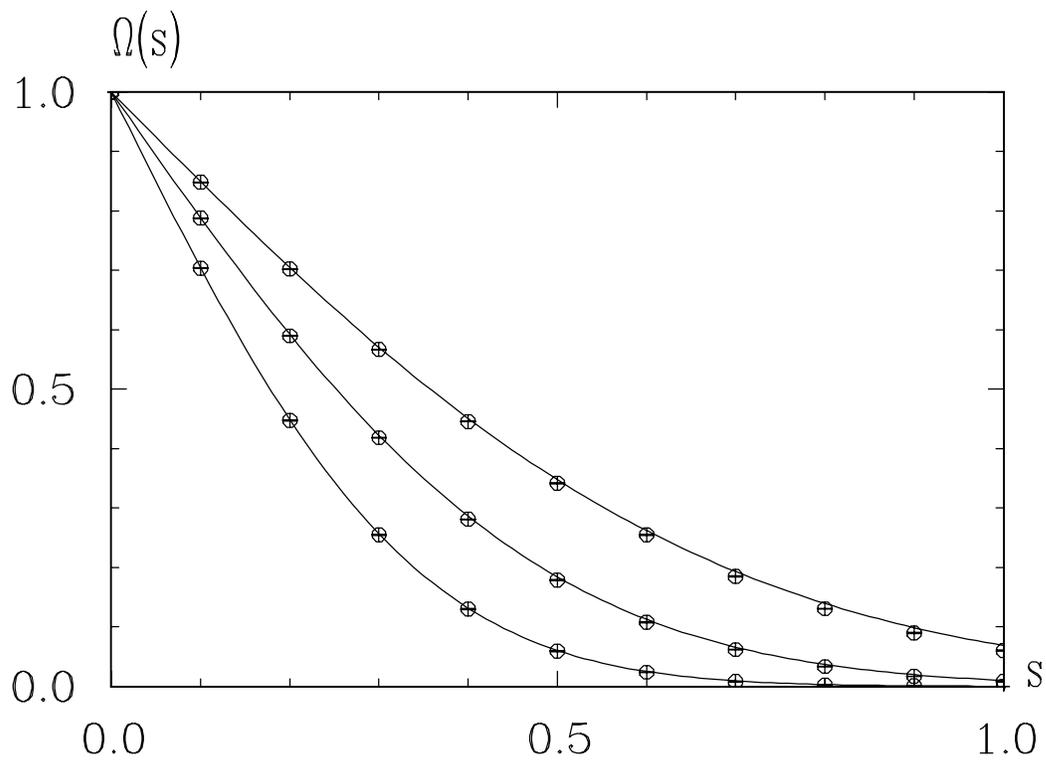}
  \end{picture}
      \caption[\Omega(s) in the 2d XY model]
              {\label{OMXY}
               \sl $\Omega(s)$ for piecewise constant
                   kernels in the two dimensional XY model,
                   $\beta=1.2$.
                   From top to bottom: $L_B=4,8,16$ on a
                   $16^2,32^2,64^2$ lattice, respectively.
                   Points with error bars: Monte Carlo results,
                   lines: analytical results}
   \end{minipage}
 \end{center}
\end{figure}
%************************************************************************

We again checked the accuracy of formula (\ref{formula}) by comparing
with Monte Carlo results at $\beta=1.2$ (which is in the massless
phase). 
The results are displayed in figure
\ref{OMXY}.
The only numerical input for the analytical formula was the
link expectation value $E$.  

\subsection{The $\phi^4$ theory in $d$ dimensions}

\label{SUBSECphi4}

Let us now turn to single-component
$d$-dimensional $\ph^4$ theory ($d=2,3,4$), defined
by the Hamiltonian
\be
\calH(\ph) = \half(\ph, -\Delta \ph)
           + \frac{m_o^2}2 \sum_x \ph_x^2
           + \frac{\lambda_o}{4!} \sum_x \ph_x^4 \, .
\ee
\begin{figure}[htbp]
 \begin{center}
   \begin{minipage}[t]{145mm}
      %GEP-File:  GEP.OMPHI44
      \begin{picture}(145,90)(0,-2)
        \hspace*{5mm} 
        \epsfig{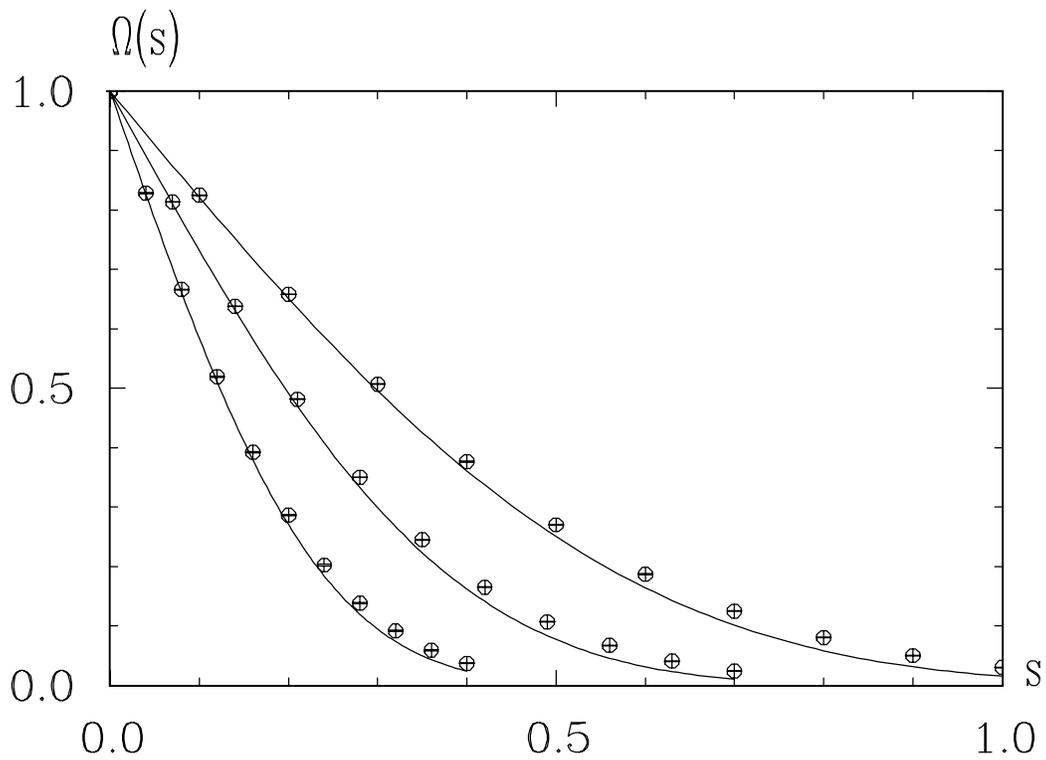}
      \end{picture}
      \caption[\Omega(s) in the  $\phi^4$ theory]
              {\label{OMPHI4}
               \sl $\Omega(s)$ for piecewise constant kernels
                   in the two dimensional $\phi^4$ theory,
                   $m_o^2=-0.56$, $\lambda_o=2.4$.
                   From top to bottom: $L_B=4,8,16$ on a
                   $16^2,32^2,64^2$ lattice, respectively.
                   Points with error bars: Monte Carlo results,
                   lines: analytical results}
   \end{minipage}
 \end{center}
\end{figure}

A nonlocal update $\ph_x \rightarrow \ph_x
 + s\psi_x$ leads to the 
energy change
$$
\Delta\calH
=\calH(\ph+s\psi)-\calH(\ph)
=       s^2 \half(\psi, -\Delta \psi)
             + s (\psi, -\Delta \ph)
$$
\be
           + \frac{m_o^2}2 \sum_x \left[ s^2\psi_x^2 + 2 s\psi_x \ph_x
             \right]
           + \frac{\lambda_o}{4!} \sum_x \left[
           s^4\psi_x^4 +4s^3\psi_x^3 \ph_x 
           +6s^2\psi_x^2\ph_x^2 +4s\psi_x \ph_x^3\right]  
\, .
\ee
Since $\calH(-\ph) = \calH(\ph)$, the expectation values
$\langle \phi_x^3 \rangle$ and
$\langle \phi_x \rangle$ vanish on finite lattices.
Therefore we find for $h_1 = \langle \Delta \calH \rangle$ 
\be
h_1 = s^2 \, \left\{
\half \alpha + \left[ \frac{m_o^2}{2} + \frac{\lambda_o}{4}
P \right] \sum_x  \psi_x^2 \right\}
+ s^4 \, \frac{\lambda_o}{4!} \sum_x \psi_x^4 \, ,
\ee
where $P=\langle \ph_x^2 \rangle$.  
Recall that $\sum_x \psi_x^2$
increases with $L^2_B$, independent of $d$, whereas $\sum_x \psi_x^4$
scales like $L^{4-d}_B$, for smooth and for piecewise constant kernels
(cf.\ the discussion of the different kernels
in section \ref{SECfree}).
We conclude that also in this model we have to face rapidly decreasing
acceptance rates when the blocks become large.  As in the case of the
Sine Gordon model, \mbox{$s$ has} to be rescaled like $1/L_B$ in order to
maintain constant acceptance rates.

Therefore there is little hope that multigrid algorithms of the type
considered here can overcome CSD in the
single-component $\ph^4$ model.  
In numerical simulations of two dimensional
$\ph^4$ theory, a dynamical critical behavior is found that is
consistent with $z \approx 2$ for piecewise constant and for smooth
kernels \cite{sokalprl,sokalrev,linn}.  In four dimensions, there is no
definite estimate for $z$ \cite{phifour}.

Figure \ref{OMPHI4} shows a comparison of Monte Carlo results for
two dimensional $\ph^4$ theory with the theoretical prediction based on
the numerical evaluation of $P$.  The simulations were performed in the
symmetric phase at $m_o^2 = - 0.56$ and $\lambda_o = 2.4$.  The
correlation length at this point is $\xi \approx 15$ \cite{linn}.

\bigskip

The quantitative comparison of 
the approximation formula (\ref{formula}) with numerical 
results for acceptance rates justified our approximations 
in three different models.
For the remainder of this section we 
assume that the approximation formula 
is valid. We now discuss the kinematical behavior
of two other spin models that are
asymptotically free in two dimensions. 

\subsection{The two dimensional $O(N)$ nonlinear $\sigma$-model}
\label{SUBSECON}

The two dimensional $O(N)$ nonlinear $\sigma$-model
is defined by the partition function
\be
Z = \int \prod_x \left( d^N s_x \,  \delta(s_x^2-1) \right)
\exp \bigl( \beta \sum_{<x,y>} s_x \cdot s_y \bigr) \ ,
\ee
where the $s_x$ 
are $N$-component real unit vector spin variables.
The case $N=2$ corresponds to the XY model.
For $N>2$ the $O(N)$ nonlinear $\sigma$-model
is in a disordered phase for all values of $\beta$.
In the limit $\beta \rightarrow \infty$,
the correlation length  
diverges exponentially in $\beta$.
The critical behavior of the model is described the
perturbative renormalization group which implies asymptotic freedom.

Nonlocal Metropolis updates can be defined 
in close analogy to the XY model by updating in
$U(1)$ subgroups of $O(N)$. 
Let us consider the special case where the updates
act on the first two components
of the $s_x$ variables by 
\be\label{update_ON}
s_x \ \rightarrow  \ s_{x}' = R_x s_x \ ,
\ee
with

\be
 s_{x} = \left(
\begin{array}{c}
s_x^{(1)} \\
s_x^{(2)} \\
s_x^{(3)} \\
\vdots \\
s_x^{(N)} \\
\end{array}
\right) \; , \; \;
 R_x = \left(
\begin{array}{ccccc}
\cos(s \psi_x) & -\sin(s \psi_x) & 0 & \cdots & 0\\
\sin(s \psi_x) &  \cos(s \psi_x) & 0 & \cdots & 0\\
0 &  0 & 1 &  & \\
\vdots & \vdots  &  & \ddots  & \\
0 &  0 &  &  & 1\\
\end{array} \right) 
\ .
\ee
Again $\psi$ obeys the normalization
condition (\ref{normpsi}).
Other directions of the $U(1)$ subgroup in $O(N)$ can be 
obtained by 
applying a randomly chosen global $O(N)$ rotation to
the configuration before updating according to
eq.\ (\ref{update_ON}).

The nonlocal updating proposal (\ref{update_ON}) changes
the Hamiltonian by
$$
\Delta \calH = -\beta \sum_{<x,y>} 
\left\{ (R_x s_x)\cdot (R_y s_y) -s_x\cdot s_y \right\}
\hspace{5.9cm}
$$
$$
= -\beta \sum_{<x,y>}
\left\{ s_x^{(1)} s_y^{(1)} \left[\cos(s(\psi_x-\psi_y)) -1
\right]
 + s_x^{(1)}  s_y^{(2)}  \sin(s(\psi_x-\psi_y)) \right.
$$
\be \hspace{3.2cm}
 - \left. s_x^{(2)}   s_y^{(1)}  \sin(s(\psi_x-\psi_y)) 
 + s_x^{(2)}  s_y^{(2)} 
\left[\cos(s(\psi_x-\psi_y)) -1\right] \right\} \ .
\ee
The Hamiltonian of the $O(N)$ model is invariant under
a global reflection of a fixed component
$s_x^{(i)} \rightarrow -s_x^{(i)}$. 
Therefore all expectation values
which change sign under such a transformation vanish.
Using this,
we obtain for $h_1 = \langle \Delta \calH \rangle$ 
\be
h_1 = \beta E_2 \sum_{<x,y>}
\lbrack 1 - \cos( s( \psi_x - \psi_y) )\rbrack \, ,
\ee
with
\be
E_2 = \langle  s_x^{(1)}s_y^{(1)}+ s_x^{(2)}s_y^{(2)} \rangle 
= \frac{2}{N}\langle  s_x \cdot s_y \rangle 
\ ,
\ee
where $x$ and $y$ are nearest neighbors.
For piecewise constant kernels,
$h_1$ is proportional to $L_B$.
For smooth kernels $h_1$ will become independent of $L_B$ for large
enough blocks. For small $s$,
\be
h_1 \approx \half s^2 \beta E_2 \sum_{<x,y>} (\psi_x -\psi_y)^2
= \half s^2 \beta E_2 \, \alpha \, ,
\ee
with $\alpha=(\psi,-\Delta \psi)$. This quantity becomes independent
of $L_B$ for large block sizes.

As in the XY model,
the scale dependence of acceptance rates in the
$O(N)$ nonlinear\\  $\sigma$-model with $N>2$ is like
in massless free field theory.
This behavior of the acceptance rates was already observed 
numerically in multigrid
Monte Carlo simulations of the $O(3)$ nonlinear \\
$\sigma$-model in two dimensions
with smooth and piecewise constant kernels \cite{hmm}. 
Using a V-cycle and smooth interpolation in a unigrid 
implementation, 
a dynamical critical
exponent $z \approx 0.2$ was observed.
With a V-cycle and piecewise constant interpolation
the result was $z \approx 1.3$. These values for $z$ are very similar
to the results for $z$ in massless free field theory
(cf.\ section~\ref{SECfree}).

In the $O(4)$ model, $z \approx 0.6$ was found~\cite{sokalO4},
using piecewise constant interpolation and a W-cycle in a recursive 
multigrid scheme.

One can do an analogous discussion 
for nonlinear $\sigma$-models
with global  $SU(N) \times SU(N)$ invariance,
leading to a similar prediction for the scale dependence of
the acceptance rates. 
In numerical experiments with a multigrid algorithm for the
$SU(3) \times SU(3)$ nonlinear $\sigma$-model $z \approx 0.2$ was 
found~\cite{hmsun}.

\subsection{The two dimensional $CP^{N-1}$ model}
\label{SUBSECCPN}

A different class of nonlinear $\sigma$-models in two dimensions
are the $CP^{N-1}$ models.
The complex projective space $CP^{N-1}$ is a complex manifold with
the topological structure of a $2N-1$ dimensional sphere,
where points related by a $U(1)$ transformation are identified.
A simple realization of a lattice action with a 
local $U(1)$ invariance is
\be
\calH(z) \ = \ 2 \beta \sum_{<x,y>} 
\left[ 1 - \vert \overline{ z_x} \cdot z_y \vert^2 \right]
\ee
where the $z_x$ 
are $N$-component complex unit vectors.
The $CP^1$  model is equivalent
 to the $O(3)$ nonlinear $\sigma$-model.
For all $N$, the $CP^{N-1}$ models are 
in a disordered phase at all values of $\beta$.
As for all nonlinear $\sigma$-models
in two dimensions,
the critical behavior of the model is described by the
perturbative renormalization group which implies asymptotic freedom
for $\beta \rightarrow \infty$.
In this limit 
the correlation length  
diverges exponentially in $\beta$.

Nonlocal Metropolis updates 
are formulated by updating in an embedded two dimensional $XY$ model. 
Consider the special case where the updates
act on the real and imaginary part of the first component
of the $z_x$ variables by 
\be\label{update_CPN}
z_x \ \rightarrow  \ z_{x}' = R_x z_x \ ,
\ee
with

\be
 z_{x} = \left(
\begin{array}{c}
z_x^{(1)} \\
z_x^{(2)} \\
\vdots \\
z_x^{(N)} \\
\end{array}
\right) \; , \; \;
 R_x = \left(
\begin{array}{cccc}
\exp({is \psi_x}) & 0 & \cdots & 0\\
0 &  1 &   & 0\\
\vdots &   &  \ddots  & \\
0 &  0 &  &  1\\
\end{array} \right) 
\ .
\ee
As above, $\psi$ is normalized according to eq.\ (\ref{normpsi}).
Other orientations of the embedded $U(1)$ subgroup can be 
obtained by acting on other components of the $z_x$-variables,
where the component is selected randomly.

The energy change of such an update is
$$
\Delta\calH \ = \ 
- 2 \beta \sum_{<x,y>} 
\left[ 
\vert \overline{R_x z_x} \cdot R_y z_y \vert^2 
- \vert \overline{ z_x} \cdot z_y \vert^2 
\right] \hspace{3.4cm}
$$
\be
= - 4 \beta \sum_{<x,y>} 
\mbox{Re}\left\{ 
\overline{z_x^{(1)}}z_y^{(1)}\sum_{i=2}^{N} z_x^{(i)} \overline{z_y^{(i)}} 
\left[\mye^{-is(\psi_x-\psi_y)} - 1 \right]
\right\} \ .
\ee
The Hamiltonian of the $CP^{N-1}$ model is invariant under 
a global complex conjugation
$z_x \rightarrow \overline{z_x}$.
Therefore the expectation value
\be
Q = \left\langle 
\overline{z_x^{(1)}}z_y^{(1)}\sum_{i=2}^{N} z_x^{(i)} \overline{z_y^{(i)}} 
\right\rangle  \ , 
\ee
where $x$ and $y$ are nearest neighbors, is real.
Using this property,
we find for
the average energy change of the update (\ref{update_CPN})
\be
h_1 = 4\beta Q \sum_{<x,y>}
\lbrack 1 - \cos( s( \psi_x - \psi_y) )\rbrack \ .
\ee
For piecewise constant kernels,
$h_1$ is proportional to $L_B$.
For smooth kernels $h_1$ will become independent of $L_B$ for large
enough blocks. For small $s$,
\be
h_1  \approx 2 s^2 \beta Q \sum_{<x,y>}( \psi_x - \psi_y)^2 
= 2 s^2 \beta Q \, \alpha \ . 
\ee
As above, $\alpha=(\psi,-\Delta \psi)$. $\alpha$ becomes independent
of $L_B$ for large block sizes.

As in the XY model and in the $O(N)$ nonlinear $\sigma$-model,
the scale dependence of acceptance rates in the
$CP^{N-1}$ model is like
in massless free field theory.

In multigrid Monte Carlo simulations
of the $CP^{3}$ model
with  smooth interpolation and a $V$-cycle in a unigrid 
implementation, 
a dynamical critical
exponent $z \approx 0.2$ was found~\cite{hmsun}.

\subsection{Summary}

A detailed quantitative analysis of the scale dependence 
of Metropolis updates was performed.
Our approximation formula has proven to be quite precise in 
three different models.
The results for all models are consistent with the following
rule:

\begin{quote}
{\sl Sufficiently high acceptance rates for a complete elimination
of CSD can only be expected if
$ h_1 = \langle \calH(\ph+s \psi) - \calH(\ph) \rangle $
contains no relevant operator in $\psi$. }
\end{quote}

As we have seen above, the typical candidate for a relevant operator
in $\psi$ is always  an algorithmic mass term
of the type $\sim s^2 \sum_x \psi_x^2$ with degree of relevance
$r=2$.

This rule is formulated for smooth kernels.
For piecewise constant kernels, it has to be modified.
There, the kinetic term $\alpha \propto  L_B$ is relevant as well.
At least in free field theory this disadvantage can be compensated for
by using a W-cycle.
Apart from this modification the rule carries over to the case
of piecewise constant kernels.

With the help of this rule it is possible to 
predict whether a given method will have a chance to eliminate CSD.

Let us reformulate the rule in heuristic way.
Recall the normalization condition (\ref{normpsi}) for the interpolation 
kernels $\psi$
$$
 L^{(d-2)/2}_B L^{-d}_B \sum_{x \in x'} \psi_x =  \delta_{x',x_o'} \, .
$$
The factor $L^{(d-2)/2}_B$ in front of the average 
$L^{-d}_B \sum_{x \in x'} (\dots)$ was chosen such that
the behavior of different terms in $\psi$ was analogous to the
perturbative renormalization group.
We observed that 
\be
\sum_x \psi_x^2 \ \sim \ L_B^2 \ 
\ee
for piecewise constant and for smooth interpolation.
The quantity $\alpha$ behaved like
\be
\alpha = (\psi, -\Delta \psi)  \ \sim \ const
\ee
for smooth interpolation and 
\be
\alpha = (\psi, -\Delta \psi)  \ \sim \ L_B
\ee
for piecewise constant interpolation,
independent of the dimension $d$.
See also the definition of the
degree of relevance at the end of section \ref{SECfree}.

Let us switch to dimensionless kernels 
\be
\hat{\psi}_x \ = \ L_B^{(d-2)/2} \psi_x \ ,
\ee
which are now normalized according to
\be
L^{-d}_B \sum_{x \in x'} \hat{\psi}_x =  \delta_{x',x_o'} \, .
\ee
E.g.\ a d-dimensional 
piecewise constant interpolation kernel in this normalization 
is just 
\be
 \hat{\psi}^{const}_x =\left\{
\begin{array}{ll}
1 &\mbox{for}\; x \in x_o' \\
0 &\mbox{for}\; x \not \in x_o' \; .\\
\end{array}
\right.
\ee
Then we find
\be
\sum_x \hat{\psi}_x^2 \ \sim \ L_B^d \ ,
\ee
for piecewise constant and for smooth interpolation.
A kinetic term in the kernels $\hat{\psi}$ now behaves as
\be
\hat{\alpha} = (\hat{\psi}, -\Delta \hat{\psi})  \ \sim \ L_B^{d-2}
\ee
for smooth interpolation and
\be
\hat{\alpha} = (\hat{\psi}, -\Delta \hat{\psi})  \ \sim \ L_B^{d-1}
\ee
for piecewise constant interpolation.

Assume that we use piecewise constant interpolation kernels $\hat{\psi}$.
Then the terms that appear in the average energy change 
$ h_1 = \langle \calH(\ph+s \hat{\psi}) - \calH(\ph) \rangle $
of a nonlocal update can be interpreted as follows:
\begin{itemize}
\item
a kinetic term $(\hat{\psi}, -\Delta \hat{\psi})$ is a {\em surface term},
i.\ e.\ the energy costs of a nonlocal update are proportional 
to the surface of the block
\item
a mass term $ \sim \sum_x \hat{\psi}^2$ is a {\em volume term},
i.\ e.\ the energy costs of a nonlocal update are proportional 
to the volume of the block
\end{itemize}
There are two different ways how multigrid algorithms can deal
with a surface term:
First, one can put more emphasis on the updating of coarse levels
by using a W-cycle.
Second, one can switch from piecewise constant interpolation 
to smooth interpolation. Then the energy change at the surface of a block
is spread uniformly over the entire block, leading to 
$(\hat{\psi}, -\Delta \hat{\psi}) \sim L_B^{(d-2)}.$

However, up to now there is no way how to overcome 
the effect of a volume term that causes CSD in
multigrid Monte Carlo simulations of critical models.

The different role of surface and volume terms seems to be a 
general feature of nonlocal update methods that multigrid 
algorithms share with cluster algorithms.
In cluster algorithms nonlocal moves are performed
by stochastically growing  a cluster of spins and then
flipping all spins in the cluster simultaneously.
In ref. \cite{embedding} it is argued that if the energy cost
of this global flip is a
``bulk term'' 
that is proportional to the volume of the cluster, 
a cluster algorithm is expected to work badly. 

The natural way to construct nonlocal piecewise constant
updates that only have surface energy costs is to look for
a global symmetry of the Hamiltonian of the model under consideration.
Then the idea is to apply the corresponding symmetry transformation 
locally on blocks.
E.\ g.\ in the massless  free field theory, the global symmetry
is the shift symmetry $\phi_x \rightarrow \phi_x + const$,
which is then applied locally on blocks.
In the $XY$-model, the global $O(2)$-symmetry of the model is applied
locally by performing piecewise constant $O(2)$-rotations on blocks.

To summarize this discussion we give an intuitive Leitmotiv for
the development of new multigrid methods:

\begin{quote} 
{\em A piecewise constant update of a nonlocal domain should have
energy costs proportional to the surface of the domain,
but not energy costs proportional to the volume of the domain. }
\end{quote}
  
We will use this heuristic criterion in the
development of multigrid algorithms for nonabelian gauge theories.
The intuitive Leitmotiv can then be made precise by performing
a quantitative acceptance analysis.
 
\clearpage

\section{Multigrid Simulation of the Sine Gordon model}

\label{SECsg}

\setcounter{equation}{0}

\subsection{Motivation}

In this section we report on a 
multigrid Monte Carlo study of the 
Sine Gordon model in two dimensions. 
We have two questions in mind:

First, we want to check the theoretical prediction
that a W-cycle (cycle control parameter $\gamma = 2$) 
with piecewise constant interpolation
will not eliminate CSD 
in the rough (massless) phase of the Sine Gordon model.
Recall the analysis of this model in the previous section:
The Sine Gordon model was an example
for a critical model where
the expansion of 
$\langle {\cal H}(\phi+\psi)\rangle$ 
contained a relevant (mass) term.
If such a term is present, 
Metropolis step sizes $\varepsilon(L_B)$ on block lattices
with increasing block size $L_B$ have to be
scaled down like $\varepsilon(L_B) \sim 1/L_B$ in order to obtain
block size independent acceptance rates.
This strong decrease of step sizes on large blocks
will lead to CSD. 

Second, we want to ask the question whether one can circumvent
 this slowing down caused by
too small steps on large blocks by accumulating many of these steps
randomly. 

We applied an intuitive 
random walk argument
for free field theory in section \ref{SECfree}.
There this reasoning explained why one can compensate
for the disadvantage of $\varepsilon(L_B) \sim 1/\sqrt{L_B}$
with piecewise constant interpolation by using a W-cycle
with $\gamma = 2$.
We can use the same argument when the Metropolis step sizes 
decrease even stronger:

A constant accumulated
step size on all length scales 
could in principle be achieved in the following way:
For step sizes scaling down like
$\varepsilon(L_B) \sim 1/L_B$ and a coarsening by a factor of two,
the Metropolis step size on a next coarser grid
is too small by a factor of two.
If we assume that subsequent update steps within a 
multigrid cycle are statistically independent and that these 
steps accumulate in a random-walk like way,
we can expect to compensate for this decrease by increasing
the number of updates on the next coarser grid by a factor of four.
This can be achieved by a higher cycle with  
cycle control parameter
$\gamma = 4$.
%%%%%%%%%%%%%%%%%%%%%%%%%%%%%%%%%%%%%%%%%%%%%%%%%%%%%%%%%%%%%%%%%%%%%%%%%%%%%%%
\setlength{\unitlength}{0.5mm}
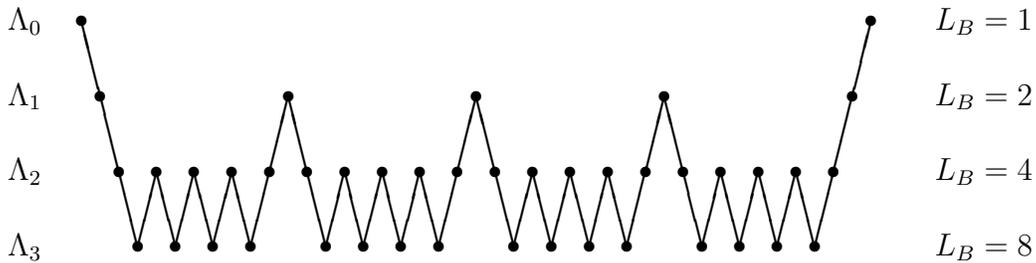
\begin{figure}[htb]
\begin{center} \begin{picture}(270,100)(10,0)
%
% Lattice-Names
%
\thicklines
\put(10,65){\makebox(10,10){$\Lambda_0$}}
\put(10,45){\makebox(10,10){$\Lambda_1$}}
\put(10,25){\makebox(10,10){$\Lambda_2$}}
\put(10, 5){\makebox(10,10){$\Lambda_3$}}
\put(265,65){\makebox(10,10){$L_B = 1$}}
\put(265,45){\makebox(10,10){$L_B = 2$}}
\put(265,25){\makebox(10,10){$L_B = 4$}}
\put(265, 5){\makebox(10,10){$L_B = 8$}}
%
% gamma=4-cycle
%
\put(45,10){\line(-1,4){15}}
\multiput(45,10)(10,0){3}{\line( 1,4){5}}
\multiput(55,10)(10,0){3}{\line( -1,4){5}}
\put(75,10){\line(1,4){10}}
\put(95,10){\line(-1,4){10}}
\multiput(95,10)(10,0){3}{\line( 1,4){5}}
\multiput(105,10)(10,0){3}{\line( -1,4){5}}
\put(125,10){\line(1,4){10}}
\put(145,10){\line(-1,4){10}}
\multiput(145,10)(10,0){3}{\line( 1,4){5}}
\multiput(155,10)(10,0){3}{\line( -1,4){5}}
\put(175,10){\line(1,4){10}}
\put(195,10){\line(-1,4){10}}
\multiput(195,10)(10,0){3}{\line( 1,4){5}}
\multiput(205,10)(10,0){3}{\line( -1,4){5}}
\put(225,10){\line(1,4){15}}

\multiput(45,10)(10,0){4}{\circle*{3}}
\multiput(95,10)(10,0){4}{\circle*{3}}
\multiput(145,10)(10,0){4}{\circle*{3}}
\multiput(195,10)(10,0){4}{\circle*{3}}
\multiput(40,30)(10,0){20}{\circle*{3}}
\multiput(35,50)(50,0){5}{\circle*{3}}
\multiput(30,70)(210,0){2}{\circle*{3}}

\end{picture} \end{center}
\caption{ \sl Higher cycle with $\gamma=4$ \label{highcycle}}
\end{figure}
\addtolength{\unitlength}{0.5mm}
%%%%%%%%%%%%%%%%%%%%%%%%%%%%%%%%%%%%%%%%%%%%%%%%%%%%%%%%%%%%%%%%%%%%%%%%%%%%%%%
By applying a higher cycle with $\gamma > 1$,
$\gamma$ times more updates on each coarser
lattice are performed.
An example for a higher cycle with $\gamma = 4$ is given in
figure \ref{highcycle}.

For a recursive multigrid algorithm, the
computational effort is
$\sim L^d$ for $\gamma < 2^d$ and $\sim L^d \log L$ for $\gamma = 2^d$ in $d$
dimensions \cite{sokalrev}.
Therefore a higher cycle with $\gamma = 4$ is practical for
$d > 2$ and borderline practical
for $d = 2$.

In summary, if
the random-walk picture is correct,
we can expect to overcome the
problem of strongly decreasing step sizes on large scales
by doing more work on coarser lattices.

\subsection{The multigrid algorithm}

The Hamiltonian of the two dimensional Sine Gordon model is 
\begin{equation} \label{sghamiltonian}
  {\cal H}(\phi) = \frac{1}{2 \beta}
                  (\phi, -\Delta \phi)
                    - \zeta \sum_x \cos 2\pi \phi_x \ 
\end{equation}
on an $L \times L $ lattice $\Lambda_0$.
Our simulations are organized as follows:
In order to allow for high cycle control parameters
$\gamma$,
we use a recursive multigrid algorithm,
piecewise constant interpolation and a coarsening with a factor of two,
as illustrated in figure \ref{blocking}.
Let us describe the 
details of the recursive multigrid algorithm.
An update to the current fine grid field $\phi_x$ is 
given by
\be \label{displacement}
\phi_x \rightarrow \phi_x ' = \phi_x + \Phi_{x'} \
 \ \mbox{if $x \in x'$}\ .
\ee
The point $x$ on the fine grid is contained in the $2 \times 2$ block
$x'$ on the coarse grid, and $\Phi_{x'}$ denotes the displacement 
field\footnote{Although we use the same letter $\Phi$,
the displacement field as used here
and the block spin field as introduced in section \ref{SECunigrid}
differ by a constant}
corresponding to the block $x'$.

The conditional coarse grid Hamiltonian is calculated by substituting
eq.\ (\ref{displacement}) into eq.\ (\ref{sghamiltonian}).
Its general form on the block lattice $\Lambda_k$ is
\be
  \calH_k(\Phi) = \frac{\alpha}{2}
                  \sum_{\langle x',y' \rangle \in \Lambda_k}
 (\Phi_{x'} - \Phi_{y'})^2
 - \sum_{x' \in \Lambda_k} f_{x'} \Phi_{x'}
- \zeta \sum_{x' \in \Lambda_k} \left( a_{x'} \cos 2\pi \Phi_{x'} +
   b_{x'} \sin 2\pi \Phi_{x'} \right) \ .
\ee
Note that the fundamental Hamiltonian $\calH$ on the fine grid 
$\Lambda_0$ is a special case of this general coarse grid Hamiltonian
$\calH_k$ on $\Lambda_k$ with 
\bea
\alpha &=& 1/\beta \ , \\
f_x &=& 0 \ \ \mbox{for all $x \in \Lambda_0$} \ ,\\
a_x &=& 1 \ \ \mbox{for all $x \in \Lambda_0$} \ ,\\
b_x &=& 0 \ \ \mbox{for all $x \in \Lambda_0$} \ .
\eea
The set of coefficients $ \{\alpha', f',a',b' \}$
of $\calH_{k+1}$ on the next coarser grid 
$\Lambda_{k+1}$ can be calculated from 
the set of coefficients $\{\alpha, f, a, b \}$
on $\Lambda_k$ by recursion:  
\bea
\alpha' &=& 2 \alpha \ , \\
f_{x'}' &=& \sum_{x \in x'} 
\left[f_x - 2\alpha \sum_{y n.n. x}(\phi_x - \phi_y) \right] \ , \\
a_{x'}' &=& \sum_{x \in x'} 
\left( \mbox{} 
a_{x} \cos 2\pi \phi_{x} + b_{x} \sin 2\pi \phi_{x} \right) \ , \\
b_{x'}' &=& \sum_{x \in x'} 
\left(  - a_{x} \sin 2\pi \phi_{x} + b_{x} \cos 2\pi \phi_{x} \right) \ .
\eea
Here $\phi_x$ denotes the configuration on $\Lambda_k$ when we
switch to the next coarser grid.

For the updating of the variables on $\Lambda_k$ within the multigrid cycle, 
we choose a sweep of single hit Metropolis updates before
the coarsening and after the interpolation, respectively.
The maximum Metropolis step size $\varepsilon(L_B)$
is scaled down like $1/L_B$.
Then, 
acceptance rates of approximately
$50\%$ are observed on all block lattices,
in accordance with the theoretical  
analysis of section \ref{SECappl}.
We tested the implementation of the algorithm
by comparison with exact results in the Gaussian model ($\zeta = 0$)
and against Monte Carlo data 
obtained on small lattices by a cluster algorithm for
the Sine Gordon model \cite{klausmartin}.

\subsection{Simulation and results}

We study the dynamical critical behavior of the different versions of the
algorithm in the rough phase, where
the correlation length is infinite and the physical length scale
is set by the linear size of the lattice $L$.
Thus, we expect the autocorrelation time $\tau$ to
diverge with the dynamical critical exponent $z$ like
$\tau \sim L^z$.

As observables we take the energy
\be
E = \frac{1}{L^{2}}\sum_{\langle x,y \rangle}
(\phi_x - \phi_y)^2
\ee
 and the
surface thickness $\sigma^2$ as defined in eq.\ (\ref{thickness}).

The simulations were performed at $\beta = 1.0$, $\zeta = 0.5$.
This is deep in the rough phase.
Note that in the limit $\zeta \rightarrow \infty$ which corresponds to the
discrete Gaussian model
the critical coupling is $\beta_c = 0.7524(8)$
\cite{rough}.
Starting from an ordered configuration, measurements
were taken after equilibration at each visit of the finest lattice.
The run parameters and results are summarized in 
table \ref{tab:num_results1} for the W-cycle
and  in table \ref{tab:num_results2} for the higher cycle
with $\gamma = 4$.
%The typical logarithmic divergence of $\sigma^2$ 
%in the rough phase can be seen in figure \ref{rough}. 

%%%%%%%%%%%%%%%%%%%%%%%%%%%%%%%%%%%%%%%%%%%%%%%%%%%%%%%%%%%%%%%%%%%%%%%%%%%%%%%
\begin{table}[htbp]
 \centering
 \caption[dummy]{\label{tab:num_results1} 
         Numerical results for the 
         W-cycle ($\gamma = 2$) in the 2-d
         Sine Gordon model on
         $L \times L$ lattices
         in the rough phase, $\beta= 1.0$,
         $\zeta = 0.5$}
 \vspace{2ex}
\begin{tabular}{|r|r|c|l|r|l|r@{}l|}
\hline\str
$L\;$ &
statistics & 
discarded & 
\multicolumn{1}{c|}{$E$} & 
\multicolumn{1}{c|}{$\tau_{int,E}$} &
\multicolumn{1}{c|}{$\sigma^2$} &
\multicolumn{2}{c|}{$\tau_{int,\sigma^2}$} 
\\[.5ex] \hline \hline \str
 4  & 25\,000 & 2\,000 &0.934(4) & 0.90(3) &
 0.268(1) & 0&.96(3)
\tabhline
 8  & 50\,000 & 2\,000 &0.986(1) & 0.97(2) &
 0.3809(9)& 1&.35(3)
\tabhline
 16 & 100\,000& 2\,000 &0.9956(5)& 1.04(2) &
 0.4896(7)& 2&.70(8)
\tabhline
 32 & 300\,000& 2\,000 &0.9987(2)& 1.03(1)&
 0.5996(7)& 8&.54(19)
\tabhline
 64 & 500\,000& 2\,000 &0.99945(5)&1.04(1)&
 0.7105(10)&30&.5(1.0)
\tabhline
 128& 500\,000& 4\,000& 0.99966(3)&1.04(1)&
0.8218(19)&113&.7(6.9)
\\[.3ex] \hline
\end{tabular}
%\end{table}
%%%%%%%%%%%%%%%%%%%%%%%%%%%%%%%%%%%%%%%%%%%%%%%%%%%%%%%%%%%%%%%%%%%%%%%%%%%%%%%
\bigskip
\bigskip
%%%%%%%%%%%%%%%%%%%%%%%%%%%%%%%%%%%%%%%%%%%%%%%%%%%%%%%%%%%%%%%%%%%%%%%%%%%%%%%
%\begin{table}[htbp]
% \centering
 \caption[dummy]{\label{tab:num_results2}
         Numerical results for the 
         higher cycle with $\gamma = 4$ 
         in the 2-d
         Sine Gordon model on
         $L \times L$ lattices
         in the rough phase, $\beta= 1.0$,
         $\zeta = 0.5$} 
 \vspace{2ex}
\begin{tabular}{|r|r|c|l|r|l|r@{}l|}
\hline\str
$L\;$ &
statistics & 
discarded & 
\multicolumn{1}{c|}{$E$} & 
\multicolumn{1}{c|}{$\tau_{int,E}$} &
\multicolumn{1}{c|}{$\sigma^2$} &
\multicolumn{2}{c|}{$\tau_{int,\sigma^2}$} 
\\[.5ex] \hline \hline \str
 4 & 25\,000 & 2\,000 &0.940(4) & 0.89(3) &
 0.268(1) & 0&.91(3)
\tabhline
 8 &  25\,000 & 2\,000 &0.986(2) & 0.94(3) & 
0.380(1) & 1&.14(4)
\tabhline 
 16 & 25\,000 & 2\,000 &0.9965(10)&0.94(3) & 
0.488(1) & 1&.67(6) 
\tabhline
 32 & 100\,000& 2\,000&0.9985(3) & 0.95(1) & 
0.5997(9)& 4&.15(11)
\tabhline
 64 &  300\,000& 2\,000&0.99945(7)& 0.96(1) & 
0.7113(9)& 14&.2(4) 
\tabhline
 128 & 300\,000& 2\,000&0.99962(4)& 0.95(1) &
0.8213(18)&58&.2(3.3)
\\[.3ex] \hline
\end{tabular} \end{table}

From the autocorrelation function for the observable $A$
\begin{equation}
\rho_A(t)\,=\,
\frac{\langle A_s A_{s+t} \rangle
-\langle A \rangle^2} 
{\langle A^2 \rangle
-\langle A \rangle^2} \ ,
\end{equation}
we compute the integrated autocorrelation time
\begin{equation}
\tau_{int,A}\,=\, \frac{1}{2} + \sum_{t=1}^{\infty} \rho_A(t) \ .
\end{equation}
The asymptotic decay of the autocorrelation function 
$\rho_A(t)$ is described by the exponential autocorrelation time
$\tau_{exp,A}$:
\be
\rho_A(t) \ \sim \ \exp(-{t}/{\tau_{exp,A}}) \ \ 
\mbox{for} \ \ t \rightarrow \infty \ .
\ee
If $\rho_A(t)$ is a single exponential,
$\tau_{int,A} \approx \tau_{exp,A}$.

\begin{figure}[htbp]
 \begin{center}
  \begin{minipage}[t]{145mm} 
   \begin{picture}(145,95)(45,130)
    \epsfig{file=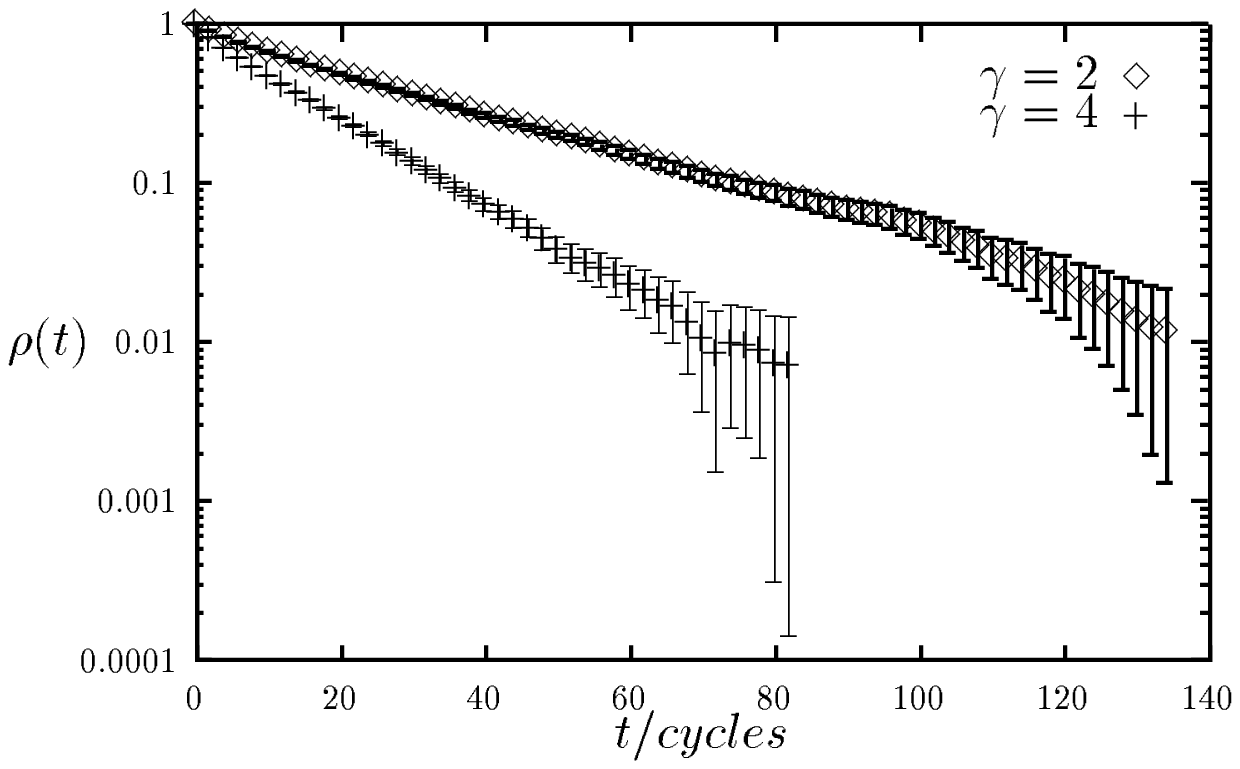,height=343mm}
   \end{picture}
   \begin{picture}(145,95)(45,130)
    \epsfig{file=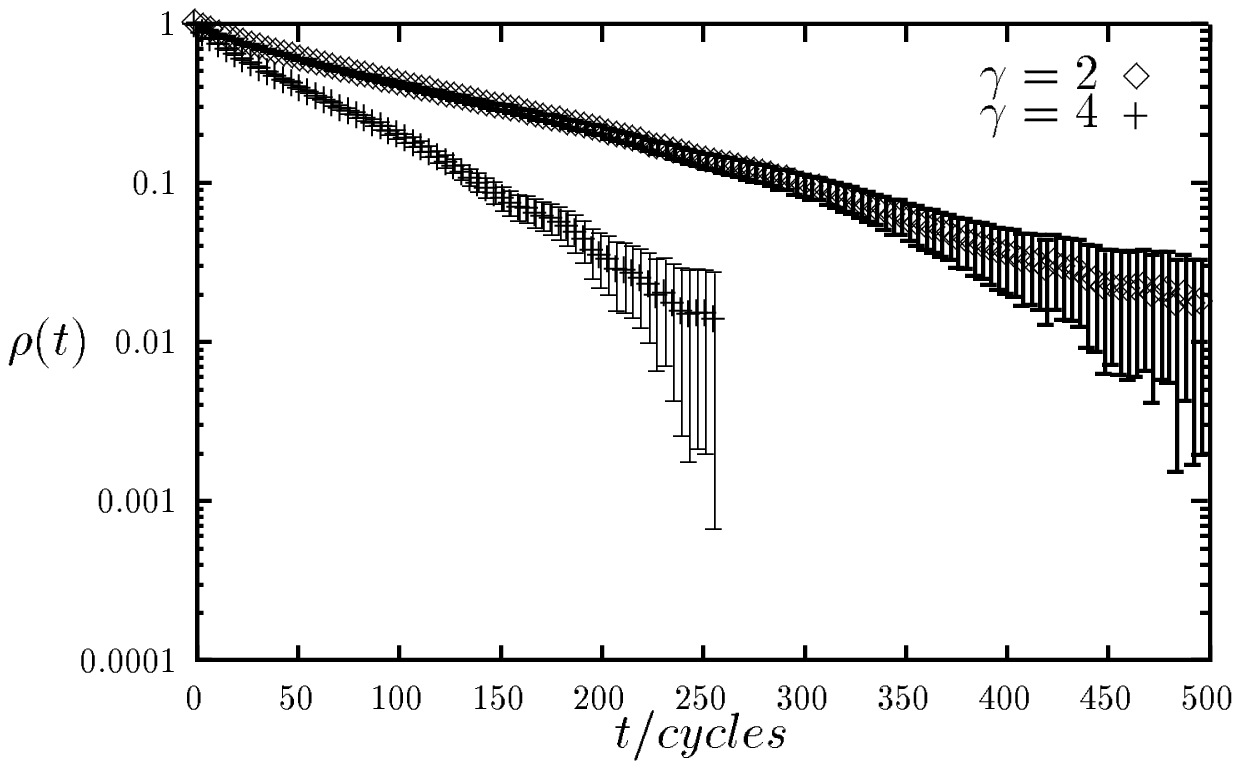,height=343mm}
   \end{picture}
   \caption[\Omega(s) in the 2d Sine Gordon model]
   {\label{sgtau}
   \sl Autocorrelation functions $\rho(t)$ for the surface thickness
   $\sigma^2$ in the rough phase of the two dimensional Sine
   Gordon model, $\beta=1.0$, $\zeta=0.5$.  Top: $64^2$ lattice,
   bottom: $128^2$ lattice. $\rho(t)$ is shown for the
   W-cycle ($\gamma = 2$) and the higher cycle ($\gamma = 4$)
   respectively.}
  \end{minipage}
 \end{center}
\end{figure}

The errors on $\tau_{int,A}$
were calculated by a window method \cite{madras}.
For the self consistent 
truncation window we took $4 \tau_{int}$, which is sufficiently 
large if the autocorrelation function shows an exponential
decay. This is indeed the case, see  
figure \ref{sgtau}.
 
\begin{figure}[htbp]
 \begin{center}
  \begin{minipage}[t]{145mm} 
   \begin{picture}(145,95)(45,85)
    \epsfig{file=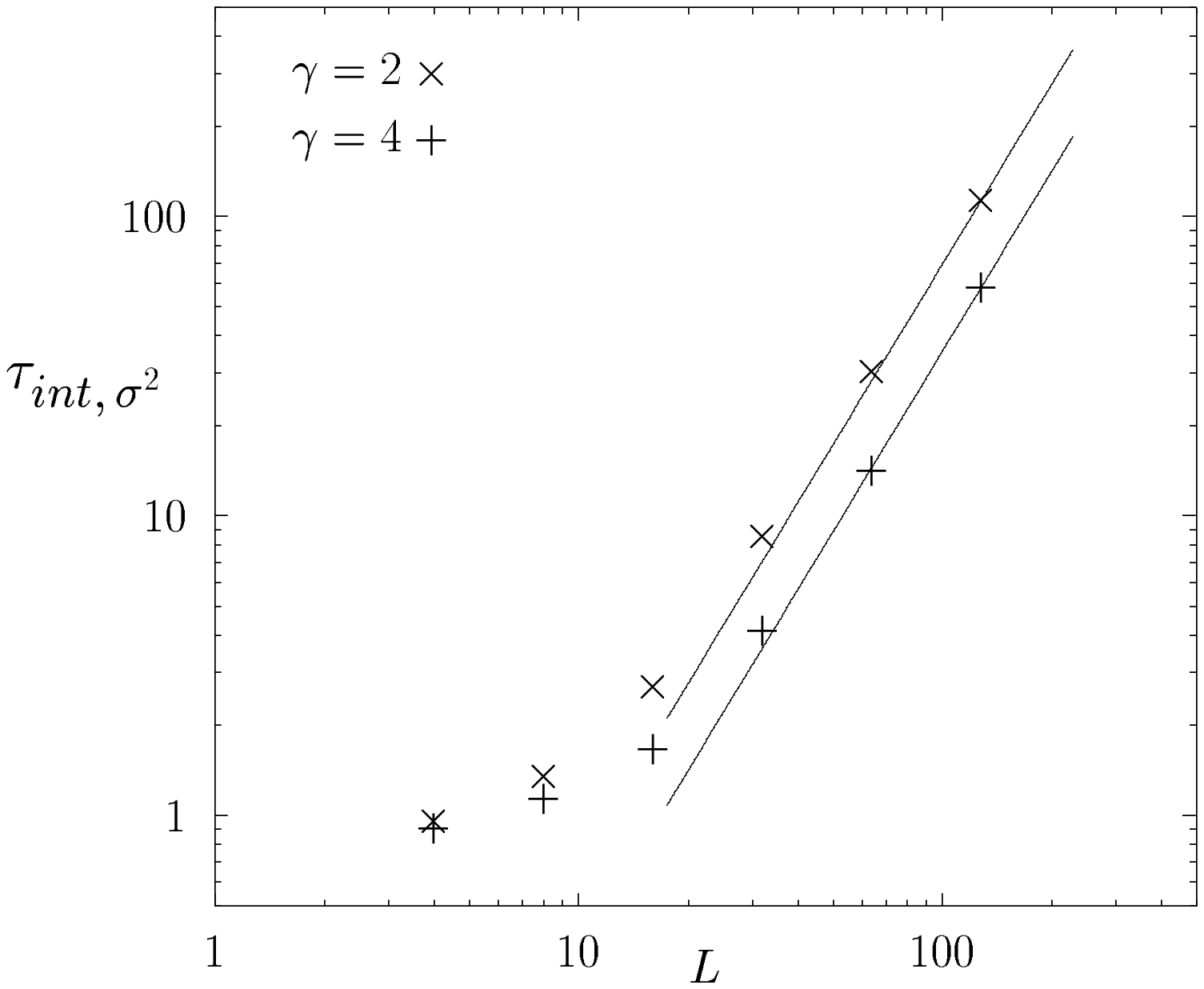,height=300mm}
   \end{picture}
   \caption[\Omega(s) in the 2d Sine Gordon model]
   {\label{fig:tau} \sl
   Dependence of the integrated autocorrelation time for the
   surface thickness $\sigma^2$ on the lattice size $L$
   in the rough phase of the 2-d Sine Gordon model, $\beta= 1.0$,
   $\zeta = 0.5$. Errors are smaller than the symbols used.
   The lines correspond to $z=2$.}
  \end{minipage}
 \end{center}
\end{figure}

The numerical results for the autocorrelation times $\tau_{int}$
are listed in table \ref{tab:num_results1}
for $\gamma = 2$ and in table \ref{tab:num_results2} for $\gamma = 4$.
The values for $\tau_{int}$ are measured in the number of visits on
the finest lattice.
Note that our runs are longer than $10\,000$ $\tau_{int}$
(longer than $4\,000$ $\tau_{int}$ on the $128^2$ lattice).
Figure \ref{fig:tau} shows the dependence of $\tau_{int,\sigma^2}$ on 
$L$ for $\gamma=2$ and $\gamma=4$. For comparison, we plotted lines
which correspond to $z=2$.

If we fit our data for the autocorrelation time of
the surface thickness in the range
$32 \leq L \leq 128$ with the
Ansatz $\tau_{int,\sigma^2} = c L^z$, we obtain $z = 1.86(4)$ with
$\chi^2/\mbox{dof} = 0.22$ for the W-cycle ($\gamma = 2$),
and $z = 1.86(4)$ with $\chi^2/\mbox{dof} = 4.6$ for the
higher cycle ($\gamma = 4$).
The uncertainty in $z$ is dominated by the relative error
of $\tau_{int}$ on the largest lattice, which is about $6\%$ in
both cases.
We therefore estimate
\begin{eqnarray}
z&=&1.9(1) \;\;\;\; \mbox{for the W-cycle ($\gamma=2$)} \ , \nonumber \\
z&=&1.9(1) \;\;\;\; \mbox{for the higher cycle ($\gamma=4$)} \ . \nonumber
\end{eqnarray}

%%%%%%%%%%%%%%%%%%%%%%%%%%%%%%%%%%%%%%%%%%%%%%%%%%%%%%%%%%%%%%%%%%%%%%%%%%%%%%%

\subsection{Summary}

In this section we reported on a simulation of the two dimensional
Sine Gordon model with a multigrid Monte Carlo algorithm.
Using a recursive multigrid implementation with piecewise 
constant interpolation we studied the dynamical critical 
behavior of the two variants of the algorithm
($\gamma =2$ and $\gamma = 4$) in the rough phase.
From this study we confirm that,
as already predicted in section \ref{SECappl}, CSD
is not reduced by a W-cycle with piecewise constant interpolation.
According to the acceptance analysis, we would expect the same result 
with smooth interpolation.
Moreover, the results clearly show
that the attempt to compensate for decreasing acceptance rates
on large blocks by choosing a higher cycle with
$\gamma= 4$ does not improve the dynamical 
critical behavior of the algorithm. 
We conclude that a random-walk argument as stated above
is not correct in the case of the Sine Gordon model. 

\clearpage

\section{Multigrid algorithms for lattice
gauge fields in two dimensions}

\label{SECSU22acc}

\setcounter{equation}{0}

In this section, we start the discussion of multigrid algorithms
for lattice gauge theory.
We begin with a review of
the method by Laursen, Smit and Vink for abelian gauge
fields in two dimensions. 
An new difficulty is to
change from an abelian to a nonabelian gauge group. 
The concepts that are needed for nonlocal updates of 
nonabelian gauge
fields in in two dimensional $SU(2)$ lattice gauge theory
are discussed.
We introduce a new updating procedure, the time slice blocking, 
and analyze its kinematics. 

Let us introduce the notations for lattice gauge theory in $d$ dimensions.
We consider partition functions
\be
Z = \int \prod_{x,\mu} d U_{x,\mu} \,
\exp\bigl( - \calH(U) \bigr) \, .
\ee

The link variables $U_{x,\mu}$ take values in the gauge
group $U(1)$ or $SU(N)$, and $dU$ denotes the corresponding
invariant Haar measure.
The standard Wilson action $\calH(U)$ is given by
\be \label{wilson-action}
{\cal H}(U)\,=\,\beta\sum_{\cal P} \bigl[ 1 -
\foN \mbox{Re} \, \Tr \, U_{\cal P} \bigr] \ \ .
\ee
The sum in (\ref{wilson-action}) is over all plaquettes in the
lattice. The $U_{\cal P}$ are path ordered
products around plaquettes ${\cal P}$,
\be
U_{\cal P} \,=\, \Um{x}\Un{x+\hat{\mu}}
                 \Um{x+\hat{\nu}}^{*}\Un{x}^{*}\ \ .
\ee
$U^*$ denotes the hermitean conjugate (= inverse) of $U$.

\subsection{Abelian gauge fields: $U(1)$ in two dimensions}

\subsubsection{The algorithm of Laursen, Smit and Vink}

A first multigrid Monte Carlo algorithm for abelian gauge fields 
was implemented and tested by
Laursen, Smit and Vink for $U(1)$ lattice gauge theory 
in two dimensions
\cite{laursen}. It was inspired by a deterministic multigrid 
procedure for the minimization of a Hamiltonian of the form
arising in lattice gauge theory \cite{sokalrev}.
As an example and a starting point for the 
following discussion of nonabelian gauge fields we review
their algorithm in the unigrid language.

In the case of gauge group $U(1)$ the link variables
are parameterized with an angle
$ -\pi \leq \theta_{x,\mu} < \pi$ through
\be
\Um{x}\,=\,\exp( i \theta_{x,\mu}) \ \ .
\ee
%%%%%%%%%%%%%%%%%%%%%%%%%%%%%%%%%%%%%%%%%%%%%%%%%%%%%%%%%%%%%%%%%%%
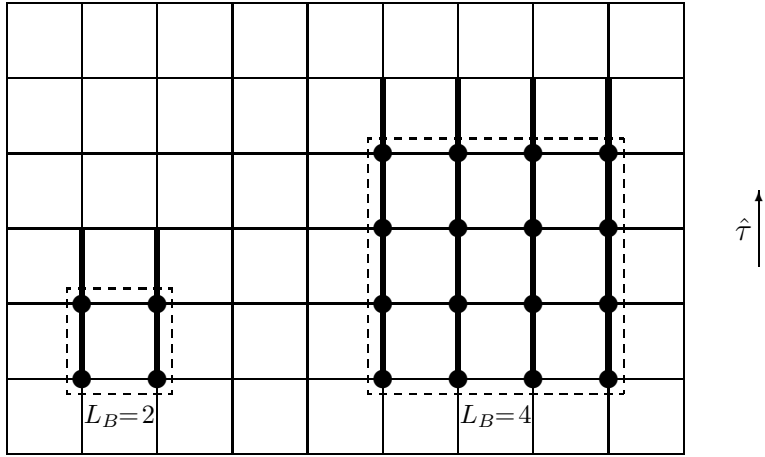
\begin{figure}
\begin{center}
\begin{picture}(160,50)(-30,0)
\multiput(0,0)(0,10){7}{\line(1,0){90}}
\multiput(0,0)(10,0){10}{\line(0,1){60}}
\put(100,25){\vector(0,1){10}}
\put(93,25){\makebox(10,10){$\hat{\tau}$}}
\put(8,8){\dashbox{1}(14,14)}
\put(10,00){\makebox(10,10){\footnotesize $L_B \!\! = \! 2$}}
\put(48,8){\dashbox{1}(34,34)}
\put(60,00){\makebox(10,10){\footnotesize $L_B \!\! = \! 4$}}
\linethickness{0.7mm}
\multiput(10,10)(10,0){2}{\line(0,1){20}}
\multiput(10,10)(10,0){2}{\circle*{2.5}}
\multiput(10,20)(10,0){2}{\circle*{2.5}}
\multiput(50,10)(10,0){4}{\line(0,1){40}}
\multiput(50,10)(10,0){4}{\circle*{2.5}}
\multiput(50,20)(10,0){4}{\circle*{2.5}}
\multiput(50,30)(10,0){4}{\circle*{2.5}}
\multiput(50,40)(10,0){4}{\circle*{2.5}}
\end{picture} \end{center}
\caption[dummy]{\label{2dblocking} 
                \sl Blocking of link variables
                for two dimensional $U(1)$ gauge fields}
\end{figure}
%%%%%%%%%%%%%%%%%%%%%%%%%%%%%%%%%%%%%%%%%%%%%%%%%%%%%%%%%%%%%%%%%%%

Nonlocal updates are defined 
as illustrated in figure \ref{2dblocking}:
One chooses a square block $x_o'$ of size $L_B^2$
and a direction $\tau$ with  $\tau = 1$ or $2$.
During the update, $\tau$ will be kept fixed.
All the link variables $\Ut{x}$ attached to sites $x$ inside the block
$x_o'$ are proposed to be ``rotated'' simultaneously:
\be 
\Ut{x} \rightarrow \exp(is\psi_x) \Ut{x} \, ,
\ee
or in terms of the link angles
\be 
\theta_{x,\tau} \rightarrow  \theta_{x,\tau} + s\psi_x  \, .
\ee
$\psi$ denotes an interpolation kernel
as introduced in section \ref{SECkernels}.
In the reported work a two dimensional piecewise constant kernel
\be
 \psi^{const}_x=\left\{
\begin{array}{ll}
1 &\mbox{for}\; x \in x_o' \\
0 &\mbox{for}\; x \not \in x_o'
\end{array}
\right.
\ee
was taken. For this choice of interpolation, a recursive multigrid
implementation with a $W$-cycle is feasible \cite{laursen}.
Another interesting feature of
this implementation is 
(in contrast to the time slice blocking below)
that both directions of link variables
pointing from the block $x_o'$ in the $\tau = 1$ and in the $\tau = 2$ 
direction can be updated on a coarse layer without going back to 
the finest lattice.

The reported result for the dynamical 
critical exponent $z$ for the simulation with a $W$-cycle is
$z \approx 0.2$ with respect to the
autocorrelation times of Polyakov loop observables.
However, it was also observed that the multigrid algorithm is not able
to move efficiently between different topological sectors.
This leads to 
autocorrelation times  $\tau \sim \exp(c\beta)$  with $c \approx 1.9$
for the topological charge. 
If the physical size of the lattice is fixed, this will lead to
$\tau \sim \exp(c' L)$, i.e.\ super-CSD
increasing exponentially in $L$.
The exponent quoted above should therefore 
be interpreted with some care. 

Let us analyze the kinematical behavior of this algorithm 
with the help of eq.\ (\ref{formula})
$$
\Omega(s) \approx \erfc \left(\half \sqrt{h_1} \right) \ .
$$
For $h_1 = \langle \Delta \calH \rangle$ we find
in the case of piecewise constant interpolation
\be \label{U1_const}
h_1 \,=\, 2 \beta P L_B [ 1 - \cos(s)] \  = \ \beta P L_B s^2 + O(s^4)
\ee
with $P = \EW{ \Tr U_{\cal P} }$.
The derivation of this expression
is described in appendix \ref{APPacc}.
In order to obtain acceptance rates
independent of the block size, we have to rescale the amplitudes for the 
nonlocal updates like $s \sim 1/\sqrt{L_B}$,
as in massless 
free field theory with piecewise constant interpolation. 
With respect to this, the behavior of this multigrid algorithm in
the $U(1)$ lattice gauge theory in two dimensions is similar to a 
multigrid algorithm in
massless free field theory. 
Therefore the reported acceleration for Polyakov loop observables
can be understood by our analysis.

On the other hand 
the problems with topological quantities are again an
example  (cf.\ the discussion of the two dimensional $XY$-model 
in the vortex phase in section \ref{SECappl})  for the fact
that good acceptance rates alone are not sufficient to overcome CSD.

\subsubsection{Comments and possible modifications}

Let us denote
the time slice of lattice sites
with ${\tau}$-component $t$ as
$\Lambda_t^{\tau} =
\left\{ x \in \Lambda_0 \, \vert \, x_{\tau} = t \right\}$.
Here the name ``time'' direction has no physical meaning.
We use this word to label the fixed direction of 
link variables that are updated simultaneously.
Of course, the time direction in an update algorithm
will be changed periodically from $\tau = 1$ to $\tau = 2$.
In the following, we will denote the time direction with $\tau$
and the spatial direction(s) different from $\tau$ with $\mu$.

\begin{figure} \begin{center}
\begin{picture}(160,50)(-30,0)
\multiput(0,0)(0,10){5}{\line(1,0){90}}
\multiput(0,0)(10,0){10}{\line(0,1){40}}
\put(100,15){\vector(0,1){10}}
\put(93,15){\makebox(10,10){$\hat{\tau}$}}
\put(-10,5){\makebox(10,10){$\Lambda_1^{\tau}$}}
\put(-10,15){\makebox(10,10){$\Lambda_2^{\tau}$}}
\linethickness{0.7mm}
\multiput(10,10)(10,0){8}{\line(0,1){20}}
\multiput(10,10)(10,0){8}{\circle*{2.5}}
\multiput(10,20)(10,0){8}{\circle{2.5}}
\end{picture}
\end{center}
\caption[dummy]{\label{decouple} 
                \sl Decoupling of time slices 
                $\Lambda_1^{\tau}$ and
                $\Lambda_2^{\tau}$. }
\end{figure}
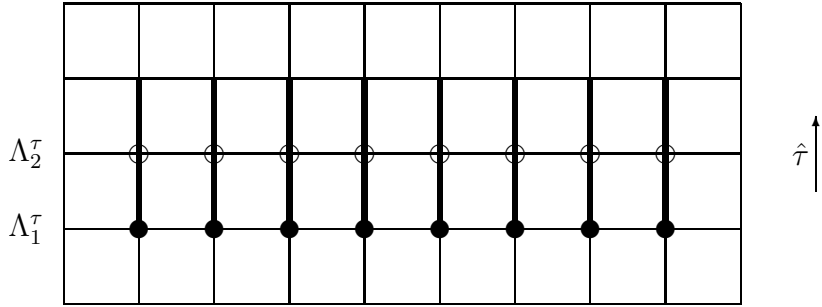

In the unigrid picture a general feature of nonlocal updating
schemes in lattice gauge theory is transparent:
As long as we restrict the possible nonlocal changes in the
configuration 
to link variables $U_{x,\tau}$ of a fixed time direction $\tau$
and keep all other variables $U_{x,\mu}$  
with $\mu \neq \tau$ unchanged, adjacent time slices decouple.
This is illustrated in figure \ref{decouple}.

%%%%%%%%%%%%%%%%%%%%%%%%%%%%%%%%%%%%%%%%%%%%%%%%%%%%%%%%%%%%%%%%%%%
%%%%%%%%%%%%%%%%%%%%%%%%%%%%%%%%%%%%%%%%%%%%%%%%%%%%%%%%%%%%%%%%%%%

Here, link variables $\Ut{x}$ pointing
from sites in two different adjacent
time slices $\Lambda_1^{\tau}$ and
$\Lambda_2^{\tau}$ are shown.
The point is that there are only plaquette terms in the Hamiltonian
that contain link variables $U_{x,\tau}$ pointing from sites
$x$ in the same
time slice $\Lamt$, i.e.\ either in $\Lambda_1^{\tau}$ or 
$\Lambda_2^{\tau}$. 
This means that the link variables pointing from 
sites in $\Lambda_1^{\tau}$ and from sites in
$\Lambda_2^{\tau}$ are independent
 as long as all other
$\Um{x}$ with $\mu \neq \tau$ are fixed. 

A possible modification of the algorithm 
is the use of smooth interpolation kernels 
$\psi$.
If we repeat the acceptance analysis
with the help of formula (\ref{formula})
for a general kernel $\psi$ we obtain for 
$h_1 = \langle \Delta \calH \rangle$ (cf. appendix \ref{APPacc})
\be \label{gaugeu1}
h_1\,=\,\beta P 
\sum_{\stackrel{{\mbox{\scriptsize $x \in \Lambda_0$}}}{\mu \neq \tau}}
\left[ 1-\cos\bigl( s (\psi_{x+\hat{\mu}}-\psi_x)\bigr) \right] \ \ .
\ee
The sum does not include contributions from links
which point in the time-direction $\tau$.
Therefore $h_1$ is a sum of independent contributions from
time slices $\Lamt$ orthogonal to the $\tau$-direction.
This is a consequence of the independence of time slices.
In particular, no smoothness
of kernels is needed in the $\tau$-direction.
From now on we choose $\psi_x$ to be constant in that direction
and assume that the support of $\psi$ in $\tau$-direction
is restricted to the $L_B$ time slices $\Lamt$ 
that contain the square block $x_o'$.
Then we find 
\be
h_1\,=\,
\beta P L_B 
\sum_{\stackrel{{\mbox{\scriptsize  $x \in \Lamt$}}}{\mu \neq \tau}}
\left[ 1-\cos\bigl( s (\psi_{x+\hat{\mu}}-\psi_x) \bigr)\right] \ \ .
\ee
For small $s$, this can be approximated by
\be
h_1\,\approx \,
\half s^2 \beta P L_B 
\sum_{\stackrel{{\mbox{\scriptsize  $x \in \Lamt$}}}{\mu \neq \tau}}
( \psi_{x+\hat\mu} - \psi_{x} )^2 =
\half s^2 \beta P \alpha_{1} \ \ .
\ee
with $\alpha_{1} = (\psi',-\Delta \psi') $.
Here, $\psi'$ denotes the kernel $\psi$ restricted to the
one dimensional time slice $\Lamt$.
A factor of $L_B^{1/2}$ is absorbed in 
$\psi'$ such that $\psi'$ is properly normalized
as a one dimensional kernel.

This means that for smooth interpolation
we can obtain acceptance rates
independent of the block size by choosing the amplitudes for the 
nonlocal updates like $s \sim const$.
The kinematical behavior of this version
of the multigrid algorithm is similar to a 
multigrid algorithm with smooth interpolation in
massless free field theory. 
Therefore we expect a successful acceleration 
for Polyakov loop observables also for this version.

In summary, the independence property of time slices
has several consequences:
First, interpolation kernels $\psi$ do not need to be
smooth in the time direction.
Second, since updates of link variables 
in different time slices are
statistically independent,
two different nonlocal update schemes are possible:
the time slice blocking and the square blocking. 
In the time slice blocking method updates of one dimensional 
blocks of size $L_B$ are performed
on separate time slices $\Lamt$ in sequence.
In the square blocking scheme as used by Laursen, Smit and Vink
one builds $L_B \times L_B$ 
blocks out of ``staples'' of $L_B$ one dimensional
blocks of size $L_B$ and performs the updates on this
square block simultaneously.
The analysis of
the kinematics is the same for both schemes.
For simplicity we are going to adopt the time slice blocking in the
following.

An important point is that the decoupling of
time slices is independent of the gauge group
and carries over to higher dimensions. We are going to use this
fact as  a basic ingredient 
of nonlocal updating schemes for nonabelian gauge fields.

%%%%%%%%%%%%%%%%%%%%%%%%%%%%%%%%%%%%%%%%%%%%%%%%%%%%%%%%%%%%%%%%%%%
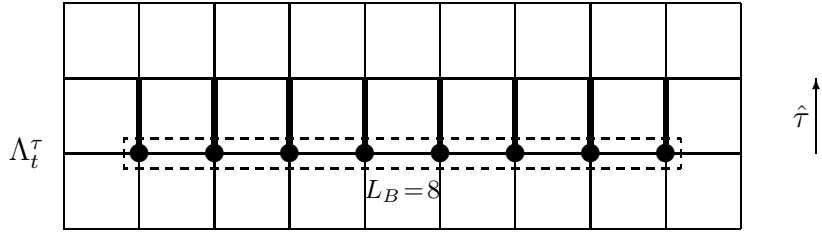
\begin{figure} \begin{center}
\begin{picture}(80,30)(0,0)
\multiput(0,0)(0,10){4}{\line(1,0){90}}
\multiput(0,0)(10,0){10}{\line(0,1){30}}
\put(100,10){\vector(0,1){10}}
\put(93,10){\makebox(10,10){$\hat{\tau}$}}
\put(40,0){\makebox(10,10){\footnotesize $L_B \! = \! 8$}}
\put(8,8){\dashbox{1}(74,4)}
\put(-10,5){\makebox(10,10){$\Lamt$}}
\linethickness{0.7mm}
\multiput(10,10)(10,0){8}{\line(0,1){10}}
\multiput(10,10)(10,0){8}{\circle*{2.5}}
%\multiput(10,20)(10,0){8}{\circle{2.5}}
\end{picture}
\end{center} 
\caption[dummy]{\label{tdtsb} 
                \sl Geometry of the time slice blocking:
                The marked link variables are updated simultaneously.
                The bottom of the block is indicated by a dashed line.}
\end{figure}
%%%%%%%%%%%%%%%%%%%%%%%%%%%%%%%%%%%%%%%%%%%%%%%%%%%%%%%%%%%%%%%%%%%

\subsection{Nonabelian gauge fields: $SU(2)$ in two dimensions}

\subsubsection{The nonabelian character of the gauge field}

Our method for a nonlocal updating procedure for nonabelian
$SU(2)$ gauge fields in two dimensions 
is based on the time slice blocking.
It is illustrated in figure \ref{tdtsb}.
We start the discussion with a naive generalization of the updates
in the abelian case:
Choose a one dimensional block $x_o'$ of size $L_B$ within a 
time slice $\Lamt$ and update all link variables $\Ut{x}$ pointing
from this block in the $\tau$-direction 
\be \label{2dsilly}
\Ut{x} \rightarrow  \Ut{x}' = R_x \Ut{x} \ \
\mbox{for all} \  \ x \in x_o' \  ,
\ee
where the ``rotation'' matrices $R_x$ are in $SU(2)$.
We parametrize them as
\be
R_x(\vn,s) = \cos( s \psi_x /2 )
 + i \sin( s \psi_x /2) \,
\vn \!\cdot\! \vs \, ,
\ee
where $\vn$ denotes a three dimensional real unit vector,
and the $\sigma_i$ are Pauli matrices.
$\vn$ will be taken randomly from the three dimensional unit sphere,
and $\psi$ will have support on the one dimensional block $x_o'$. 

The simplest version is a piecewise constant ``rotation'' 
where $R_x$ is a rotation matrix $R$ independent of $x$.
Let us examine how a plaquette in the interior of the block
(as illustrated in figure \ref{plaive}) changes under this update:
%%%%%%%%%%%%%%%%%%%%%%%%%%%%%%%%%%%%%%%%%%%%%%%%%%%%%%%%%%%%%%%%%%%
\begin{figure} \begin{center}
\begin{picture}(80,20)(0,0)
% first plaquette %%%%%%%%%%%%%%%%%
\multiput(10,10)(0,20){2}{\line(1,0){20}}
\multiput(10,10)(20,0){2}{\line(0,1){20}}
\multiput(10,10)(0,20){2}{\vector(1,0){11}}
\multiput(10,10)(20,0){2}{\vector(0,1){11}}
\put(15,0){\makebox(10,10){$U_1$}}
\put(15,30){\makebox(10,10){$U_3$}}
\put(0,15){\makebox(10,10){$U_4$}}
\put(30,15){\makebox(10,10){$U_2$}}
\put(15,15){\makebox(10,10){old}}

% second plaquette %%%%%%%%%%%%%%%%%
\multiput(60,10)(0,20){2}{\line(1,0){20}}
\multiput(60,10)(20,0){2}{\line(0,1){20}}
\multiput(60,10)(0,20){2}{\vector(1,0){11}}
\multiput(60,10)(20,0){2}{\vector(0,1){11}}
\put(65,0){\makebox(10,10){$U_1$}}
\put(65,30){\makebox(10,10){$U_3$}}
\put(50,15){\makebox(10,10){$RU_4$}}
\put(80,15){\makebox(10,10){$RU_2$}}
%\put(55,5){\makebox(10,10){$R$}}
%\put(75,5){\makebox(10,10){$R$}}
\put(65,15){\makebox(10,10){new}}

\put(40,20){\vector(1,0){10}}
\put(100,15){\vector(0,1){10}}
\put(93,15){\makebox(10,10){$\hat{\tau}$}}

\end{picture}
\end{center} 
\caption[dummy]{\label{plaive} 
                \sl Change of plaquette during naive nonlocal updating}
\end{figure}
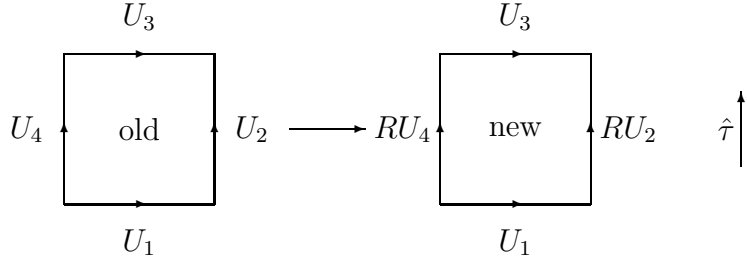
%%%%%%%%%%%%%%%%%%%%%%%%%%%%%%%%%%%%%%%%%%%%%%%%%%%%%%%%%%%%%%%%%%%
$$
P_{old} = \half \Tr( U_1 U_2 U_3^* U_4^* ) \  \rightarrow \
$$
\be
P_{new} = \half \Tr( U_1 R U_2 U_3^* (R U_4)^* ) 
= \frac{1}{2} \Tr(  R^* U_1 R U_2 U_3^* U_4^* ) 
\ee
with the notations as in figure \ref{plaive}.
Let denote the links 
\be
( x,x+\mu) \ \mbox{for all} \ x,x+\mu \in x_o' 
\ee
as the links in the bottom of the block,
as marked in figure \ref{tdtsb}.
In the abelian case the rotation matrix $R$ and 
the link variable in the bottom of the block $U_1$ would commute,
and we would have $P_{old} = P_{new}$, \mbox{i.e.\ } the plaquettes
in the interior of the block would remain unchanged.
 
But since our gauge group is nonabelian
the resulting change of the plaquette by this naive generalization
would lead to energy costs proportional to the volume of the block.
 
Recall the intuitive Leitmotiv for the development of new multigrid
methods from section~\ref{SECappl}:
{\em A piecewise constant update of a nonlocal domain should have
energy costs proportional to the surface of the domain,
but not energy costs proportional to the volume of the domain.}

Energy costs proportional to the volume
of the block lead to an algorithmic mass term that suppresses the 
amplitudes of the nonlocal moves on large scales.
Therefore the naive generalization will suffer from CSD.

This does not come as a surprise.  Due to the gauge invariance
of the model, the link variables do not have a gauge invariant meaning.
Therefore the rotation $R$ that is constant over the block for a given
gauge can be arbitrarily rough and disordered after a gauge
transformation. It is therefore natural to assume that the rotation
matrices have to be chosen in a gauge covariant way.

\subsubsection{Gauge covariant time slice blocking algorithm}

We use the additional 
gauge degrees of freedom
and generalize (\ref{2dsilly}) to
\be \label{2dclever}
\Ut{x} \rightarrow  \Ut{x}' =  R_x(g) \Ut{x} \ \
\mbox{for all} \ \ x \in x_o' \  ,
\ee
with $R_x(g) = g_x^* R_x g_x$ and $g$-matrices $g_x \in SU(2)$.
In the abelian case we obtain nothing new,
because $g_x$ and $R_x$ commute.
In the nonabelian case a plaquette in the interior of the block 
changes under a piecewise constant rotation $R$ according to
\be \label{2dchange}
P_{old} = \half \Tr( U_1 U_2 U_3^* U_4^* ) \  \rightarrow \
P_{new} = 
 \half \Tr(  R^* U_1^g R U_2^g U_3^{g *} U_4^{g *} )  \ .
\ee
The gauge transformed link variables
are  given by $\Um{x}^g = g_x \Um{x} g_{x + \mu}^*$.
If we choose the \mbox{$g$-matrices} in such a way that the 
link variable $U_1^g$
in the bottom of the block 
is equal to unity, it  will commute with an arbitrary rotation 
matrix $R$. Then the plaquettes
in the interior of the block stay unchanged 
as in the abelian case.

From this discussion we are led to the following gauge condition:
Choose the $g$-matrices in the bottom of the 
one dimensional block $x_o'$ such that
\be \label{block_axial_gauge}
 \Um{x}^g = g_x \Um{x} g_{x+\hat{\mu}}^* = 1 \ \
\mbox{for all} \ \ (x,x+\hat{\mu}) \in x_o' \ .
\ee
(The bottom of the block is indicated in figure \ref{tdtsb}.) 
In other words: All gauge transformed link variables 
in the one dimensional bottom of the block should be equal to unity.
We denote this gauge condition as block axial gauge.
Note that we still have the freedom of a constant gauge transformation
within each block $x'$,
\be
g_x \rightarrow h_{x'} g_x \ \ \mbox{for all} \ \  x \in x' \ .
\ee

Although we use the term ``gauge condition'' here,
we do not intend to
perform the gauge transformation $g$. We use the concept of
gauging only to define covariant rotations $R_x(g) = g_x^* R_x g_x$.

The gauge transformation properties of these updates are as follows:
If we apply an arbitrary gauge transformation $h$ to the gauge field $U$
\be
\Um{x} \ \rightarrow \ \Um{x}^h =  h_x \Um{x} h_{x+\hat{\mu}}^* \,
\ee
the $g$-matrices transform like
\be
g_x \ \rightarrow \ g_x^h = g_x h_x^* \,
\ee
and the covariant rotation matrix $R_x(g) = g_x^* R_x g_x$
transforms according to the adjoint representation:
\be
R_x(g)  \ \rightarrow \ (R_x(g))^h  = h_x R_x(g) h_x^* \ .
\ee
As a consequence,
if we apply the updates to the gauge transformed configuration $U^h$ 
$$
\Ut{x}^h  \ \rightarrow \ (\Ut{x}^h)' = (R_x(g))^h \Ut{x}^h 
=  h_x R_x(g) h_x^* h_x \Ut{x} h_{x+\hat{\tau}}^* 
$$
\be
= h_x R_x(g) \Ut{x} h_{x+\hat{\tau}}^* 
= h_x \Ut{x}' h_{x+\hat{\tau}}^*
= (\Ut{x}')^h \ ,
\ee
the updating commutes with a gauge transformation
$h$ and is therefore gauge covariant.

Let us summarize the steps of the time slice blocking scheme for
$SU(2)$ in the unigrid language:
\begin{enumerate}

\item Choose a block $x_o'$ of size $L_B$ that is contained
in the slice $\Lamt$. All link variables $\Ut{x}$ pointing from
sites $x$ inside the block in $\tau$-direction will be moved
simultaneously.

\item Find the gauge transformation $g$ defined by the
block axial gauge condition (\ref{block_axial_gauge})
such that $\Um{x}^g =1$ for all link variables 
in the bottom of the block.

\item Propose new link variables $\Ut{x}'$ by
\be
\Ut{x} \rightarrow \Ut{x}' = R_x(g) \, \Ut{x} \, ,
\ee
with $R_x(g) = g_x^* R_x g_x$ and
\be
R_x(\vn,s) = \cos( s \psi_x /2 )
 + i \sin( s \psi_x /2) \,
\vn \!\cdot\! \vs \, .
\ee
$s$ is a uniformly distributed random number from the
interval $[-\varepsilon,\varepsilon]$, $\vn$ is a
vector selected randomly from the three dimensional unit sphere,
and $\psi$ is a one dimensional kernel.

\item Calculate the associated change of the Hamiltonian  $\Delta \calH$
and accept the proposed link variables with probability $ \min
\lbrack 1,\exp(-\Delta \calH) \rbrack $.
\end{enumerate}

The detailed balance condition is fulfilled
by this updating scheme:  For the naive version with $g = 1$ it is
straightforward to show that the detailed balance condition holds, since
the rotation matrices $R_x$ are chosen according to a probability
distribution which is symmetric around unity.

If we now take $g$ according to some gauge condition, we have to be
careful that we get the same $g$ before and after the move $\Ut{x}
\rightarrow \Ut{x}'$ is performed.  
Otherwise this move would not be reversible.  In
other words:  The gauge condition yielding $g$ must not depend on
$\Ut{x}$.  This is indeed the case, since only link variables $\Um{x}$
with $\mu \neq \tau$ enter in the block axial gauge condition. 

The details of an implementation and simulation
of the covariant time slice blocking algorithm 
for $SU(2)$ gauge fields will be described in section \ref{SECSU22sim}.

\subsubsection{Acceptance analysis of the proposal}

The energy change associated with the update proposal (\ref{2dclever}) is
\be
\Delta \calH = - \frac{\beta}2
\sum_{\calP} \Tr \bigl( U_{\calP}' - U_{\calP} \bigr)
= - \frac{\beta}2 
\sum_{\stackrel{{\mbox{\scriptsize $x \in \Lamt$}}}{\mu \neq \tau}}   
\Tr \bigl\{
( R_x(g)^* \Um{x} R(g)_{x+\hat\mu} - \Um{x} ) H_{x,\mu}^* \bigr\} \, ,
\ee
where $H_{x,\mu}^* = \Ut{x+\hat\mu} \Um{x+\hat\tau}^* \Ut{x}^*$,
and $\psi$ stands for a one dimensional interpolation kernel.
The relevant quantity for the acceptance rates is
$h_1 = \EW{\Delta \calH}$.
If we assume that the $g$-matrices are chosen according to the
block axial gauge condition and that 
$\psi$ vanishes outside the block $x_o'$ 
we find 
$$ 
h_1 = \beta P \, 
\sum_{\stackrel{{\mbox{\scriptsize $x \in \Lamt$}}}{\mu \neq \tau}}  
\left[ 1- \cos \left( s(\psi_x-\psi_{x+\hat{\mu}})/2 \right) \right] 
$$
\be \label{2dgauge_deltah}
=  
\frac{s^2}{8} \beta P \, 
\sum_{\stackrel{{\mbox{\scriptsize $x \in \Lamt$}}}{\mu \neq \tau}}
 \left(\psi_x-\psi_{x+\hat{\mu}} 
\right)^2 + O(s^4) 
= \frac{s^2}{8} \beta P \alpha_1 + O(s^4)
\ee
with $P = \EW{ \half \Tr U_{\cal P} }$ and
$\alpha_{1} = (\psi,-\Delta \psi) $. 
The derivation of eq.\ (\ref{2dgauge_deltah})
is described in appendix~\ref{APPacc}.
Remember that 
for the time slice blocking in two dimensions 
$\psi$ is a one dimensional kernel.

Thus the kinematical behavior of this method is the same as
that of the massless Gaussian model in one dimension.
In contrast to $U(1)$ lattice gauge theory in two dimensions
we do not expect any additional topological problems to occur
for $SU(2)$. Therefore we expect a successful acceleration
of the simulation by the proposed time slice blocking algorithm.
This prediction will be verified in in section \ref{SECSU22sim}.

\subsection{Summary}

In this section we discussed multigrid algorithms
for lattice gauge theory in two dimensions.
The successful acceleration of Polyakov loop observables in
the simulation of abelian gauge fields in two dimensions
by Laursen, Smit and Vink can be understood by the kinematical analysis.

An important observation is the statistical decoupling of 
adjacent time slices as long as only link variables
in the time direction are updated. The
time slice blocking algorithm is based on this property.
The statistical independence of adjacent time slices 
is independent of the gauge group and of the dimensionality.

The nonabelian character of the gauge field was discussed. 
We proposed the gauge covariant time slice blocking
for $SU(2)$ in two dimensions that is particularly adapted
to the nonabelian character.
From the kinematical analysis we expect 
a strong reduction of CSD in a 
simulation of the proposed algorithm.

\clearpage

\section{Multigrid Monte Carlo simulation of $SU(2)$
         lattice gauge fields in two dimensions} 

\label{SECSU22sim}

\setcounter{equation}{0}

In this section we simulate $SU(2)$ lattice gauge theory
in two dimensions by a multigrid algorithm. 
The acceptance analysis of the previous section predicted
that CSD can be eliminated by the time slice blocking.
To verify this statement, we perform
numerical experiments on systems with
lattice sizes up to $256^2$. 
The details of the implementation of the time slice blocking are
discussed. Then the run parameters and numerical results are reported.

\subsection{Implementation of the time slice blocking}

For a concrete implementation of the time slice blocking method as 
introduced in section~\ref{SECSU22acc} we choose a unigrid algorithm 
with piecewise linear interpolation and a V-cycle. 

\subsubsection{Smooth nonlocal heat bath updates in $U(1)$ subgroups}

Although the analysis of nonlocal updates in section \ref{SECSU22acc} 
was performed in terms of a Metropolis version,
we are going to use a heat bath version of the nonlocal moves in our 
simulation.
We think that the change of the update method from
Metropolis to heat bath will not affect the dynamical critical
behavior in a substantial way.
The advantage of the heat bath updating is that there are no
tuneable parameters such as the Metropolis step size $\varepsilon(L_B)$.

A heat bath implementation of nonlocal time slice blocking updates
is possible if we use one dimensional piecewise linear
interpolation and update in $U(1)$ subgroups of $SU(2)$.
This will be described in the following.

Compared to section \ref{SECSU22acc},
we formulate the piecewise linear interpolation in a different language:
Assume that the 
$g$-matrices defined according to the block axial gauge condition 
(\ref{block_axial_gauge})
have been applied as a gauge transformation in the bottom of the block
$x_o'$. Then the gauged link variables 
in the bottom of the block are equal to unity.
Now a piecewise linear block update is formulated
by multiplying the $L_B$ gauged link variables $\Ut{x}^g$ 
pointing in $\tau$-direction
from left to right by the $SU(2)$-matrices
$R, R^2, R^3, \dots R^{L_B/2}, R^{L_B/2}, R^{L_B/2-1}, \dots, R$.
$R$ is given by
\be
R(\vn,\theta) = \cos( \theta )
 + i \sin( \theta) \,
\vn \!\cdot\! \vs \, ,
\ee
where the randomly chosen three-dimensional vector $\vn$ 
specifies the direction of a $U(1)$ subgroup in $SU(2)$.
All changes of plaquettes that are generated 
by this update can be written in the form
\be \label{U1hbplaq}
 \Tr(U_{\cal P}') = \Tr (R V) = 
\Tr(V) \cos(\theta) + \Tr(i \vn \!\cdot\! \vs V) \sin(\theta) \ ,
\ee
with the $SU(2)$ matrix $V = U_{\cal P}^g$ or $V = U_{\cal P}^{g *}$.
By summing over all 
changed plaquettes (\ref{U1hbplaq}) we obtain an overall change 
in the Hamiltonian
of the form
\be
 \calH(U')  = a \cos(\theta) + b \sin(\theta) + const \ ,
\ee
with real constants $a$ and $b$. To generate 
$U(1)$ random numbers distributed according to the distribution
\be
 \mbox{dprob}(\theta) \propto 
\mbox{e}^{a \cos(\theta)+b \sin(\theta)} d\theta
\ee
we 
use the fast vectorizable method of Hattori and Nakajima \cite{hattori}.

\subsubsection{Sequence of updates}

The sequence of updates is organized as follows:
We start with a V-cycle of time slice blocking updates
on the links $\Ut{x}$ pointing in the $\tau = 1$ direction.
The largest block size is $L/2$ on a $L \times L$ lattice. 
This means that the sequence of updated block sizes is 
$L_B = 2, 4, \dots L/2, L/2, L/4 \dots 2$.
When all time slices have been updated by a V-cycle,
we perform a sweep of local $SU(2)$ heat bath updates
through all links on the lattice.
Here we use the ``incomplete'' 
Kennedy-Pendleton \cite{kennedypendleton} heat bath algorithm
with one trial per
link. 
This means that in a sweep through the lattice not
all links  but only a very high percentage 
of them are updated. The advantage is that 
scalar operations on a vector computer are avoided
\cite{fredenhagenmarcu}. 
Then we do a V-cycle on all links pointing in the 
$\tau = 2$ direction and again a local heat bath sweep. 
This sequence is repeated periodically.

Measurements are performed after each local heat bath sweep.
To avoid effects from fixed block boundaries,
we use stochastically overlapping blocks \cite{hmsun} 
by applying a random 
translation before each V-cycle.  

\subsubsection{Axial gauge}

In order to save computer time we use a slight modification
of the gauge condition.
Recall the block axial gauge (\ref{block_axial_gauge}): 
Take the $g$-matrices in the bottom of a block $x_o'$ such that
\be
 \Um{x}^g = g_x \Um{x} g_{x+\hat{\mu}}^* = 1 \ \
\mbox{for all} \ \ (x,x+\hat{\mu}) \in x_o' \ .
\ee

We modify it to the axial gauge: 
Take the $g$-matrices in the bottom of a time slice $\Lamt$ such that
\be
 \Um{x}^g = g_x \Um{x} g_{x+\hat{\mu}}^* = 1 \ \
\mbox{for} \ \ (x,x+\hat{\mu}) \in \Lamt, \; x_{\mu} = 1,\dots L-1 \ .
\ee
In other words: all spatial links but the last link 
are gauged to one within the bottom of a time slice.

The computational advantage of this slight change in the gauge condition
is that for the block axial gauge the $g$-matrices have to be 
calculated for each block lattice individually.
For the axial gauge the $g$-matrices
are the same for all block lattices and they do not change during the
updating on different block lattices with different $L_B$.
Therefore we have to calculate them only once before
performing the entire V-cycle.  
The gauge covariance properties of the update is not affected
by this modification.  

\subsection{Simulation and results}

\subsubsection{Observables}
The observables measured are square Wilson loops
\be
W(I,I) = \langle \half \Tr (U(C_{I,I}) \rangle \ ,
\ee
where $U(C_{I,I})$ is the parallel transporter 
around a rectangular Wilson loop $C_{I,I}$ of size $I \times I$.
On an $L \times L$-lattice we measure $W(1,1),  
W(2,2), W(4,4), W(8,8),\dots, W(L/2,L/2)$.
Another important class of 
quantities is built up from timelike Polyakov loops.
A Polyakov loop at the one dimensional spatial point $r$ 
is defined by
\be
P_{r} = \half \Tr \prod_{t=1}^{L} U_{(r,t),2} \ .
\ee
We measure the lattice averaged Polyakov loop
\be
\bar{P} = \left\langle \frac{1}{L}\sum_{r=1}^L P_{r} \right\rangle \ ,
\ee
and the lattice averaged Polyakov loop squared
\be
\bar{P}^2 = \left\langle \left(\frac{1}{L}
\sum_{r=1}^L P_{r} \right)^2 \right\rangle \ .
\ee

\subsubsection{Run parameters and results}

In order to investigate the dynamical critical behavior of the
multigrid algorithm, we simulate a sequence of lattices with 
fixed physical size $L \approx 10 \xi$,
where the correlation length $\xi$ is related to the string tension
by $\kappa$ by $\xi = 1/\sqrt{\kappa}$.
Then we have to use lattice sizes and $\beta$ values such that
$L^2/\beta$ is constant.
If we choose this large ratio of $L/ \xi$,
finite size effects are negligible.
The detailed run parameters are given in table 
\ref{tab:SU22param}. The quoted correlation length is calculated by the 
exact solution \cite{bailandrouffe} in the infinite volume limit.
The main results of this solution are summarized in appendix \ref{APPSU22x}.
We started our runs from ordered configurations 
with all link variables set equal to unity. After equilibration,
measurements were taken after each local heat bath sweep 
through the lattice.
%%%%%%%%%%%%%%%%%%%%%%%%%%%%%%%%%%%%%%%%%%%%%%%%%%%%%%%%%%%%%%%%%%%%%%%%%%%%%%%
\begin{table}[htbp]
 \centering
 \caption[dummy]{\label{tab:SU22param}
         Run parameters for the multigrid Monte Carlo
         simulation of two dimensional $SU(2)$ lattice gauge theory.}
  \vspace{2ex}
\begin{tabular}{|c||c|c|c|c|c|}
\hline\str
$\beta$ & 4& 16 & 64 & 256 & 1024
\tabhline
$L$ & 16& 32 & 64 & 128 & 256
\tabhline
$\xi$ & 1.55 & 3.22 & 6.51 & 13.05 & 26.12
\\[.3ex] \hline
\end{tabular} \end{table}
%%%%%%%%%%%%%%%%%%%%%%%%%%%%%%%%%%%%%%%%%%%%%%%%%%%%%%%%%%%%%%%%%%%%%%%%%%%%%%%

The static results of the simulation 
are given in table \ref{tab:SU22static}.
All our results for the Wilson loops are consistent with the
exact solution (\ref{tdwilson}) in the infinite volume limit.
Here only results for 
$W(\frac{L}{16},\frac{L}{16})$ and $W(\frac{L}{8},\frac{L}{8})$
are quoted, which are the loops of about the size
of a correlation length squared. 
We observed that these 
loop sizes have the largest 
autocorrelation times among the Wilson loops.

\begin{figure}[htbp]
 \begin{center}
  \begin{minipage}[t]{145mm} 
   \begin{picture}(145,95)(45,130)
    \epsfig{file=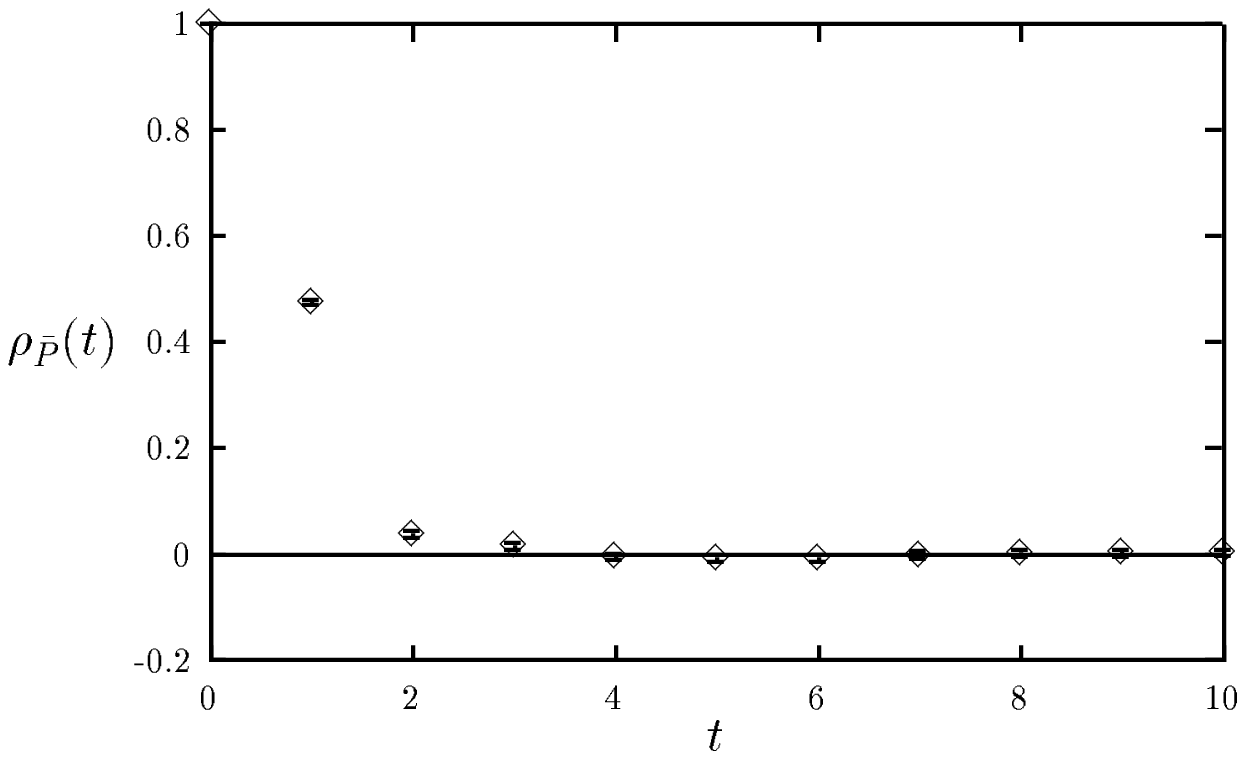,height=343mm}
   \end{picture}
   \begin{picture}(145,95)(45,130)
    \epsfig{file=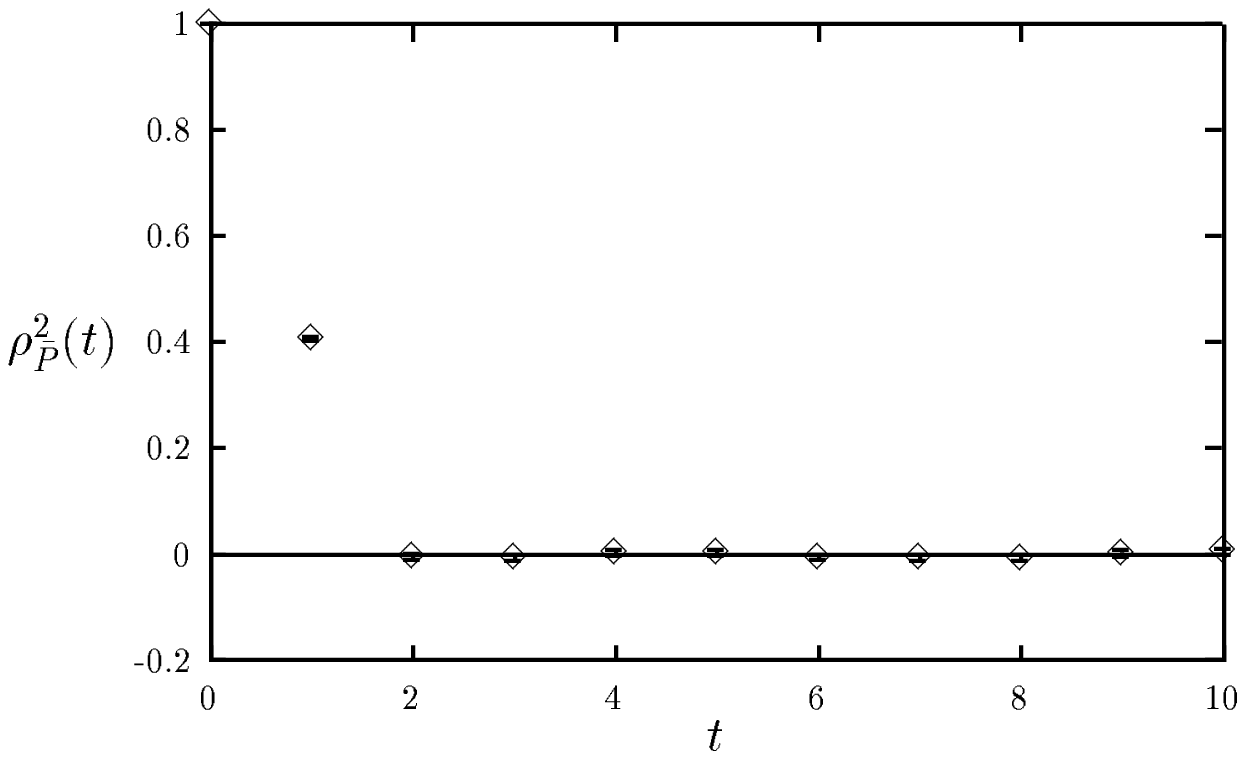,height=343mm}
   \end{picture}
   \caption[$\rho(t)$ in the two dimensional $SU(2)$ lattice gauge theory]
   {\label{SU22rho}
   \sl Autocorrelation functions $\rho(t)$ 
   in the two dimensional $SU(2)$ pure lattice gauge theory,
   $\beta=1024$ on a $256^2$ lattice.
   Top: $\rho(t)$ for $\bar{P}$,
   bottom: $\rho(t)$ for $\bar{P}^2$.}
   \end{minipage}
 \end{center}
\end{figure}

In general we found a very fast 
decorrelation of subsequent configurations in the Markov chain.
Typically the autocorrelation function 
$\rho(t)$ dropped to zero within errors
after $3 - 5$ measurements
(see figure \ref{SU22rho}). Therefore it was impossible to look
for an exponential regime in the decay of $\rho(t)$. 
We tried to estimate integrated autocorrelation 
times $\tau_{int}$
with a self consistent truncation window of 
$ 4 \tau_{int}$ according to the method of ref. \cite{madras}.
The results for the integrated autocorrelation times
are given in table \ref{tab:SU22dynamic}.
Estimates for $\tau_{int}$ are only quoted if we observed 
that $\rho(t)$ was positive in the entire interval
from $t = 0$ to $t = 4 \tau_{int}$.
Note that $\tau_{int}$ is defined such that 
$\tau_{int} = 0.5$ in the case of complete decorrelation.
All our runs are longer than $30\,000 \tau_{int}$.

In summary,
all $\tau$'s are smaller or consistent with one in the range of
parameters studied, with a very weak tendency to increase
with increasing lattice size.
Due to the ambiguities of the estimation of $\tau$
in the situation of almost complete decorrelation,
we do not want to give an estimate for $z$ here.
We only state that the results indicate that CSD is 
almost completely eliminated by the time slice blocking algorithm.

%%%%%%%%%%%%%%%%%%%%%%%%%%%%%%%%%%%%%%%%%%%%%%%%%%%%%%%%%%%%%%%%%%%%%%%%%%%%%%%
\begin{table}[htbp]
 \centering
 \caption[dummy]{\label{tab:SU22static}
         Static observables in the 
         two dimensional $SU(2)$ lattice gauge theory on
         $L \times L$ lattices,
         $L/\xi \approx 10$.}
 \vspace{2ex}
\begin{tabular}{|r|r|r|l|l|r|c|}
\hline\str
$L\;$ &
statistics & 
discarded & 
\multicolumn{1}{c|}{${W(\frac{L}{16},\frac{L}{16})}$} & 
\multicolumn{1}{c|}{${W(\frac{L}{8},\frac{L}{8})}$} &
\multicolumn{1}{c|}{$\bar{P}$} &
\multicolumn{1}{c|}{${\bar{P}^2}$} 
\\[.5ex] \hline \hline \str
 16 & 100\,000 & 10\,000 & 0.65816(10) & 0.18751(14) &
 0.0002(4) & 0.01564(7) 
\tabhline
 32 & 100\,000 & 10\,000 & 0.67917(4) & 0.21287(12) &
 0.0002(3) & 0.00855(4) 
\tabhline
 64 & 50\,000 & 10\,000 & 0.68525(6) & 0.2204(2) &
 -0.0003(5) & 0.00610(5)
\tabhline
128 & 40\,000 & 5\,000 & 0.68688(6) & 0.2222(3) &
 0.0008(9) & 0.00545(8)
\tabhline
256 & 40\,000 & 5\,000 & 0.68720(7) & 0.2232(2) &
 0.0001(6) & 0.00519(4)
\\[.3ex] \hline
\end{tabular} \end{table}
%%%%%%%%%%%%%%%%%%%%%%%%%%%%%%%%%%%%%%%%%%%%%%%%%%%%%%%%%%%%%%%%%%%%%%%%%%%%%%%

%%%%%%%%%%%%%%%%%%%%%%%%%%%%%%%%%%%%%%%%%%%%%%%%%%%%%%%%%%%%%%%%%%%%%%%%%%%%%%%
\begin{table}[htbp]
 \centering
 \caption[dummy]{\label{tab:SU22dynamic}
         Integrated autocorrelation times 
         $\tau_{int}$ for the 
         two dimensional $SU(2)$ lattice gauge theory on
         $L \times L$ lattices,
         $L/\xi \approx 10$.
         If no value is given, we found
         almost complete decorrelation.}
 \vspace{2ex}
\begin{tabular}{|r|r|r|c|c|c|c|c|}
\hline\str
$L\;$ &
statistics & 
discarded & 
\multicolumn{1}{c|}{$\tau_{int,W(\frac{L}{16},\frac{L}{16})}$} & 
\multicolumn{1}{c|}{$\tau_{int,W(\frac{L}{8},\frac{L}{8})}$} &
\multicolumn{1}{c|}{$\tau_{int,\bar{P}}$} &
\multicolumn{1}{c|}{$\tau_{int,\bar{P}^2}$} 
\\[.5ex] \hline \hline \str
 16 & 100\,000 & 10\,000 & 0.54(1) & - &
 - & - 
\tabhline
 32 & 100\,000 & 10\,000 & - & 0.60(1) &
 - & - 
\tabhline
 64 & 50\,000 & 10\,000 & 0.67(1) & 0.70(1) &
 0.71(1) & 0.59(1)
\tabhline
128 & 40\,000 & 5\,000 & 0.76(2) & 0.74(2) &
 0.92(2) & -
\tabhline
256 & 40\,000 & 5\,000 & 0.88(3) & 0.83(2) &
 1.01(3) & -
\\[.3ex] \hline
\end{tabular} \end{table}
%%%%%%%%%%%%%%%%%%%%%%%%%%%%%%%%%%%%%%%%%%%%%%%%%%%%%%%%%%%%%%%%%%%%%%%%%%%%%%%

\subsection{Summary}

In this section we reported on a multigrid Monte Carlo simulation
of $SU(2)$ lattice gauge theory in two dimensions.
The time slice blocking algorithm 
with piecewise linear interpolation was implemented 
in the unigrid style with a V-cycle. We used a heat bath 
version in randomly chosen $U(1)$ subgroups. The simulations
were performed on physically large lattices of the size 
$10 \xi \times 10 \xi$.

All observed integrated autocorrelation times
are found to be smaller or consistent with one on lattice sizes up to
$256^2$, with a very weak tendency to increase
with increasing lattice size.
Therefore we conclude that the time slice blocking algorithm
eliminates CSD almost completely.

It is fair to say that we used
a special feature of two dimensional lattice gauge theory:
In the infinite volume limit or for open boundary conditions
one can decouple two dimensional lattice gauge theory to a set 
of independent one dimensional spin models 
by choosing the axial gauge. 
Although we use periodic boundary conditions here, our method
is very much in the spirit of updating on independent time slices.

It will be discussed in the next section
whether this concept is still successful if we generalize 
it to four dimensions.

\clearpage

\section{Multigrid methods for lattice gauge fields in four dimensions}

\label{SECSU24acc}

\setcounter{equation}{0}

In this section we generalize the multigrid procedure for
pure lattice gauge theory from two dimensions to four dimensions.
We start with a discussion of the abelian case.

For nonabelian gauge fields in four dimensions additional 
difficulties arise because of the more complicated geometrical structure 
of nonlocal updates in four dimensions compared to two dimensions.
We study the behavior of acceptance rates with increasing block size $L_B$
and analyze the special features of the 
algorithm in the four dimensional case in detail.

\subsection{The abelian case}
\label{SECabel}

The generalization of the multigrid algorithm for $U(1)$ gauge fields of
Laursen, Smit and Vink (as discussed in section \ref{SECSU22acc})
from two dimensions to higher dimensions is straightforward:

Nonlocal updates are formulated as follows:
One chooses a hypercubic block $x_o'$ of size $L_B^d$
and a direction $\tau$ with $1 \leq \tau \leq d$.
All the link variables $\Ut{x}$ attached to sites $x$ inside the block
$x_o'$ are proposed to be changed simultaneously:
\be
\Ut{x} \rightarrow \exp(is\psi_x)\Ut{x} \, ,
\ee
or in terms of the link angles
\be 
\theta_{x,\tau} \rightarrow  \theta_{x,\tau} + s\psi_x  \, .
\ee
The kernel $\psi$ obeys the normalization condition
(\ref{normpsi}).
The canonical dimension of the kernel $\psi$ and of the angle
$\theta$ is that of a vector potential,
i.e.\ $(2-d)/2$ in $d$ dimensions.
The simplest choice for $\psi$ is again a piecewise constant kernel,
\be
 \psi^{const}_x=\left\{
\begin{array}{ll}
{L_B}^{(2-d)/2} &\mbox{for}\; x \in x_o' \\
0&\mbox{for}\; x \not \in x_o' \; .\\
\end{array}
\right.
\ee

Let us now study the acceptance rates for these update proposals
with the help of formula (\ref{formula}). We consider
general kernels $\psi$.
For $h_1 = \langle \Delta \calH \rangle$ we find 
(cf. appendix \ref{APPacc})
\be 
h_1\,=\,
\beta P \sum_{x \in \Lambda_0}
\sum_{\mu \neq \tau}
\left[ 1-\cos\bigl( s (\psi_{x+\hat{\mu}}-\psi_x)\bigr) \right] \ \ ,
\ee
with $P = \EW{ \Tr U_{\cal P} }$.

Because of the statistical independence of adjacent time slices we 
choose $\psi_x$ to be constant in the $\tau$-direction,
with the support of $\psi$ in $\tau$-direction
restricted to the block $x_o'$. Then we obtain
\be
h_1\,=\,
\beta P L_B \sum_{x \in \Lamt}
\sum_{\mu \neq \tau}
\left[ 1-\cos\bigl( s (\psi_{x+\hat{\mu}}-\psi_x) \bigr)\right] \ \ .
\ee
For small $s$, this can be approximated by
\be
h_1\,\approx \,
\half s^2 \beta P L_B \sum_{x \in \Lamt}
\sum_{\mu \neq \tau}
( \psi_{x+\hat\mu} - \psi_{x} )^2 =
\half s^2 \beta P \alpha_{d-1} \ \ ,
\ee
with $\alpha_{d-1} = (\psi',-\Delta \psi') $.
As in two dimensions $\psi'$ denotes the kernel $\psi$, restricted to the
$d-1$ dimensional time slice $\Lamt$. We absorbed
a factor $L_B^{1/2}$ in $\psi'$ (now normalized
as a \mbox{$d-1$} dimensional kernel).

In the special case of piecewise constant kernels we find for $h_1$
\be 
h_1 \,=\, 2(d-1) \beta P L_B^{d-1}[ 1 - \cos(s L_B^{(2-d)/2})] \, .
\ee
From the kinematical point of view, the behavior of acceptance rates in
the $U(1)$ lattice gauge theory in $d$ dimensions is the same as in
massless free field theory.

This  multigrid algorithm  with constant interpolation
and a W-cycle was studied for the simulation of
four dimensional $U(1)$ theory by Laursen and Vink \cite{newgauge}.
In this model there is a deconfinement phase transition between a 
confinement phase and a Coulomb phase. The transition is
believed to be of weakly first order \cite{U14firstorder}.
The authors report on an acceleration of the
Monte Carlo dynamics 
in both phases close to the transition point.
This can be understood by our kinematical analysis.
However the very low frequence of changes 
from one phase to the other 
on larger lattices in the simulation with a local algorithm 
could not be accelerated by the nonlocal updating.
Like the topological problems in the $XY$-model
and in the two dimensional $U(1)$ model,
the problem of metastability can not be resolved 
from a purely kinematical point of view.

\subsection{The nonabelian case: gauge group $SU(2)$}
\label{SUBSECsu2}

\subsubsection{Covariant time slice blocking for $SU(2)$ in four dimensions}
\label{SUBSUBSECcov}

%%%%%%%%%%%%%%%%%%%%%%%%%%%%%%%%%%%%%%%%%%%%%%%%%%%%%%%%%%%%%%%%%%%
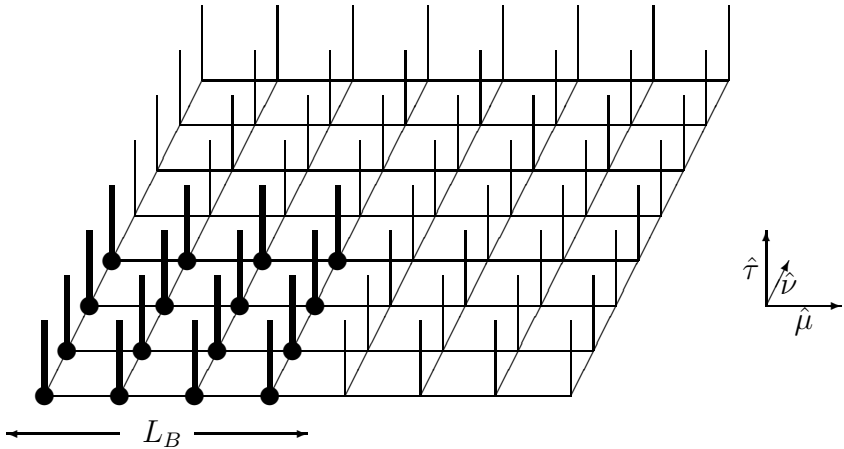
\begin{figure}
\begin{center}
\begin{picture}(160,55)(0,5)
\multiput(10,10)(3,6){8}{\line(1,0){70}}
\multiput(10,10)(10,0){8}{\line(1,2){21}}
\multiput(10,10)(3,6){8}{\line(0,1){10}}
\multiput(20,10)(3,6){8}{\line(0,1){10}}
\multiput(30,10)(3,6){8}{\line(0,1){10}}
\multiput(40,10)(3,6){8}{\line(0,1){10}}
\multiput(50,10)(3,6){8}{\line(0,1){10}}
\multiput(60,10)(3,6){8}{\line(0,1){10}}
\multiput(70,10)(3,6){8}{\line(0,1){10}}
\multiput(80,10)(3,6){8}{\line(0,1){10}}
\put(20,0){\makebox(10,10){ $L_B$}}
\put(20,5){\vector(-1,0){15}}
\put(30,5){\vector(1,0){15}}
\put(106,22){\vector(0,1){10}}
\put(106,22){\vector(1,0){10}}
\put(106,22){\vector(1,2){3}}
\put(99,22){\makebox(10,10){$\hat{\tau}$}}
\put(106,15){\makebox(10,10){$\hat{\mu}$}}
\put(104,20){\makebox(10,10){$\hat{\nu}$}}

\linethickness{0.7mm}
\multiput(10,10)(3,6){4}{\line(0,1){10}}
\multiput(20,10)(3,6){4}{\line(0,1){10}}
\multiput(30,10)(3,6){4}{\line(0,1){10}}
\multiput(40,10)(3,6){4}{\line(0,1){10}}
\multiput(10,10)(3,6){4}{\circle*{2.5}}
\multiput(20,10)(3,6){4}{\circle*{2.5}}
\multiput(30,10)(3,6){4}{\circle*{2.5}}
\multiput(40,10)(3,6){4}{\circle*{2.5}}
\end{picture}
\end{center}
\caption[dummy]{\label{4dblocking} 
                \sl Illustration of the geometry of time slice blocking 
                in three dimensional lattice gauge theory.}               
\end{figure}

We now discuss a generalization of the 
covariant time slice blocking algorithm of section~\ref{SECSU22acc}
from two dimensions to four dimensions.
Many steps of the two dimensional method can be translated
directly to the four dimensional case. Other features such as the nontrivial 
background field in the bottom of higher dimensional blocks will require
a refined treatment.  

Nonlocal updates can be defined as shown in figure \ref{4dblocking}
(for simplicity illustrated in three dimensions).
Let us consider a fixed time direction $\tau$ with $1 \leq \tau \leq 4$
and a three dimensional time slice $\Lamt =
\left\{ x \in \Lambda_0 \, \vert \, x_{\tau} = t \right\}$.
One chooses a cubic block $x_o'$ of size $L_B^3$ that is contained in $\Lamt$.
All the link variables $\Ut{x}$ attached to sites $x$ inside the block
$x_o'$ are proposed to be changed simultaneously:
\be \label{clever}
\Ut{x} \rightarrow \Ut{x}' =  R_x(g) \, \Ut{x} \, ,
\ee
where $R_x(g) = g_x^* R_x g_x$ and $g_x \in SU(2)$.
The rotation matrices $R_x \in SU(2)$
are again parametrized as
\be
R_x(\vn,s) = \cos( s \psi_x /2 )
 + i \sin( s \psi_x /2) \,
\vn \!\cdot\! \vs \, ,
\ee
where $\vn$ denotes a three-dimensional real unit vector,
and the $\sigma_i$ are Pauli matrices.
$\psi$ will
have support on the three dimensional block $x_o'$.

Up to now the $g$-matrices are arbitrary. 
In the two dimensional case we have chosen them according to
the block axial gauge (\ref{block_axial_gauge}):
Choose the $g$-matrices in the one dimensional 
bottom of the block $x_o'$ such that
\be
 \Um{x}^g = g_x \Um{x} g_{x+\hat{\mu}}^* = 1 \ \
 \mbox{for all} \ \ (x,x+\hat{\mu}) \in x_o' \ .
\ee
The gauge transformed link variables $\Um{x}^g$
in the bottom of the block 
were equal to unity and for the case of piecewise constant
interpolation the plaquettes
in the interior of the block stayed unchanged 
(recall the discussion following figure \ref{plaive}).
Therefore only the two plaquettes at the boundary of a 
block contributed to the energy change of a piecewise constant 
update. The energy cost was proportional to the surface
of the block, not proportional to the volume of the block. 

By choosing the block axial gauge in two dimensions we used the fact that
we always can gauge link variables along 
a one dimensional open line to unity.
Any nontrivial content of the
gauge field in the one dimensional bottom of the block
could be shifted outside of the block by applying the
block axial gauge.

However in more than two dimensions, the bottom of a block will contain closed 
loops (see figure \ref{4dblocking}).
Since a parallel transporter along a closed loop is gauge invariant,
we can not get rid of the nontrivial curvature that is contained in the
loop by any gauge transformation.
Therefore for nontrivial gauge fields
we can not find a gauge transformation $g$ such that
$U^g = 1$ for all link variables in the bottom of the block $x_o'$.
This means that for nontrivial gauge fields all timelike plaquettes
that share a link with the bottom of the block
will contribute to the average energy change of the update.

The intuitive Leitmotiv for the development of new multigrid
methods from section \ref{SECappl} was:
{\em A piecewise constant update of a nonlocal domain should have
energy costs proportional to the surface of the domain,
but not energy costs proportional to the volume of the domain.}

Unfortunately,
according to the discussion above,
the nontrivial background field in the bottom
of the block will lead to energy costs proportional to the 
volume of the block.
Let us nevertheless attempt to have as little energy costs
as possible: 

Consider the extreme case
of $\beta \rightarrow \infty$.  Then the allowed configurations are pure
gauges, i.e.\ configurations that are gauge equivalent to $\Um{x} = 1$
for all $x,\mu$.  If we choose $g$ as the transformation that
brings all links to unity, it is obvious that 
the plaquettes in the interior of the block will not be changed
by a piecewise constant update.
In particular,
to have this property, it is sufficient to gauge all links inside the
bottom of the block to unity. This
consideration leads to the following gauge condition: 
Choose $g$ as the gauge
transformation that maximizes the functional
\be\label{coulomb_gauge}
G_{C,x_0'}(U,g) = \sum_{(x,x+\hat\mu) \in x_o'} \Tr \bigl(
g_x \Um{x} g_{x+\hat\mu}^*\bigr) \, .
\ee
We call this gauge ``block Coulomb gauge''.\footnote{
Gauge conditions on the lattice and their relation to 
transversality conditions for the gauge potential $A$ are
discussed in appendix \ref{APPgauge}}
For finite $\beta$ this gauge will not bring all the links in the 
bottom of the block
to unity, but still as close to unity as possible.
Therefore the gauge field in the bottom of the block is as smooth as possible. 
This leads to a kind of minimization
of the energy costs from the interior of the block.
Note that in two dimensions the block Coulomb gauge condition
reduces to the block axial gauge (\ref{block_axial_gauge}).

The main difference compared to the two dimensional case
is that we
have to expect that the energy change of the update
will be proportional to the volume of the block $x_o'$.
This property is caused by the fact that in four dimensions 
the gauge field in the bottom of the block is smooth but nonzero.  
This will lead to an algorithmic mass term that 
grows quadratic with the block dimension $L_B$.
We are going to investigate 
the behavior of this algorithmic mass in detail below.

Let us summarize the steps of the nonlocal updating scheme for
$SU(2)$ in four dimensions:
\begin{enumerate}

\item Choose a block $x_o'$ of size $L_B^3$ that is contained
in the time slice $\Lamt$. All link variables $\Ut{x}$ pointing from
sites $x$ inside the block in $\tau$-direction will be moved
simultaneously.

\item Find the $g$-matrices defined by the
block Coulomb gauge condition
\be
G_{C,x_0'}(U,g) = \sum_{(x,x+\hat\mu) \in x_o'} \Tr \bigl(
g_x \Um{x} g_{x+\hat\mu}^*\bigr)
 \,\stackrel{\mbox{!}}{=} \, \mbox{maximal}\, .
\ee

\item Propose new link variables $\Ut{x}'$ by
\be
\Ut{x} \rightarrow \Ut{x}' = R_x(g) \, \Ut{x} \, ,
\ee
with $R_x(g) = g_x^* R_x g_x $ and
\be
R_x(\vn,s) = \cos( s \psi_x /2 )
 + i \sin( s \psi_x /2) \,
\vn \!\cdot\! \vs \, .
\ee
$s$ is a uniformly distributed random number from the
interval $[-\varepsilon,\varepsilon]$, and $\vn$ is a
vector selected randomly from the three dimensional unit sphere.

\item Calculate the associated change of the Hamiltonian  $\Delta \calH$
and accept the proposed link variables with probability $ \min
\lbrack 1,\exp(-\Delta \calH) \rbrack $.
\end{enumerate}

The argument that the detailed balance condition is fulfilled
by this updating scheme is analogous to two dimensions
(cf.\ section \ref{SECSU22acc}).
The only additional point is that although we only have an iterative
gauge fixing algorithm in four dimensions, 
we do not have to fix the gauge perfectly.  If we always use the
same procedure in finding $g$ (e.g.\ a given number of relaxation sweeps
starting from $g=1$), we will always get the same $g$ and the nonlocal
update is reversible.

The detailed implementation and simulation of nonlocal block updates
will be described in section~\ref{SECSU24sim}.  

\subsubsection{Acceptance analysis for nonlocal $SU(2)$-updates}
\label{SUBSUBSECaca}

First numerical studies revealed that there is no substantial difference
in the acceptance rates when instead of using the block Coulomb gauge
condition one uses the Coulomb gauge condition for the whole slice
$\Lamt$:
\be\label{coulomb_gauge_slice}
G_C(U,g) = \sum_{(x,x+\hat\mu) \in \Lamt} \Tr \bigl(
g_x \Um{x} g_{x+\hat\mu}^*\bigr)
 \,\stackrel{\mbox{!}}{=} \, \mbox{maximal} \, .
\ee
For a detailed discussion of Landau and Coulomb gauges on the lattice
see appendix \ref{APPgauge}.
From a practical point of view the Coulomb gauge condition is very
convenient: 
The $g$-matrices can be calculated once and then be used for all
block sizes $L_B$. In the block Coulomb gauge they would have to be 
recalculated for every individual block lattice.
In addition, the relaxation algorithm to determine
the $g$-matrices according to the Coulomb gauge condition
can be vectorized in a straightforward way.

The energy change associated with the update proposal (\ref{clever}) is
\be
\Delta \calH = - \frac{\beta}2
\sum_{\calP} \Tr \bigl( U_{\calP}' - U_{\calP} \bigr)
= - \frac{\beta}2 \sum_{x \in \Lamt}
\sum_{\mu \neq \tau} \Tr \bigl\{
( R_x^* \Um{x}^g R_{x+\hat\mu} - \Um{x}^g ) H_{x,\mu}^{g *} \bigr\} \, ,
\ee
with $H_{x,\mu}^* = \Ut{x+\hat\mu} \Um{x+\hat\tau}^* \Ut{x}^*$
and $U^g_{x,\mu} = g_x \Um{x} g_{x+\hat\mu}^*$. $H^g$ is defined analogously.
The relevant quantity for the acceptance rates is
$h_1 = \EW{\Delta \calH}$. For piecewise constant kernels
and the gauge condition (\ref{coulomb_gauge_slice}) we get
(cf. appendix \ref{APPacc})
\be \label{h1_constant}
h_1 = 3 \beta A \, (L_B -1) L_B^2 \, \sin^2(s L_B^{-1/2}/2) + \,
6 \beta P \, L_B^2 \bigl[ 1- \cos(s L_B^{-1/2}/2) \bigr] \, , 
\ee 
with
\bea \label{defA}
A &=&  \bigl\langle \half \Tr \bigl((\Um{x}^g - 
\vn \!\cdot\!  \vs \, \Um{x}^g \,
\vn\!\cdot\!\vs) H_{x,\mu}^{g\, *} \bigr) \bigr\rangle  \ ,\\
P &=& \ \ \langle \half \Tr U_{\calP} \rangle \ .
\eea

To the first term in eq.\ (\ref{h1_constant}) all links contribute that
are entirely inside the block, whereas the second term contains the
contributions of all links that have one site in common with the surface
of the block.  For small $s$, the first term behaves like $ s^2 L_B^2 $.
This is exactly the behavior of a mass term that, as we have learned in
the previous sections, can be toxic for the multigrid algorithm.
Note that the Coulomb gauge attempts to minimize 
this mass term by minimizing the quantity $A$.
We identify the square root of $\beta A$ with a ``disorder mass'' $m_D$,
\be
m_D = \sqrt{\beta A} \, .
\ee
To have a physical interpretation of $m_D$ let us discuss
the ``disorder scale'' $l_D$ that is given by the inverse of the disorder mass:
\be
l_D = \frac{1}{m_D} = \frac{1}{\sqrt{\beta A}} \, .
\ee
If the block size $L_B$ gets of the order of 
the disorder scale $l_D$,
the mass term in eq.\ (\ref{h1_constant}) becomes of order one,
and the amplitudes of the nonlocal moves become suppressed.
Now the crucial question is:
How does the scale $l_D$ behave for large $\beta$
in comparison with the physical correlation length $\xi$? 
($\xi$ is given by the inverse glueball mass or the inverse square root of the
string tension.)
If $l_D$ scaled with $\xi$ the algorithm could efficiently create
fluctuations up to the scale of $\xi$, as required by the physics of the model.
Everything would be all right if for large $\xi$
\be \label{scale}
\frac{l_D}{ \xi} \ 
\rellow{\longrightarrow}{\beta \rightarrow \infty} \ const  \ .
\ee
%(This argumentation is similar to the discussion of massive free 
%field theory in section \ref{SECfree}.)
In 4-dimensional $SU(2)$ lattice gauge theory
the correlation length is known to increase exponentially fast with
$\beta$.  Therefore, for eq.\ (\ref{scale}) to hold, we would need that
the disorder mass $m_D$ decreased exponentially fast with increasing
$\beta$. 
Formulated differently, the crucial question is whether or not
$m_D$ scales like a physical mass.

\subsubsection{Monte Carlo study of $m_D$}

We computed $m_D$ by Monte Carlo simulations for several values of
$\beta$.  To maximize $G_C$ we used a standard Gauss-Seidel relaxation
algorithm vectorized over a checkerboard structure.  The relaxation
procedure consists in going through the lattice and
minimizing the gauge functional (\ref{coulomb_gauge_slice}) locally.
%updating the actual
%$g$ configuration according to
%\be
%g_x \rightarrow g_x' = \frac{V_x^*}{\sqrt{\mbox{det} V_x}} \, ,
%\ee
%where the $2 \times 2$ matrix $V_x$ is given by
%\be
%V_x = \sum_{\mu \neq \tau}
%\bigl( \Um{x} g_{x+\hat\mu}^* +
%\Um{x-\hat\mu}^* g_{x-\hat\mu}^* \bigr) \, .
%\ee
%
For production runs it would be advantageous to use an accelerated gauge
fixing algorithm such as overrelaxation or 
multigrid \cite{mandula,mgaugefix}.
In the Monte Carlo studies reported in this section, we always used 50
Gau\ss -Seidel sweeps to determine $g$.  Note that by this procedure
$G_C$ is not entirely maximized, especially on very large lattices where
the relaxation algorithm suffers from CSD.  However,
for the detailed balance to be fulfilled, we only need that one uses
always the same number of relaxation sweeps.  Several tests revealed
that increasing the number of relaxation sweeps beyond 50 did not affect
the acceptance rates in a substantial way.  In our implementation, 50
Gau\ss -Seidel sweeps over all slices of a given direction $\tau$
required the same CPU time (on a CRAY Y-MP) as four Creutz heat bath
$SU(2)$ update sweeps.

%%%%%%%%%%%%%%%%%%%%%%%%%%%%%%%%%%%%%%%%%%%%%%%%%%%%%%%%%%%%%%%%%%%%%%%
\begin{figure}[htbp]
 \begin{center}
  \begin{minipage}[t]{145mm}
      % GEP-File:  GEP.OMSU2.l20.bg26.time101
       \begin{picture}(145,95)(0,0)
        \epsfig{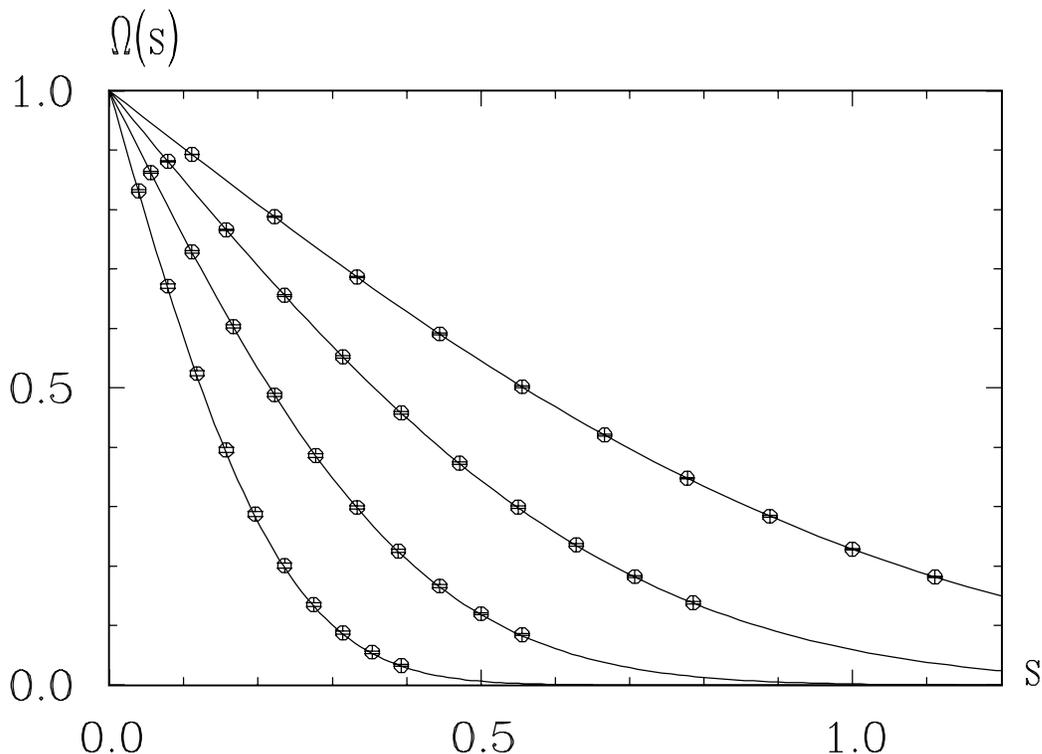}
       \end{picture}
      \caption[\Omega(s) in the 4d SU(2) lattice gauge theory]
              {\label{SU2}
               \sl $\Omega(s)$ in 4-dimensional $SU(2)$ lattice
                   gauge theory using piecewise constant kernels,
                   $\beta=2.6$ on a $20^4$-lattice.
                   From top to bottom: $L_B=2,4,8,16$.
                   Points with error bars: Monte Carlo results,
                   lines: analytical results using $m_D$ and $P$
                 from Monte Carlo (errors smaller than line width).}
  \end{minipage}
 \end{center}
\end{figure}
%%%%%%%%%%%%%%%%%%%%%%%%%%%%%%%%%%%%%%%%%%%%%%%%%%%%%%%%%%%%%%%%%%%%%%%

We checked the validity of the acceptance formula (\ref{formula})
using Monte Carlo estimates for $m_D$ and $P$.
Figure \ref{SU2} shows results for $\beta = 2.6$ on a $20^4$ lattice.
The results perfectly justify the usage of the approximation formula.
It is therefore sufficient to study the behavior of the quantities $m_D$
and $P$.  Our Monte Carlo results are presented in table \ref{tab3}.
The last column gives the statistics in sweeps (equilibration sweeps are
not counted here).
We used a mixture of four microcanonical
overrelaxation sweeps followed by a single Creutz heat bath sweep.
Measurements (including the determination of $g$) were performed every
25 sweeps.

%%%%%%%%%%%%%%%%%%%%%%%%%%%%%%%%%%%%%%%%%%%%%%%%%%%%%%%%%%%%%%%%%%%%%%%%%
\begin{table}
 \centering
 \caption[dummy]{\label{tab3} Monte Carlo results for $m_D$ and $P$ }
 \vspace{2ex}
\begin{tabular}{|c|c|c|c|c|}
\hline\str
 lattice size & $\beta$  &  $m_D$ &  $P$   & statistics
\\[.5ex] \hline \hline \str
 $8^4 $  & $2.4$ & 0.507(2)   &  0.6305(3)  & 10\,000
\tabhline
 $12^4$  & $2.4$ & 0.4957(4)  &  0.6300(2)   & 10\,000
\tabhline
 $16^4$  & $2.4$ & 0.4955(2)  &  0.62996(5)  & 10\,000
\\[.3ex] \hline \hline \str
 $8^4$  & $2.6$ &  0.497(4)  &   0.6703(1)   & 30\,000
\tabhline
 $12^4$  & $2.6$ & 0.465(2)   &  0.6702(1)   & 20\,000
\tabhline
 $16^4$  & $2.6$ & 0.4644(3)  &  0.67004(5)   & 10\,000
\tabhline
 $20^4$  & $2.6$ & 0.4650(2)  &  0.67008(5)  & 5\,000
\\[.3ex] \hline
\end{tabular} \end{table}
%%%%%%%%%%%%%%%%%%%%%%%%%%%%%%%%%%%%%%%%%%%%%%%%%%%%%%%%%%%%%%%%%%%%%%%%%

In table \ref{tab4} we display the ratios of the disorder mass $m_D$
with two physical masses, the square root of the string tension $\kappa$
and the lowest glue ball mass $m_{0^+}$.  The estimates for the physical
masses are taken from ref.\ \cite{michael}.  The results show that the
disorder mass is nearly independent of $\beta$ in the range studied,
whereas the physical masses decrease by roughly a factor of two.
Thus, $m_D$ is not scaling like a physical mass for the couplings
studied here.
We conclude from this that for large blocks the term quadratic in $L_B$
will strongly suppress the acceptance rates even
when the ratio of correlation length and block size $L_B$ is kept
constant.

%%%%%%%%%%%%%%%%%%%%%%%%%%%%%%%%%%%%%%%%%%%%%%%%%%%%%%%%%%%%%%%%%%%%%%%%%
\begin{table}
 \centering
 \caption[dummy]{\label{tab4} Comparison of $m_D$ with physical masses}
 \vspace{2ex}
\begin{tabular}{|c|c|c|c|c|c|c|}
\hline\str
 lattice size & $\beta$  &  $m_D$ &  $\sqrt{\kappa}$ & $m_{0^+}$ &
 $m_D/\sqrt{\kappa}$ & $m_D/m_{0^+}$
\\[.5ex] \hline \hline \str
 $16^4$  & $2.4$ & $0.4955(2)$  & $0.258(2)$  & $0.94(3)$ &
 $1.92$  & $0.53$
\tabhline
 $20^4$  & $2.6$ & 0.4650(2)  &  0.125(4)   & 0.52(3) &
 $3.72$  & $0.89$
\\[.3ex] \hline
\end{tabular} \end{table}
%%%%%%%%%%%%%%%%%%%%%%%%%%%%%%%%%%%%%%%%%%%%%%%%%%%%%%%%%%%%%%%%%%%%%%%%%

From this kinematical analysis it is clear that we can not expect
that CSD will be eliminated by such nonlocal updates.

Let us give a plausibility argument for the large $\beta$ behavior 
of $m_D = \sqrt{\beta A}$:
From the definition (\ref{defA}) of $A$ it is clear that $A$ vanishes for
large $\beta$
because $\Um{x}^g$ goes to unity in this limit.
 Since $A$ is a quantity that is defined on the scale of the plaquette,
it is dominated by the local disorder.
$A$ has nothing to do with collective excitations of the gauge
field that are responsible for the formation of glueballs with a mass that 
decreases exponentially in $\beta$. 
The leading weak coupling behavior of the plaquette is \cite{creutz}
\be
P \ = \ 1 - \frac{3}{4\beta} + O \left(\frac{1}{\beta^2} \right) \ .
\ee
Therefore 
it is natural to expect a weak coupling 
behavior for $A$ like 
\be
A \ = \ \frac{c}{\beta} + O \left(\frac{1}{\beta^2} \right) \ ,
\ee
with a constant $c$.
If this conjecture was true we would get
\be
m_D \ =  \ \sqrt{\beta A} \
\rellow{\longrightarrow}{\beta \rightarrow \infty} \ const  \ .
\ee
In four dimensions the range of $\beta$ values 
and lattice sizes studied here is
too small to check the large $\beta$ behavior of $m_D$ in detail.
Let us study the analogous situation in two dimensions.

\subsubsection{Digression: Monte Carlo study of $m_D$ in two dimensions}

We want to study the kinematical effect of a remaining nontrivial
background field 
$\Um{x}$ for $\mu \neq \tau$  in the bottom of a block
in two dimensions.
This is of course an artificial complication of the situation in two dimensions.
We always can choose the block coulomb gauge there, 
as we did in the previous sections.
Nevertheless we can use the two dimensional case as a simple example
where the region of large $\beta$ is easily accessible,
in contrast to four dimensions.
In analogy to the four dimensional case we are going to take the $g$-matrices
within a time slice according to the Coulomb gauge condition
\be\label{2dcoulomb_gauge_slice}
G_C(U,g) = \sum_{(x,x+\hat\mu) \in \Lamt} \Tr \bigl(
g_x \Um{x} g_{x+\hat\mu}^*\bigr)
 \,\stackrel{\mbox{!}}{=} \, \mbox{maximal} \ .
\ee 
In two dimensions this maximization has a simple solution:
Take the $g$-matrices such that 
\be
\Um{x}^g = g_x \Um{x} g^*_{x+\hat{\mu}} = P_t^{1/L} \ 
\mbox{for all} \ x \in \Lamt \ .
\ee
$P_t$ is the Polyakov loop in space direction,
\be
P_t = 
\prod^L_{\stackrel{ {\mbox{\scriptsize $x_{\mu}=1$ }} }{x_{\tau}=t} } 
\Um{x}
\ee
on a $L \times L$ lattice, and $P_t^{1/L}$ is the closest root
to unity.

By taking the Coulomb gauge condition in two dimensions
we now have a smooth but nontrivial background field in the 
bottom of the block. Let us study the effect on the kinematics.
For piecewise constant interpolation we have
(cf. appendix \ref{APPacc})
\be
h_1 =  2 \beta P  
\left[ 1- \cos \left( s L_B^{1/2} \right) \right] +
\beta  A (L_B-1)    
\sin^2 \left( s L_B^{1/2} \right) \ ,
\ee
with 
\be
A = \bigl\langle \half \Tr \bigl((\Um{x}^g - \vn \!\cdot\! \vs \, \Um{x}^g \,
 \vn \!\cdot\! \vs)
H_{x,\mu}^{g *} \bigr) \bigr\rangle \ ,
\ee
where $g$ is taken according to eq.\ (\ref{2dcoulomb_gauge_slice}).

We computed $A$ and $m_D = \sqrt{\beta A}$ 
by Monte Carlo simulations in the region of large
$\beta$ close to the continuum.  
Our Monte Carlo results are presented in table \ref{2dcoulombA}.
The statistics are counted in sweeps (we started from 
equilibrated configurations).
We used a mixture of four microcanonical
overrelaxation sweeps followed by a single heat bath sweep.
Measurements were performed every
5 sweeps. 

%%%%%%%%%%%%%%%%%%%%%%%%%%%%%%%%%%%%%%%%%%%%%%%%%%%%%%%%%%%%%%%%%%%%%%%%%
\begin{figure}[htbp]
 \begin{center}
  \begin{minipage}[t]{145mm}
   \begin{picture}(145,100)(30,110)
    \epsfig{file=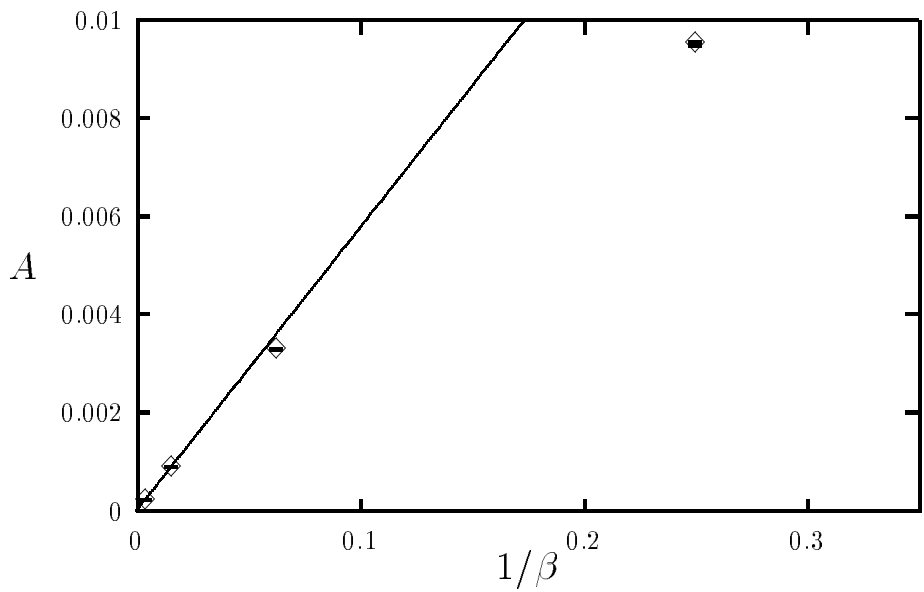,height=300mm}
   \end{picture}
   \caption[$A$ as a function of $1/\beta$]
   {\label{fig:Abeta} \sl
   Dependence of the quantity A on $1/\beta$ in two dimensional
   $SU(2)$ lattice gauge theory.
   Diamonds: Monte Carlo results.
   The line corresponds to $A \sim 1/\beta$.}
  \end{minipage}
 \end{center}
\end{figure}
%%%%%%%%%%%%%%%%%%%%%%%%%%%%%%%%%%%%%%%%%%%%%%%%%%%%%%%%%%%%%%%%%%%%%%%%%

The large $\beta$ behavior of $A$ is shown in figure \ref{fig:Abeta}.
It is consistent with $A \sim 1/\beta$.
In table \ref{2dcoulombmD} we display the ratios of the disorder mass $m_D$
with the physical mass, the square root of the string tension $\kappa$.
The values for the string tension are taken from the exact solution 
in the infinite volume limit (appendix \ref{APPSU22x}).
The results show that the
disorder mass in two 
dimension goes to a constant for large 
$\beta$, 
whereas the physical mass decreases like $1/\sqrt{\beta}$. 
Therefore, $m_D$ is definitely not scaling like a physical mass in the 
continuum limit in two dimensions.

We conclude from this that for large blocks the term quadratic in $L_B$
will strongly suppress the acceptance rates 
also in two dimensions.
Therefore the smooth but nontrivial background field in the coulomb gauge
that is still left in the one dimensional bottom of the time slice
will cause CSD in this variant of the algorithm.

In analogy to this result in two dimensions we expect the same
behavior of $m_D$ in the large $\beta$ limit in four dimensions.

%%%%%%%%%%%%%%%%%%%%%%%%%%%%%%%%%%%%%%%%%%%%%%%%%%%%%%%%%%%%%%%%%%%%%%%%%
\begin{table}
 \centering
 \caption[dummy]{\label{2dcoulombA} 
                 Monte Carlo results for $P$, $A$ and $m_D$
                 in two dimensions }
 \vspace{2ex}
\begin{tabular}{|c|c|l|l|c|c|}
\hline\str
lattice size  & $\beta$ &
\multicolumn{1}{c|}{$P$} &
\multicolumn{1}{c|}{$A$} &
\multicolumn{1}{c|}{$m_D$} & statistics
\\[.5ex] \hline \hline \str
  $8^2$  & 1 & 0.2400(4) & 0.01357(15) & 0.1165(7) & 20\,000
\tabhline
 $16^2$  & 4 & 0.65791(14) & 0.00951(3) & 0.1950(4) & 20\,000
\tabhline
 $32^2$  & 16 & 0.90783(2) & 0.003284(6) & 0.2292(2) & 20\,000
\tabhline
 $64^2$  & 64 & 0.976655(3) & 0.000887(1) & 0.2383(2) & 20\,000
\tabhline
 $128^2$  & 256 & 0.9941462(3) & 0.000225(2) & 0.2400(1) & 20\,000
\\[.3ex] \hline
\end{tabular} \end{table}
%%%%%%%%%%%%%%%%%%%%%%%%%%%%%%%%%%%%%%%%%%%%%%%%%%%%%%%%%%%%%%%%%%%%%%%%%

%%%%%%%%%%%%%%%%%%%%%%%%%%%%%%%%%%%%%%%%%%%%%%%%%%%%%%%%%%%%%%%%%%%%%%%%%
\begin{table}
 \centering
 \caption[dummy]{\label{2dcoulombmD} 
                 Comparison of $m_D$ with physical masses
                 in two dimensions}
 \vspace{2ex}
\begin{tabular}{|c|c|c|c|c|c|}
\hline\str
 lattice size & $\beta$  &  $\xi$ & $m_D$  & $\sqrt{\kappa}$ & $m_D/\sqrt{\kappa}$
\\[.5ex] \hline \hline \str
  $8^2$  & 1 & 0.8373 & 0.1165(7) & 1.1942 & 0.0976
\tabhline
 $16^2$  & 4 & 1.5458 & 0.1950(4) & 0.6469 & 0.301
\tabhline
 $32^2$  & 16 & 3.2155 & 0.2292(2) & 0.3110 & 0.737
\tabhline
 $64^2$  & 64 & 6.5065 & 0.2383(2) & 0.1536 & 1.55
\tabhline
 $128^2$  & 256 & 13.0511 & 0.2400(1) & 0.0766 & 3.13
\\[.3ex] \hline
\end{tabular} \end{table}
%%%%%%%%%%%%%%%%%%%%%%%%%%%%%%%%%%%%%%%%%%%%%%%%%%%%%%%%%%%%%%%%%%%%%%%%%

\subsubsection{Comparison of volume and surface effects in four dimensions}

By the kinematical analysis in four dimensions we found 
that CSD has to be expected because of a mass $m_D$ generated by
the local disorder in the bottom of the blocks.
However, one could hope that the value of the unwanted mass term
is so small that it is only harmful at very large values of $\beta$
on huge lattices.
Let us examine the effect of this term in more detail.  

Recall  that
$h_1$ is built up from two contributions.  The first contribution is
that related to the gauge field disorder inside the block and is
quantitatively represented by the mass $m_D$.  The second contribution
is associated with the block surface.  The latter can of course be
made smaller by using smooth kernels $\psi$ instead of the piecewise
constant kernels discussed so far. However, the disorder
term cannot be expected to become smaller for smooth kernels (see
below).  In figure \ref{BORD} we plotted separately
the two contributions to $h_1$
\bea
h_{1,A} &=& 3 \beta
A \, (L_B -1) L_B^2 \, \sin^2(s L_B^{-1/2}/2) \, ,
\nonumber \\
h_{1,P} &=& 6 \beta P
 \, L_B^2 \bigl( 1- \cos(s L_B^{-1/2}/2) \bigr) \, ,
\eea
for $\beta=2.6$ and block size $L_B=8$ on a $20^4$ lattice.  The plot
shows that already for this block size the disorder contribution is by
no means small -- it is comparable to the surface effect.  It
is therefore not clear that one could achieve any significant
improvement by using smooth kernels.  To investigate this in more
detail, we derive an expression for $h_1$ (cf.\ appendix \ref{APPacc}),
valid for smooth kernels as
well:
\be
h_1 = \frac{3 \beta}4 s^2 A
                     \sum_{x \in \Lamt} \psi_x^2+
  \frac{\beta}{8}s^2 (P-A) \alpha_3 + O(s^4) \, .
\ee
Since $\sum \psi_x^2 \sim L_B^2$, we get essentially the
same behavior for the disorder contribution as in the case
of piecewise constant kernels.

%%%%%%%%%%%%%%%%%%%%%%%%%%%%%%%%%%%%%%%%%%%%%%%%%%%%%%%%%%%%%%%%%%%%%%%
\begin{figure}[htbp]
 \begin{center}
  \begin{minipage}[htbp]{145mm}
      % GEP-File:  GEP.OMSU2.l20.bg26.border
       \begin{picture}(145,95)(0,0)
        \epsfig{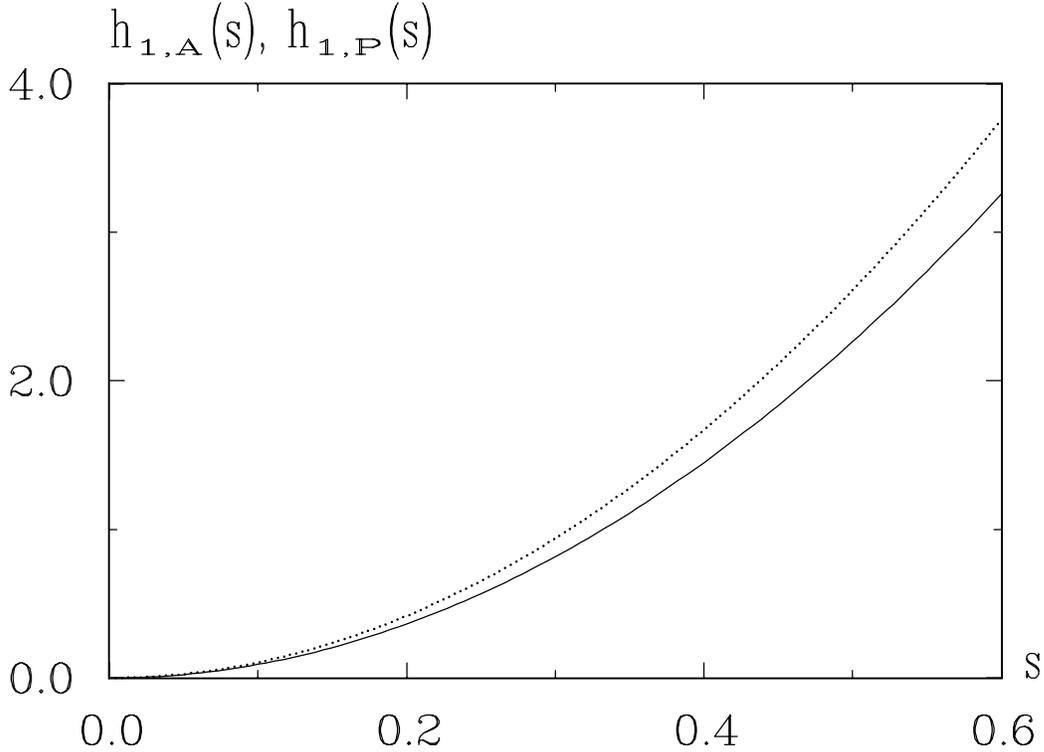}
       \end{picture} 
      \caption[ Comparison of disorder and boundary effects]
              {\label{BORD}
               \sl Comparison of disorder and surface effects for
                   four dimensional $SU(2)$ lattice gauge theory using
                   piecewise constant kernels on an $8^3$-block,
                   $\beta=2.6$ on a $20^4$-lattice.
                   Solid line: $h_{1,A}(s)$ (disorder effects),
                   dashed line: $h_{1,P}(s)$ (surface effects).}
  \end{minipage}
 \end{center}
\end{figure}
%%%%%%%%%%%%%%%%%%%%%%%%%%%%%%%%%%%%%%%%%%%%%%%%%%%%%%%%%%%%%%%%%%%%%%%

\begin{figure}[h]
 \begin{center}
  \begin{minipage}[hbpt]{145mm}
      % GEP-File:  GEP.OMSU2.l20.bg26.border
       \begin{picture}(145,95)(0,0)
        \epsfig{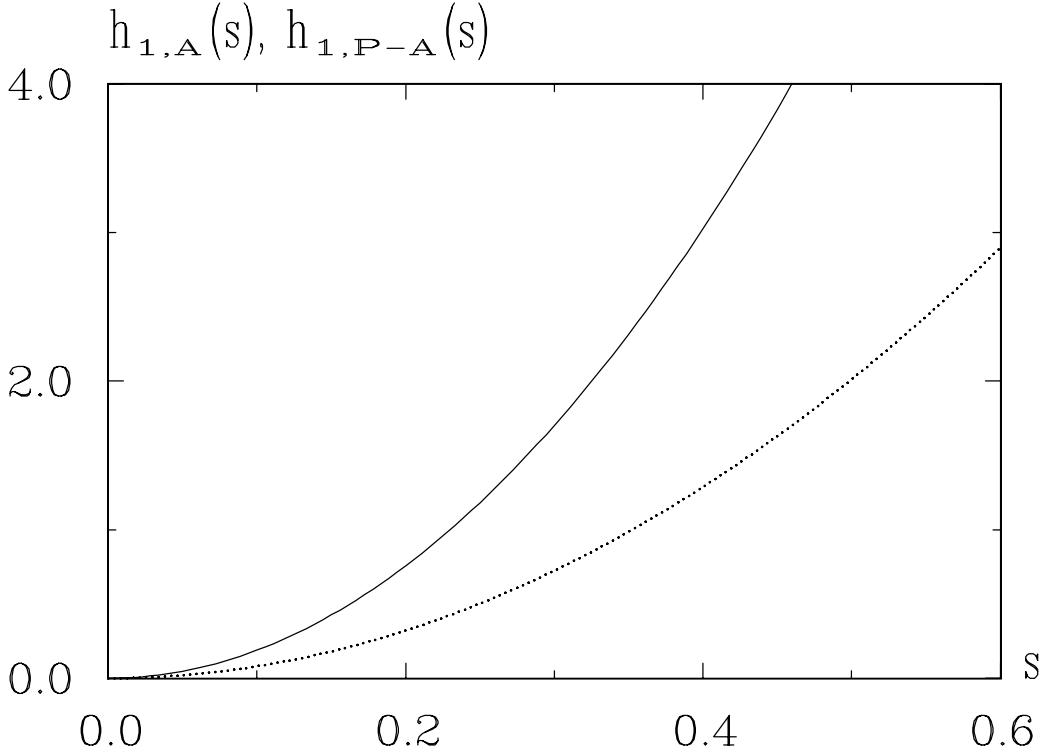}
       \end{picture} 
      \caption[ Comparison of disorder and boundary effects]
              {\label{BORS}
               \sl Comparison of disorder and surface effects for
                   the four dimensional $SU(2)$ lattice gauge theory using
                   smooth $\psi^{sine}$ kernels on a $8^3$-block,
                   $\beta=2.6$ on a $20^4$-lattice,
                   quadratic approximation used.
                   Solid line: $h_{1,A}(s)$ (disorder effects),
                   dashed line: $h_{1,P-A}(s)$ (surface effects).}
  \end{minipage}
 \end{center}
\end{figure}

We checked this quadratic approximation against the exact
expression (\ref{h1_constant}) in the case of piecewise constant
kernels.
For $L_B = 8$ the deviations are already negligible in the range of
$s$-values displayed in figure \ref{BORD}.
For smooth $\psi^{sine}$ kernels we show separately in figure \ref{BORS}
the two contributions to $h_1$
\be
h_{1,A} = \frac{3 \beta}4 s^2 A
                     \sum_{x \in \Lamt} \psi_x^2 \, ,\;\;\;
h_{1,P-A}= \frac{\beta}{8}s^2 (P-A) \alpha_3  \, ,
\ee
for $\beta=2.6$ and block size $L_B=8$ on a $20^4$ lattice.
We observe that the surface effects are lowered by the smooth kernels,
but the disorder contribution is even higher than for piecewise constant
kernels.

%%%%%%%%%%%%%%%%%%%%%%%%%%%%%%%%%%%%%%%%%%%%%%%%%%%%%%%%%%%%%%%%%%%%%%%
%%%%%%%%%%%%%%%%%%%%%%%%%%%%%%%%%%%%%%%%%%%%%%%%%%%%%%%%%%%%%%%%%%%%%%%

Piecewise constant kernels have the practical feature that once the
change of the Hamiltonian has been calculated, one can perform
multihit Metropolis updating or microcanonical overrelaxation.
In a special case, even a recursive multigrid implementation
with a W-cycle is possible (see section \ref{SECSU24sim} below).
For smooth kernels the change in the Hamiltonian would have to
be calculated again and again.  Also the advantages of smooth kernels
are not that clear on small three or four dimensional blocks.  For the
concrete simulation we will use piecewise constant kernels.

\subsubsection{Maximally abelian gauge}

Our proposal for the choice for $g$ was motivated by the desire to
minimize the quantity $A$.  We now ask whether there is a better choice
for $g$ than the $g$ determined by the Coulomb gauge condition.  For the
sake of simplicity let us take $\vn = (0,0,1)$, i.e.
$\vn \!\cdot\! \vs \, = \sigma_3$.
Then $A$ is given by
\be
 A =   \bigl\langle \half \Tr\bigl((\Um{x}^g - 
\sigma_3 \, \Um{x}^g \, \sigma_3)
H_{x,\mu}^{g\, *} \bigr) \bigr\rangle \, .
\ee
The choice of
the Coulomb gauge condition aimed at bringing $\Um{x}^g$ as close to
unity as possible.  Alternatively, one might require that $\Um{x}^g$
should be as close as possible to a $SU(2)$-matrix of the form
$a_0 + ia_3 \sigma_3$. This will also lead to
a small $A$.
The corresponding gauge transformation $g$ can be found by maximizing
the functional
\be\label{abel_gauge_slice} G_A(U,g) =
\sum_{(x,x+\hat\mu) \in \Lamt} \Tr \bigl( \sigma_3 \Um{x}^g \sigma_3
\Um{x}^{g \, *} \bigr) \, ,
\ee
leading to the maximally abelian gauge
\cite{maxabel}, here implemented only on a slice.  We computed $m_D$
also using the $g$'s resulting from this gauge condition and compared
the results with the ones obtained by using the Coulomb gauge condition.
We did not find a substantial difference.  We prefer the Coulomb gauge
condition because it does not depend on the direction $\vn$ and thus
saves computer time.

\subsection{Summary}

In this section we discussed the generalization of multigrid
Monte Carlo methods for the simulation of pure gauge fields 
from two dimensions to four dimensions.

In the abelian case the generalization is straightforward.
We have seen that the kinematical behavior of the resulting method
is similar to massless free field theory.
Therefore a considerable acceleration can be expected and was indeed 
observed in four dimensional $U(1)$ lattice gauge
theory \cite{newgauge}.
However, metastability causes additional 
problems close to the phase transition which can not be understood
in terms of our method.

In the second part of this section we have generalized the 
time slice blocking algorithm that lead to successful acceleration
of the simulation of $SU(2)$ gauge fields in two dimensions to four dimensions. 
In four dimensions, new difficulties occur that are due to the nontrivial
background field in the bottom of the blocks. 
We investigated the scale dependence of acceptance rates in detail.
Here we found that an algorithmic mass term generated by the disorder 
in the background field suppresses the acceptance rates on large blocks. 
From our kinematical analysis we can not expect that
the proposed algorithm will have a chance to 
reduce the dynamical critical exponent 
below $z \approx 2$.
However, compared to local Monte Carlo algorithms there could still be an 
acceleration of the dynamics by a constant factor,
depending on the details of the implementation.
This question will be investigated in the following section.

\clearpage

\section{Multigrid Monte Carlo simulation of 
$SU(2)$ lattice gauge fields in four dimensions} 

\label{SECSU24sim}

\setcounter{equation}{0}

In this section we study whether our nonlocal Monte Carlo updating
can accelerate the Monte Carlo dynamics of a local heat bath algorithm  
by a constant factor. We do not aim at estimating
the dynamical critical exponent $z$. 

We begin with a description of the implementation of the time slice 
blocking scheme for $SU(2)$ in four dimensions as a
recursive multigrid algorithm with piecewise constant
interpolation and a W-cycle.
Then we give a detailed report on a 
comparison of the time slice blocking algorithm 
with the local heat bath algorithm.

\subsection{Implementation of the time slice blocking}

For a concrete implementation of the multigrid Monte Carlo algorithm 
in four dimensions we choose 
the time slice blocking method: 
In a basic time slice blocking step, all links pointing from a given
three dimensional 
time slice $\Lambda^{\tau}_t$ in the $\tau$-direction are going to be updated.
First we are going to describe the implementation of such a basic 
step for a given time slice $\Lambda^{\tau}_t$.
In particular we explain a recursive multigrid procedure 
that can be constructed by updating in a global
$U(1)$ subgroup of $SU(2)$.
How time slice blocking steps are applied to
different time slices $\Lambda^{\tau}_t$ and to different orientations
$\tau$ in sequence will be explained in a second paragraph.

\subsubsection{Organization of a basic time slice blocking step}
 
We now describe the time slice blocking method for a fixed three dimensional
time slice $ \Lambda_t^{\tau} =
\left\{ x \in \Lambda_0 \, \vert \, x_{\tau} = t \right\}$.

The basic updating step is based on a
local, site dependent ``rotation'' of all link variables
pointing from sites $x \in \Lamt$ in the $\tau$-direction by
\be
\Ut{x} \rightarrow \Ut{x}' =  R_x(g) \, \Ut{x} \, ,
\ee
with $R_x(g) = g_x^* R_x g_x$ and
\be
R_x(\vn_{x},\theta_{x}) = \cos( \theta_{x} )
 + i \sin( \theta_{x}) \,
\vn_{x} \!\cdot\! \vs \, .
\ee
The gauge transformation $g$ is obtained by imposing the
Coulomb gauge condition (\ref{coulomb_gauge_slice})
on slices $\Lamt$ as defined 
in section \ref{SECSU24acc}.

We are interested in a recursive multigrid
implementation.
Therefore we use piecewise constant kernels and update
a fixed $U(1)$ subgroup of $SU(2)$ globally.
In this special case an explicit multigrid scheme can be constructed.
This feature was explained in ref.\ \cite{newgauge}
in a related context.

The restriction to a global $U(1)$ subgroup is defined
by taking the directions $\vn_{x}$ of the
embedded $U(1)$-subgroups to be independent of the site $x$
within the time slice, i.e.\
$\vn_{x} = \vn $ for all $x \in \Lamt$.
The global $\vn$ direction is taken as a random vector from
the three dimensional unit sphere.
Then a conditional Hamiltonian on the finest lattice can be calculated
by substituting
the ``rotated'' gauge field $U'$ in the fundamental Hamiltonian.
The resulting conditional Hamiltonian $\calH_0(S)$
is of the type of a generalized three dimensional $XY$-model:
\be\label{Hcond0}
- \calH_0(S) = \sum_{x \in \Lamt} \sum_{\mu \neq \tau}
   S_x^T B_{x,\mu} S_{x+\hat{\mu}} + const \ ,
\ee
where $S_x$ denotes the $XY$-spin variable 
\be
 S_x = \left(
\begin{array}{c}
\cos(\theta_x) \\
\sin(\theta_x) \\
\end{array}\right) \  ,
\ee
and $S_x^T$ its transpose.
$B_{x,\mu}$ is a link dependent real $2 \times 2$ matrix
\be
 B_{x,\mu} = \left(
\begin{array}{cc}
B_{x,\mu}^{11} &
B_{x,\mu}^{12} \\
B_{x,\mu}^{21} &
B_{x,\mu}^{22} \\
\end{array}\right)
\ee
with the matrix elements 
\bea
B_{x,\mu}^{11} &=& \frac{\beta}{2} 
\Tr \left(\Um{x}^g H_{x,\mu}^{g\,  *} \right) \ , \nonumber  \\
B_{x,\mu}^{12} &=& \frac{\beta}{2}  \Tr \left(
\Um{x}^g \,  i\vn \!\cdot\! \vs H_{x,\mu}^{g\,  *} \right) \ , 
\nonumber \\
B_{x,\mu}^{21} &=&  - \frac{\beta}{2} \Tr \left(i\vn \!\cdot\! \vs \,
\Um{x}^g  H_{x,\mu}^{g\,  *} \right)  \ ,
\nonumber  \\
B_{x,\mu}^{22} &=& - \frac{\beta}{2} \Tr \left(i\vn \!\cdot\! \vs \,
\Um{x}^g \, i\vn \!\cdot\! \vs H_{x,\mu}^{g\,  *} \right)  \ .
\\ \nonumber
\eea
%We used the notation
%$H_{x,\mu}^* = \Ut{x+\hat\mu} \Um{x+\hat\tau}^* \Ut{x}^*$ and 
%$U^g_{x,\mu} = g_x \Um{x} g_{x+\hat\mu}^*$. 
%$U^g$ is the gauge field obtained by applying a gauge transformation
%$g$ to $U$. $H^g$ is defined analogously. \RM still needed?

Now we set up a recursive multigrid procedure to update the three 
dimensional generalized $XY$-model on a time slice $\Lamt$.
The first blocking step is performed by 
dividing the fine lattice $\Lamt$ in cubic blocks $x'$
of size $2^3$. 
This defines a three dimensional block lattice $\Lambda_1$.
Then a nonlocal block update
consists of a constant rotation of all $XY$-spins $S_x$ 
within a three dimensional block $x' \in \Lambda_1$ by the block angle
$\theta_{x'}$:
\be \label{XYupdate}
S_x = \left(
\begin{array}{c}
\cos(\theta_x) \\
\sin(\theta_x) \\
\end{array} \right) \;
\rightarrow \; 
 S_x' = \left(
\begin{array}{c}
\cos(\theta_x + \theta_{x'}) \\
\sin(\theta_x + \theta_{x'}) \\
\end{array}
\right) \ \mbox{for} \  x \in x' \ .
\ee
An equivalent parametrization of this update is 
\be
S_x \; \rightarrow \; S_x' = A_x S_{x'} \ ,
\ee
where the $SO(2)$-matrix $A_x$ and the block-$XY$-spin $S_{x'}$
are given by
\be
 A_x = \left(
\begin{array}{cc}
\cos(\theta_x) & -\sin(\theta_x) \\
\sin(\theta_x) &  \cos(\theta_x) \\
\end{array} \right) \; , \; \;
 S_{x'} = \left(
\begin{array}{c}
\cos(\theta_{x'}) \\
\sin(\theta_{x'}) \\
\end{array}
\right)  \ .
\ee
By iterating this procedure one gets conditional Hamiltonians
$\calH_k(S)$ on three dimensional coarser layers $\Lambda_k$.
An important point is that in the special case considered here
$\calH_k(S)$ always stays of the form 
\be
- \calH_k(S) =
 \sum_{x \in \Lambda_k} \sum_{\mu \neq \tau} S_x^T B_{x,\mu} S_{x+\mu} 
 + \sum_{x \in \Lambda_k} S_x^T M_x S_x
 + const
\ee
with $2\times 2$ coupling matrices
$B_{x,\mu}$ and $M_x$ that depend 
on the links $(x,\mu)$ and on the sites $x$ in $\Lambda_k$ respectively.
The coupling matrices $B_{x',\mu}$ and $M_{x'}$ 
that depend on the block links $(x',\mu)$ or
blocks $x'$ on the next coarser lattice 
$\Lambda_{k+1}$ can be calculated recursively:
\bea
B_{x',\mu} & = &
 \sum_{\stackrel{\mbox{\scriptsize $x \in x'$}}{x+\hat{\mu} \in x'+\hat{\mu}}}
 A_x^T B_{x,\mu} A_{x+\hat{\mu}} 
 \nonumber \\
M_{x'} & = &
 \sum_{\stackrel{\mbox{\scriptsize$ x \in x'$}}{x+\hat{\mu} \in x'}} 
A_x^T B_{x,\mu} A_{x+\hat{\mu}} 
+ \sum_{x \in x'} A_x^T M_x A_x  \ . \\
\nonumber
\eea
Here $x' + \hat{\mu}$ denotes the next neighbor block of the 
block $x'$ in the coarse layer $\Lambda_{k+1}$.
Note that by the recursion the conditional Hamiltonian 
$\calH_k(S)$ on a coarser layer $\Lambda_k$ contains not only
next neighbor couplings of the $XY$-variables
(as the conditional Hamiltonian $\calH_0(S)$ in eq.~(\ref{Hcond0})
on the finest lattice)
 but also
site-site couplings of the type of a mass term with 
site dependent mass matrix $M_x$.
The algorithmic mass term
discussed in section \ref{SECSU24acc} that leads to a decrease of
acceptance rates on large blocks originates exactly from this 
site-site term in the conditional Hamiltonian.

This recursive blocking procedure is repeated until we end up with
the coarsest layer that consists of a single point.
(I.\ e.\ in our simulation on a $8^4$-lattice we 
are going to update on
block lattices of the size $4^3$, $2^3$ and $1^3$.)

After a blocking step to a coarser layer $\Lambda_{k+1}$ the 
initial configuration 
%of the corresponding $XY$-spins 
on this coarse layer is 
%$S_{x'}^1 = 1$, $S_{x'}^2 = 0$ 
\be
\theta_{x'} \ = \ 0 \
\Leftrightarrow \
 S_{x'} = \left(
\begin{array}{c}
1 \\
0 \\
\end{array}
\right)  \ \mbox{for all}  \ x' \in \Lambda_{k+1} \ .
\ee
On each layer we perform one sweep of
10 Hit Metropolis updates
in terms of the $\theta_x$ variables:
\be
\theta_x \rightarrow \theta_x' = \theta_x + \Delta \theta \ ,
\ee
where $\Delta \theta$ is taken to be equally distributed 
from the interval 
$[ -\varepsilon, \varepsilon]$. 
We tuned the Metropolis step size $\varepsilon$ in order to obtain
acceptance rates of $\approx 50 \%$ on each layer. 
If we return from a coarse layer $\Lambda_{k+1}$ to the next finer
layer $\Lambda_k$, the configuration $\{ S \}$ on the finer layer
is updated according to eq.~(\ref{XYupdate}).

In this implementation, the sequence of visiting the 
different three dimensional layers of a time slice is
organized according to a W-cycle. 
From the point of view of numerical work this is possible 
because of the recursive definition of the algorithm.
In principle the described implementation is not only feasible
in the time slice blocking formulation but also  
with hypercubic four dimensional blocks.

\subsubsection{Sequence of basic time slice blocking steps}

The sequence of the updating on different time slices is as follows:
On an $L^4$-lattice we visit the $L$ time slices $\Lambda^{1}_t$, 
$t = 1, \dots L$. In this way, all link variables $U_{x,1}$ on the lattice  
that point in the $1$-direction are updated in a nonlocal way
on block lattices with $L_B = 2, 4, \dots$.
Then
we perform a local Creutz heat bath sweep through all links on the lattice.
Now we change the time direction $\tau$ from $\tau = 1$ to $\tau = 2$:
We visit the $L$ time slices $\Lambda^{2}_t$, 
$t = 1, \dots L$, again followed by a sweep of local heat bath updates.
The same scheme (a visit of all time slices
and a local heat bath sweep) 
is repeated also for the  $\tau = 3$ and $\tau = 4$ direction,
such that all the link variables $U_{x,\tau}$ have been updated 
in a nonlocal manner
for all $\tau = 1,\dots 4$.
This sequence is repeated periodically.

Observables are measured after each local heat bath sweep.
In addition, we perform a random translation of the lattice after each 
local heat bath sweep in order to avoid effects from fixed block boundaries
\cite{hmsun}.

To reduce the computer time needed for gauge fixing we use a slight 
modification:
If all the link variables $U_{x,\mu}$ with $\mu \neq \tau$ 
lying in the bottom of the time slice $\Lambda^{\tau}_t$ have been gauged  
according to the Coulomb gauge condition, 
we update not only the link variables $U_{x,\tau}$ pointing 
from this time slice in the positive $\tau$-direction,
but we update also the links pointing from this time slice to the 
negative $\tau$-direction. 
Thus, only every second time slice $\Lambda^{\tau}_t$
has to be gauged.
This leads to  a decrease of the numerical work required for 
gauge fixing by a factor of two.

\subsection{Simulation and Results} 

\subsubsection{Observables} 

The observables measured are square Wilson loops
\be
W(I,I) = \langle \half \Tr (U(C_{I,I}) \rangle \ ,
\ee
where $U(C_{I,I})$ is the parallel transporter 
around a rectangular Wilson loop $C_{I,I}$ of size $I \times I$.
On the $8^4$-lattice we measure $W(1,1)$,  
$W(2,2)$, and $W(4,4)$. Another important class of 
quantities is built up from timelike Polyakov loops.
A Polyakov loop at the three dimensional spatial point $\vec{x}$ 
is defined by
\be
P_{\vec{x}} = \half \Tr \prod_{t=1}^{L} U_{(\vec{x},t),4} \ .
\ee
We measure the lattice averaged Polyakov loop
\be
\bar{P} = \left\langle \frac{1}{L^3}\sum_{\vec{x}} P_{\vec{x}} \right\rangle \ ,
\ee
the lattice averaged Polyakov loop squared
\be
\bar{P}^2 = \left\langle \left(\frac{1}{L^3}
\sum_{\vec{x}} P_{\vec{x}} \right)^2 \right\rangle \ ,
\ee
and the sign of the lattice averaged Polyakov loop
\be
\mbox{sign}(\bar{P})
 = \left\langle \mbox{sign}\left( \frac{1}{L^3}\sum_{\vec{x}} 
P_{\vec{x}} \right) \right\rangle \ .
\ee
%and the absolute value of the sign of the lattice averaged Polyakov loop
%\be
%|\mbox{sign}(P)|
%= \left\langle \left| \mbox{sign}\left( \frac{1}{L^3}\sum_{\vec{x}} 
%P_{\vec{x}} \right) \right| \right\rangle \ .
%\ee

\subsubsection{Run parameters and static results}

In order to study the acceleration by the multigrid algorithm,
we compare the autocorrelations of the nonlocal algorithm with a
standard local Creutz heat bath algorithm.
Precise measurements of autocorrelation times $\tau$ require
high statistics simulations with run lengths of at least
$1000 \tau - 10\,000 \tau$.
For reasons of computer time 
we decided to simulate on relatively small lattices.
The algorithms are compared on a $8^4$-lattice 
at $\beta = 2.2$, $2.4$ and $2.6$.
We started the local heat bath runs from ordered configurations 
(all link variables set equal to unity) and discarded
a suitable number of iterations for equilibration.
For the multigrid simulations we used
warm starts from already equilibrated configurations. 
Measurements were taken after each local heat bath sweep. 

In order to reduce the computer time needed for gauge fixing, we use
10 sweeps of overrelaxation \cite{mandula} with overrelaxation parameter
$\omega = 1.7$. The degree of maximization of the Coulomb gauge 
functional $G_C$ (\ref{coulomb_gauge_slice}) was found to be 
about the same as if we used 50 Gau\ss-Seidel relaxation sweeps.

With these run parameters the computer time needed 
by our implementation on a CRAY Y-MP for one measurement on the 
$8^4$-lattice 
by the multigrid procedure is about a factor of $2.8$ larger than 
the time needed for one measurement by the Creutz heat bath algorithm.
This factor could still be lowered by using a multigrid
method for gauge fixing~\cite{mgaugefix}.

The static results of our runs are given in table \ref{SU24static}.
The values for the heat bath and for the multigrid algorithm 
agree within errors.
Where a comparison is possible,
our results for the Wilson loop observables are consistent with 
existing data in the literature \cite{bergstehr}.

\begin{table}[htbp]
 \centering
 \caption[dummy]{\label{SU24static}
         Comparison of  
         numerical results for various static observables
         of heat bath (HB) and multigrid (MG) simulations
         on the $8^4$ lattice. }
 \vspace{2ex}
\begin{tabular}{|c|l|l|l|l|l|l|}
\hline\str
$\beta$ & \multicolumn{2}{c|}{2.2} & 
\multicolumn{2}{c|}{2.4} & 
\multicolumn{2}{c|}{2.6} 
\tabhline
algorithm  & 
\multicolumn{1}{c|}{HB} &
\multicolumn{1}{c|}{MG} &
\multicolumn{1}{c|}{HB} &
\multicolumn{1}{c|}{MG} &
\multicolumn{1}{c|}{HB} &
\multicolumn{1}{c|}{MG} 
\tabhline
statistics & 
\multicolumn{1}{c|}{100\, 000} &
\multicolumn{1}{c|}{50\, 000} &
\multicolumn{1}{c|}{100\, 000} &
\multicolumn{1}{c|}{100\, 000} &
\multicolumn{1}{c|}{100\, 000} &
\multicolumn{1}{c|}{100\, 000} 
\tabhline
discarded &
\multicolumn{1}{c|}{10\, 000} &
\multicolumn{1}{c|}{equi.} &
\multicolumn{1}{c|}{10\, 000} &
\multicolumn{1}{c|}{equi.} &
\multicolumn{1}{c|}{10\, 000} &
\multicolumn{1}{c|}{equi.} 
\tabhline
$W(1,1)$ & 0.56922(5) & 0.56917(9) &  \ 0.63024(3) & \ 0.63026(3) 
& 0.67029(2) & 0.67029(2)
\tabhline
$W(2,2)$ & 0.13628(7) & 0.13625(10) & \ 0.22319(11) & \ 0.22330(13)
& 0.28875(7) & 0.28877(5)
\tabhline
$W(4,4)$ & 0.00127(3) & 0.00132(4) & \ 0.01371(7) & \ 0.01379(6)
& 0.03762(8) & 0.03770(6)
\tabhline
$\bar{P}$ & 0.00006(13) & 0.00015(18) & -0.0020(16) & -0.00006(13)
& 0.008(8) & 0.002(6)
\tabhline
$\bar{P}^2$ & 0.000584(3) & 0.000580(4) & \ 0.00211(5) & \ 0.00202(5)
& 0.0092(2) & 0.0096(2)
\tabhline
$\mbox{sign}(\bar{P})$ & 0.0000(2) &  0.0002(3) & -0.004(3) & -0.000(2)
& 0.013(13) & 0.003(11)
\\[.3ex] \hline
\end{tabular}
\end{table}

\subsubsection{Autocorrelation times}
\begin{table}[htbp]
 \centering
 \caption[dummy]{\label{SU24dynamic} 
         Comparison of  
         the exponential autocorrelation times
         of heat bath (HB) and multigrid (MG) simulations
         for different observables
         on the $8^4$ lattice. }
         
 \vspace{2ex}
\begin{tabular}{|c|c|c|c|c|c|c|}
\hline\str
$\beta$ & \multicolumn{2}{c|}{2.2} & 
\multicolumn{2}{c|}{2.4} & 
\multicolumn{2}{c|}{2.6} 
\tabhline
algorithm  & HB & MG & HB & MG & HB & MG 
\tabhline
statistics & 100\, 000 & 50\, 000 & 100\, 000 &
 100\, 000 & 100\, 000 & 100\, 000
\tabhline
discarded & 10\, 000 & equi. & 10\, 000 & 
equi. & 10\, 000 & equi. 
\tabhline
$\tau_{W(1,1)}$ & 6.9(2.9) & 10.3(3.5) & 5.7(1.5) & 13(9) & 1.8(6) & 1.9(1.0)
\tabhline
$\tau_{W(2,2)}$ & 6.9(1.7) & 8.0(1.8) & 26(6) & 35(7) & 4.0(1.2) & 2.9(6)
\tabhline
$\tau_{W(4,4)}$ & $\approx$ 1.3 & $\approx$ 1.3 & 22(3) &
 26(3) & 10.4(1.8) & 10.3(2.4)
\tabhline
$\tau_{\bar{P}}$ & 3.3(5) & 3.8(6) & 93(13) & 67(4) & 279(45) & 274(33)
\tabhline
$\tau_{\bar{P}^2}$ & 1.0(3) & $\approx$ 1.3 & 35(4) & 32(6) & 48(8) & 46(6)
\tabhline
$\tau_{\mbox{\scriptsize sign}(\bar{P})}$ & 5.3(1.2) & 5.2(1.3) & 92(13) & 
68(6) & 
275(41) & 277(39)
\\[.3ex] \hline
\end{tabular}
\end{table}

\begin{figure}[htbp]
 \begin{center}
  \begin{minipage}[t]{145mm} 
   \begin{picture}(145,95)(45,130)
    \epsfig{file=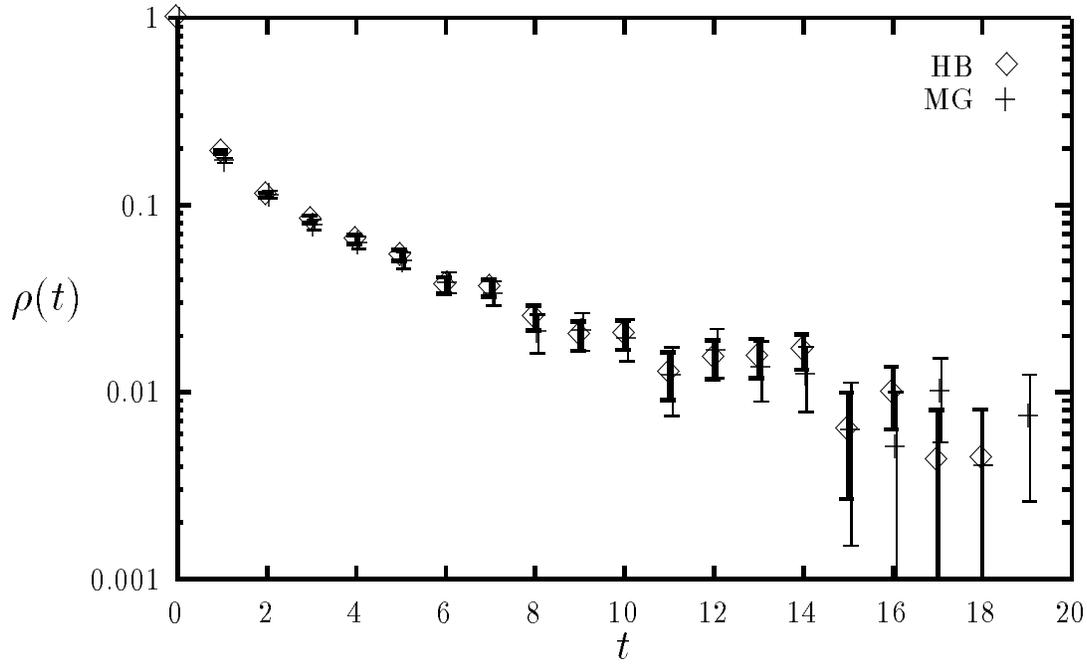,height=343mm}
   \end{picture}
   \begin{picture}(145,95)(45,130)
    \epsfig{file=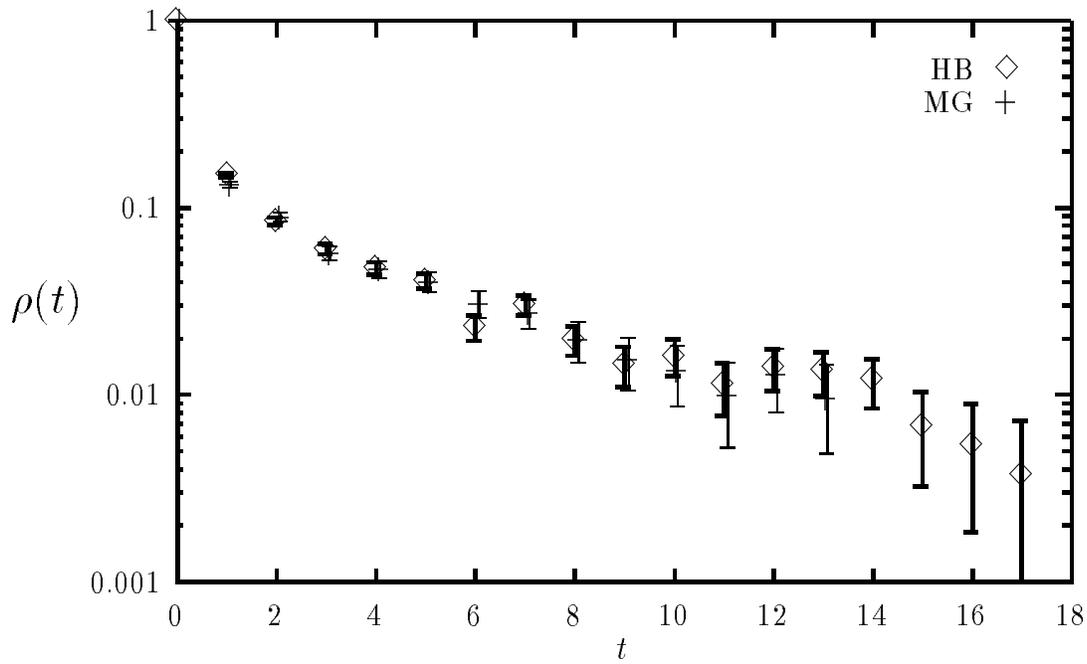,height=343mm}
   \end{picture}
   \caption[\Omega(s) in the 2d Sine Gordon model]
   {\label{tau22}
   \sl Comparison of the 
   autocorrelation functions $\rho(t)$ for the heat bath (HB)
   and multigrid (MG) algorithm for SU(2) on the $8^4$-lattice at
   $\beta=2.2$. 
   Top: Polyakov loop $\bar{P}$, 
   bottom: average sign of Polyakov loop 
   $\mbox{sign}(\bar{P})$.}
  \end{minipage}
 \end{center}
\end{figure}

\begin{figure}[htbp]
 \begin{center}
  \begin{minipage}[t]{145mm} 
   \begin{picture}(145,95)(45,130)
    \epsfig{file=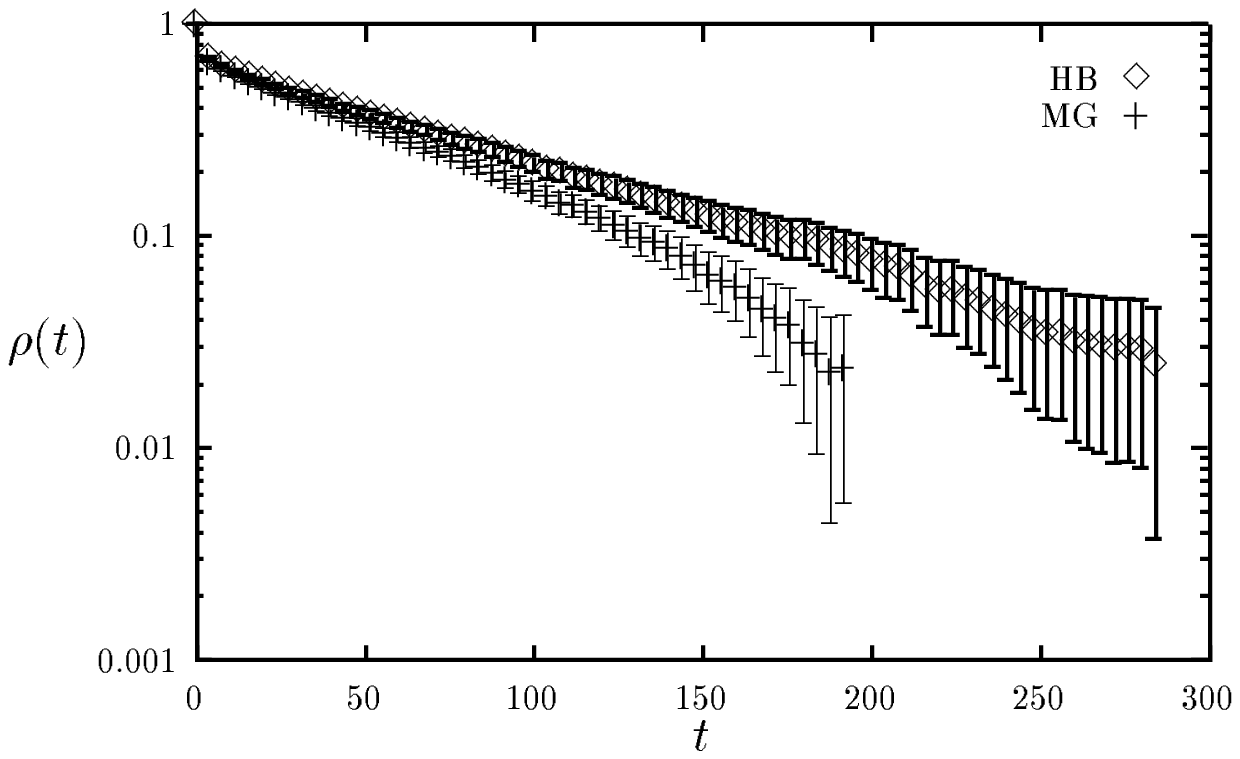,height=343mm}
   \end{picture}
   \begin{picture}(145,95)(45,130)
    \epsfig{file=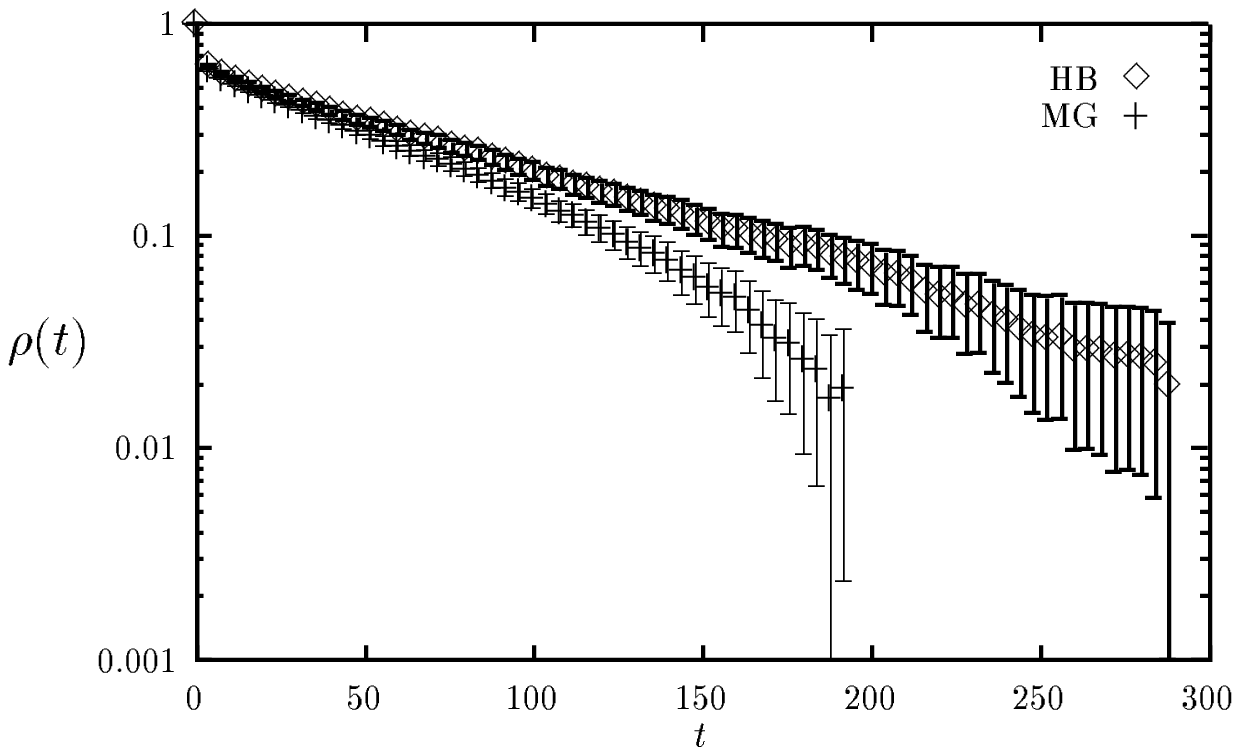,height=343mm}
   \end{picture}
   \caption[\Omega(s) in the 2d Sine Gordon model]
   {\label{tau24}
   \sl Comparison of the 
   autocorrelation functions $\rho(t)$ for the heat bath (HB)
   and multigrid (MG) algorithm for SU(2) on the $8^4$-lattice at
   $\beta=2.4$. 
   Top: Polyakov loop $\bar{P}$, 
   bottom: average sign of Polyakov loop 
   $\mbox{sign}(\bar{P})$.}
  \end{minipage}
 \end{center}
\end{figure}

\begin{figure}[htbp]
 \begin{center}
  \begin{minipage}[t]{145mm} 
   \begin{picture}(145,95)(45,130)
    \epsfig{file=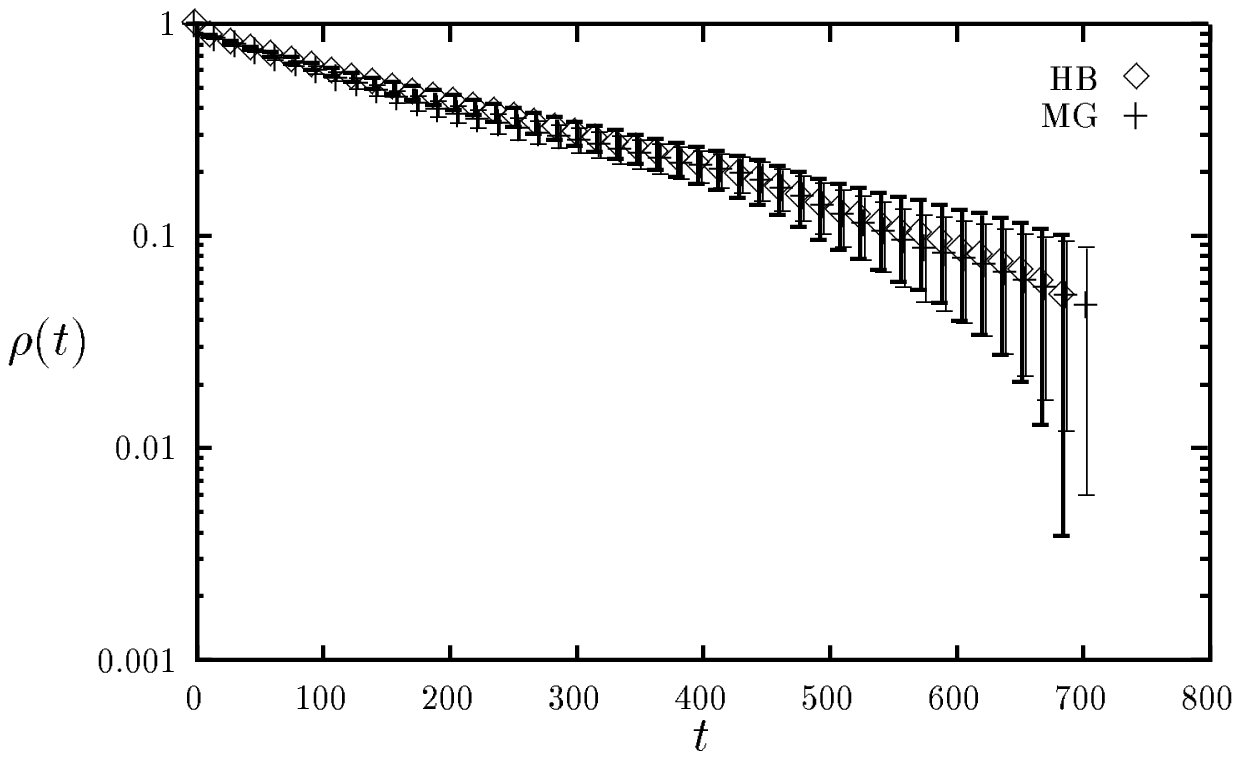,height=343mm}
   \end{picture}
   \begin{picture}(145,95)(45,130)
    \epsfig{file=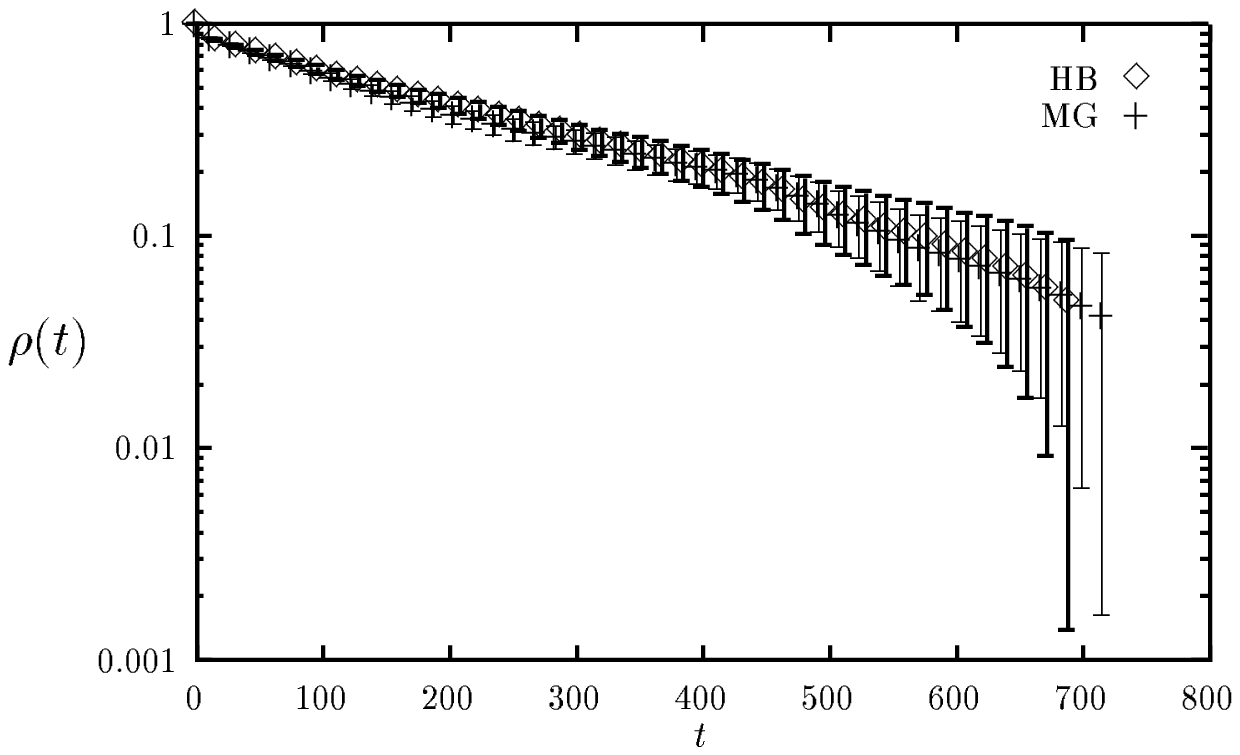,height=343mm}
   \end{picture}
   \caption[\Omega(s) in the 2d Sine Gordon model]
   {\label{tau26}
   \sl Comparison of the 
   autocorrelation functions $\rho(t)$ for the heat bath (HB)
   and multigrid (MG) algorithm for SU(2) on the $8^4$-lattice at
   $\beta=2.6$. 
   Top: Polyakov-loop $\bar{P}$, 
   bottom: average sign of Polyakov loop 
   $\mbox{sign}(\bar{P})$.}
  \end{minipage}
 \end{center}
\end{figure}

The results for the autocorrelation times are given in table \ref{SU24dynamic}.
For the observables with the longest
autocorrelation times $\bar{P}$ and $\mbox{sign}(\bar{P})$, we give a
comparison of the autocorrelation functions 
of the multigrid algorithm with the heat bath algorithm
for $\beta = 2.2$, $2.4$ and $2.6$ in figures \ref{tau22},
\ref{tau24} and \ref{tau26}. 

Note that the autocorrelation functions $\rho(t)$ do not in general show a pure
exponential decay but exhibit a crossover from a 
fast mode to a slow mode that eventually governs the asymptotic decay for large $t$.
Therefore the measurement of the integrated autocorrelation time $\tau_{int}$
with a self consistent truncation window of $4 \tau_{int}$ might not 
capture the asymptotic decay of $\rho(t)$ correctly.
In the present case we decided to extract a rough estimate
of the exponential autocorrelation time $\tau_{exp}$
and the corresponding errors
by the following procedure:
We plotted the $\rho(t)$ with error bars on
a logarithmic scale and decided at what value of $t$ 
the asymptotic exponential behavior started.
Then we drew the highest and lowest straight lines that were
compatible with the data in the asymptotic regime.
The errors are given such that the highest and the lowest value
of $\tau_{exp}$ lie at the ends of the interval result $\pm$ error.

All our results for the exponential autocorrelation times 
$\tau_{exp}$ of the multigrid and the heat bath algorithm 
are compatible within errors except from $\tau_{exp}$ for
$\bar{P}$ and $\mbox{sign}(\bar{P})$ at $\beta = 2.4$, see also figure 
\ref{tau24}.  In this case the superficial gain of the 
multigrid algorithm is a factor of $1.4$.
However the computational overhead of 
the multigrid method compared to 
the heat bath algorithm (a factor of $2.8$ in our implementation)
was not yet taken into account. So there is no net gain in computer time
also in the case of $\beta = 2.4$. 

\subsection{Summary}
 
In this section we studied an implementation 
of the multigrid Monte Carlo algorithm 
in four dimensions. We used the 
time slice blocking method and
piecewise constant interpolation with a W-cycle in a
recursive multigrid version  by updating in
$U(1)$ subgroups of $SU(2)$.
Simulations with gauge couplings $\beta = 2.2$, $2.4$ and $2.6$ were 
performed on an $8^4$-lattice.

Apart from a modest acceleration
of Polyakov loop observables at $\beta = 2.4$ by a factor of~$1.4$,
no improvement was found compared to a local heat bath algorithm.
Since the nonlocal update procedure has a computational overhead of
a factor of $2.8$ on the $8^4$-lattice (on a CRAY Y-MP), there is no
net gain but even a loss in CPU-time. 
This factor depends however on the details of the
implementation. 

Since from the theoretical analysis (section \ref{SECSU24acc})
one can not expect a lower dynamical critical exponent $z$,
there is no hope that on larger lattices the method will perform
better.
 
Possible improvements of our nonlocal updating scheme were
investigated by Gutbrod~\cite{gutbrod}.
Starting from the Coulomb gauge,
he uses an additional smoothing by taking a quadratic approximation
to the action and updating in terms of approximate eigenfunctions. 
The resulting nonlocal updates are performed as
(approximately) microcanonical overrelaxation steps.
Asymmetric lattices of size $L^3 T$ with $T >\!\!> L$
and anisotropic couplings (e.g.\ $\beta = 4.5$ in the time direction
and $\beta=3.0$ in the space directions) are used.
The results indicate a superficial
gain of a factor of $1.5 - 3$ compared to a local overrelaxation algorithm. 
However if the 
(implementation dependent) computational overhead of the nonlocal
method is taken into account, the net gain 
is marginal.

\clearpage

\section{Summary and Outlook}

\label{SECsummary}

\setcounter{equation}{0}

We presented a simple yet accurate formula that
expresses acceptance rates for nonlocal update algorithms in terms of
the average energy change 
$h_1 = \langle \Delta\calH \rangle$
of the proposed move.
The quantity $h_1$ depends only on one local observable
(or two in four dimensional $SU(2)$ lattice gauge theory) 
that is simple to compute,
e.g.\ by Monte Carlo simulations on a small lattice.

With the help
of our formula we found that
all investigated models fall into two kinematical classes:
For Sine Gordon, $\phi^4$ theory and $SU(2)$ lattice
gauge theory in four dimensions, 
$s$ had to be rescaled like $1/L_B$ for piecewise
constant and for smooth kernels in order to maintain
$L_B$-independent acceptance rates.
For massless free field theory,
the XY model, the $O(N)$ nonlinear $\sigma$-model,
the $CP^{N-1}$ model, $U(1)$ lattice gauge theory and
$SU(2)$ lattice gauge theory in two dimensions
one can achieve $s \sim const$ 
by choosing smooth kernels.

The kinematical behavior of multigrid
algorithms in interacting models
can be related to the dynamical 
critical behavior.
By comparing the behavior of the acceptance rates in interacting
models close to a critical point
with free field theory, where CSD is known to be eliminated by a
multigrid algorithm,
we found
the following rule:
For an interacting
model, sufficiently high acceptance rates for a complete elimination of
CSD can only be expected if $\, h_1 = \langle \calH(\ph+s \psi) -
\calH(\ph) \rangle $ contains no algorithmic mass term
$\sim s ^2 \sum_x \psi^2_x$.  
To our knowledge, all numerical experiments with
multigrid Monte Carlo algorithms 
that have been performed so far are consistent with this rule.
With the help of this criterion it is possible
to decide whether a given statistical model is a natural candidate 
for multigrid Monte Carlo or not.

Heuristically, the meaning of the rule can be
stated differently:
A piecewise constant update of a nonlocal domain should only have
energy costs proportional to the surface of the domain,
not energy costs proportional to the volume of the domain.

To check the prediction that too small acceptance rates on large blocks
cause CSD, we performed a multigrid Monte Carlo simulation of 
the Sine-Gordon model in two dimensions. 
We found $z = 1.9(1)$ with
piecewise constant interpolation and a W-cycle.
It was studied whether one can compensate for too small
update steps on large blocks
by accumulating many of these steps randomly.
To check this possibility, we simulated the model with a higher cycle
with $\gamma = 4$. 
We found $z = 1.9(1)$ also for this variant of the algorithm.

A multigrid algorithm for
nonabelian gauge fields, the time slice blocking, was developed.
It is based on the statistical decoupling of 
adjacent time slices as long as only link variables
in the time direction are updated.
The kinematical analysis predicted a 
strong reduction of CSD in a 
simulation of $SU(2)$ in two dimensions.
Almost complete elimination of CSD
was indeed observed in numerical experiments on
lattice sizes up to $256^2$.

The multigrid Monte Carlo algorithm was generalized
from two dimensions to four dimensions. 
In four dimensions, a 
nontrivial background field in the bottom of the blocks
caused difficulties.
We found that this background field
generates an algorithmic mass term 
that suppresses the acceptance rates on large blocks. 

We argued that the algorithm will only work well if the
algorithmic disorder mass $m_D$ 
scales like a physical mass, i.e.\ if
$m_D$ decreases exponentially fast with increasing
$\beta$. 
However, there is numerical and theoretical evidence 
that $m_D \sim const$ for large $\beta$.
Therefore the expectation is that
the proposed algorithm will not have a chance to reduce CSD.

Whether the nonlocal algorithm is at least able
to accelerate the local heat bath algorithm by a constant
factor
was investigated by simulations of $SU(2)$
lattice gauge theory on an \mbox{$8^4$-lattice}.
Superficially, a modest acceleration for 
Polyakov loop observables 
was found at $\beta = 2.4$,
however 
when the additional computational work is taken into account,
there is no net gain.
From the theoretical analysis
it can not expected that on larger lattices the method will perform
better.
In summary, it is fair to say that
for nonabelian pure gauge theory in four dimensions
the local overrelaxation algorithm is superior to nonlocal 
updating schemes up to now.

The kinematical mechanism that leads to a
failure of multigrid algorithms is well described by our analysis.
As an example, the analysis of the
time slice blocking algorithm 
for nonabelian gauge fields in four dimensions
can help to understand the additional difficulties
one has to face in four dimensions
compared to two dimensions.
We hope that a better understanding can lead to improved
multigrid algorithms that can overcome kinematical
obstructions stemming from an algorithmic mass term.

There are two possible strategies for improved algorithms:

First one could consider the 
stated rule as a (non rigorous) no-go theorem.
Then one can try to
avoid kinematical problems beforehand by
developing algorithms that do not have
energy costs proportional to the volume of a block.
A natural way is to take a global symmetry of the 
model under consideration and to apply
the corresponding symmetry transformation locally on blocks.

From this point of view it is clear that 
updates that are well adapted to the $\phi^4$ model 
(with a discrete $Z_2$ symmetry of the Hamiltonian) are not 
continuous, but discrete. In fact, the very efficient cluster
algorithm that has been developed for this model
is based on the discrete $Z_2$ symmetry~\cite{browertamayo}.
Another example where discrete updates appear to be more natural
than continuous updates
is the Sine Gordon model 
(with a discrete $\bbZ$ symmetry of the Hamiltonian).
Generalizations of the cluster algorithm 
for the discrete Gaussian model \cite{zspin} (that has the same symmetry
as the Sine Gordon model) seem to be efficient
also in this case \cite{sgcluster}.
The four dimensional nonabelian $SU(N)$ lattice gauge theory
has both a global discrete $Z_N$ symmetry and a local
$SU(N)$ gauge symmetry (which includes a global $SU(N)$ symmetry 
as a subgroup). Which symmetry could be used for
a fast nonlocal updating algorithm (if there exists any)
is an open question.

A second, more ambitious way to deal with kinematical obstructions
is not to avoid them but to attempt to overcome them
directly. Recently, very efficient methods for the acceleration of 
tunneling processes in systems close to a first order phase transition
have been developed~\cite{bergparisi}.
There, the computational effort could be reduced from an
exponential to a polynomial growth with the volume.

Since the mass term problem in models with a
discrete symmetry can also be interpreted as 
slow tunneling between different
minima of a conditional coarse grid Hamiltonian,
possible combinations of these new developments with
multigrid methods might be promising also for
reducing CSD.
However, since CSD is ``only'' a polynomial problem,
the requirement that the computational work should
be proportional to the volume
(and not to a power of the volume as
in systems close to a first order phase transition)
is quite restrictive.

\clearpage

\section*{Acknowledgements}

\addcontentsline{toc}{section}{\protect\numberline{}
         {Acknowledgements}}

First of all, I would like to thank Gerhard Mack
for his guidance and his constant support.
A~better way of supervising a thesis is hard to imagine for me.

\noindent
This work would have been impossible without Klaus Pinn.
During my ``sabbatical'' in M\"unster,
where we developed the main ideas of this work together,
I learned a lot from him. Also
during repeated visits from Hamburg, the 
``M\"unster-Hamburg-connection'' 
was very motivating and fruitful for me.

\noindent
Gernot M\"unster helped to arrange my stay in M\"unster and was
always open for discussions.

\noindent
I want to thank Bernhard Mikeska for the fruitful collaboration
on the simulation of the Sine Gordon model.

\noindent
During two visits at the Weizmann Institute I profited
much from the stimulating atmosphere in the discussions
with Achi Brandt. 

\noindent
I am also indebted to Sorin Solomon for his 
warm hospitality in Jerusalem.

\noindent
I profited from repeated conversations with
Martin B\"aker, Hermann Dilger, Hans-Gerd Evertz, Fritz Gutbrod,
Martin Hasenbusch, Arjan Hulsebos,
Thomas Kalkreuter, Mihai Marcu, Steffen Meyer, 
Alan Sokal and Ulli Wolff.

\noindent
Bernhard Mikeska and Klaus Pinn helped me a lot in the discussion
and a careful reading of the manuscript.

\bigskip

\bigskip

\noindent
Financial support by the Deutsche Forschungsgemeinschaft
and the German Israeli Foundation is acknowledged.

\noindent 
The numerical computations were performed on the NEC SX-3 in Cologne,
the Landesvektorrechner of the RWTH in Aachen, the CRAY Y-MP of the
HLRZ in J\"ulich and on hp $9000/730$ RISC workstations
at DESY.

\clearpage

\appendix

\section{Details of the acceptance analysis for gauge theories}

\setcounter{equation}{0}

\label{APPacc}

In this appendix we discuss the details of the calculation
of $h_1 = \langle \Delta \calH \rangle$ for 
gauge theories. 

\subsection{Abelian gauge fields}

The Wilson action for abelian gauge fields in $d$ dimensions 
can be written in link angles $\theta_{x,\mu}$ 
\be
\calH(\theta) \,=\,
\beta \sum_{x \in \Lambda_0}
\sum_{\mu \neq \tau}
\left[ 1-\cos\bigl(\theta_{\mu\nu}(x) ) \right] \ \ ,
\ee
with the plaquette angle 
\be
\theta_{\mu\nu}(x) = \theta_{x,\mu} + \theta_{x+\hat{\mu},\tau}
- \theta_{x+\hat{\tau},\mu} - \theta_{x,\tau} \ .
\ee
All the link angles $\theta_{x,\tau}$ that point from sites $x$ 
inside the hypercubic block
$x_o'$ in the time direction are proposed to be changed simultaneously:
\be\label{appupdate}
\theta_{x,\tau} \rightarrow  \theta_{x,\tau} + s\psi_x  \, .
\ee
The associated change of the hamiltonian is
$$
\Delta\calH \,=\,
-\beta \sum_{x \in \Lambda_0}
\sum_{\mu \neq \tau}
\left\{ \cos\left(\theta_{\mu\nu}(x)+s(\psi_{x+\hat{\mu}}-\psi_x )\right)
- \cos\left(\theta_{\mu\nu}(x)\right)
 \right\} 
$$
\be
 \,=\,
-\beta \sum_{x \in \Lambda_0}
\sum_{\mu \neq \tau}
\left\{ \cos(\theta_{\mu\nu}(x))
\left[\cos(s(\psi_{x+\hat{\mu}}-\psi_x ))-1\right]
- \sin(\theta_{\mu\nu}(x))\sin(s(\psi_{x+\hat{\mu}}-\psi_x ))
 \right\} \ \ .
\ee
Since the transformation $\theta \rightarrow -\theta$ for all
link angles is a symmetry of the Hamiltonian, whereas the
term $\sin(\theta_{\mu\nu}(x))$ changes its sign
under this transformation,
the expectation value $\langle\sin(\theta_{\mu\nu}(x))\rangle$
vanishes.
Therefore the average energy change of the nonlocal update
(\ref{appupdate}) is 
\be\label{h1abelian}
h_1\,=\,
\beta P \sum_{x \in \Lambda_0}
\sum_{\mu \neq \tau}
\left[ 1-\cos\bigl( s (\psi_{x+\hat{\mu}}-\psi_x)\bigr) \right] \ \ ,
\ee
with $P = \EW{\cos(\theta_{\mu\nu}(x)) } =\EW{ \Tr U_{\cal P} }$.
All expressions for $h_1$ that are given in section \ref{SECSU22acc}
and in section \ref{SECSU24acc} can be obtained from this result by 
specifying the number of dimensions $d$ and 
the form of the interpolation kernel $\psi$.

As an example, let us derive eq.\ (\ref{U1_const}) for abelian 
gauge theory in two dimensions:
For piecewise constant interpolation on a two dimensional block we have
$$
 \psi^{const}_x=\left\{
\begin{array}{ll}
1 &\mbox{for}\; x \in x_o' \\
0 &\mbox{for}\; x \not \in x_o'
\end{array}
\right. \ .
$$
From  eq.\ (\ref{h1abelian}) it is clear that
only $L_B$ links at the left boundary and $L_B$ links
at the right boundary of the block contribute to 
the average energy change $h_1$.
This leads to eq.\ (\ref{U1_const}):
$$
h_1 \,=\, 2 \beta P L_B [ 1 - \cos(s)] \ .
$$
 
\subsection{Nonabelian gauge fields}

The energy change of a time slice blocking 
update for $SU(2)$ in $d$
dimensions is
$$
\Delta \calH = - \frac{\beta}2
\sum_{\calP} \Tr \bigl( U_{\calP}' - U_{\calP} \bigr)
= - \frac{\beta}2 
\sum_{x \in \Lamt}\sum_{\mu \neq \tau} \Tr \bigl\{
( R_x(g)^* \Um{x} R(g)_{x+\hat\mu} - \Um{x} ) H_{x,\mu}^* \bigr\}
$$
\be
= - \frac{\beta}2 
\sum_{x \in \Lamt}
\sum_{\mu \neq \tau} \Tr \bigl\{
( R_x^* \Um{x}^g R_{x+\hat\mu} - \Um{x}^g ) H_{x,\mu}^{g *} \bigr\} \, ,
\ee
with $H_{x,\mu}^* = \Ut{x+\hat\mu} \Um{x+\hat\tau}^* \Ut{x}^*$
and $U^g_{x,\mu} = g_x \Um{x} g_{x+\hat\mu}^*$. $H^g$ is defined analogously.
$\psi$ denotes a \mbox{$d-1$}~dimensional interpolation kernel, and
the rotation matrices $R_x \in SU(2)$ are given by
\be
R_x(\vn,s) = \cos( s \psi_x /2 )
 + i \sin( s \psi_x /2) \,
\vn \!\cdot\! \vs \, .
\ee
Therefore we obtain
\bea
\Delta \calH =  - \frac{\beta}2 
\sum_{x \in \Lamt}\sum_{\mu \neq \tau}&\left\{ \str \right. &
\Tr \left(\Um{x}^g H_{x,\mu}^{g\,  *} \right)
[\cos(s\psi_x/2) \cos(s\psi_{x+\hat{\mu}}/2) - 1]\nonumber\\
&-&
\Tr \left(i\vn \!\cdot\! \vs \,\Um{x}^g  H_{x,\mu}^{g\,  *} \right)
\sin(s\psi_x/2)\cos(s\psi_{x+\hat{\mu}}/2)\strr \nonumber \\
&+& 
 \Tr \left(\Um{x}^g \,  i\vn \!\cdot\! \vs H_{x,\mu}^{g\,  *} \right)
 \strr
\cos(s\psi_x/2)\sin(s\psi_{x+\hat{\mu}}/2)\nonumber\\
& -&\left. \Tr \left(i\vn \!\cdot\! \vs \,
\Um{x}^g \, i\vn \!\cdot\! \vs H_{x,\mu}^{g\,  *} \right)
\sin(s\psi_x/2)\sin(s\psi_{x+\hat{\mu}}/2)
\str \right\}\strr \ .
\eea
Now we will show that to $h_1 = \EW{\Delta\calH}$ only
terms contribute that are even in $s$.
The point is that all gauge conditions that are 
used to define the $g$-matrices specify $g$ only up to a global
gauge transformation:
\be
g_x \rightarrow h g_x \ \ \mbox{for all} \ \  x \in \Lambda_0 \ .
\ee
If we average the expectation value
\be
B = 
\left\langle
\Tr \left(i\vn \!\cdot\! \vs \,\Um{x}^g  H_{x,\mu}^{g\,  *} 
\right) \right\rangle
=\left\langle
\Tr \left(i\vn \!\cdot\! \vs \,g_x \Um{x}^{ }  H_{x,\mu} g_x^*
\right) \right\rangle
\ee
over this global symmetry of the Hamiltonian, we get 
\be
B = \int dh \left\langle
\Tr \left(i\vn \!\cdot\! \vs \, h \,g_x \Um{x}  H_{x,\mu} g_x^*
\,h^*\,\right) \right\rangle\ = \ 0 \ .
\ee
Here we used an identity for $SU(2)$ matrices:
\be
\int dh\, \Tr(A\, h\, B\, h^*) \ = \ \half \, \Tr(A)\,\Tr(B) \ ,
\ee
and that the Pauli matrices are traceless.

Therefore the studied updating proposal leads to an average energy change 
\bea
h_1 
=   \beta
\sum_{x \in \Lamt}\sum_{\mu \neq \tau}&\left\{ \str \right.
&\left\langle \half \Tr \left(\Um{x} H_{x,\mu}^ * \right)\right\rangle
[1\  -\  \cos(s\psi_x/2)\cos(s\psi_{x+\hat{\mu}}/2)
]\nonumber\\
&+&\left\langle \half  \Tr \left(i\vn \!\cdot\! \vs \,
\Um{x}^g \, i\vn \!\cdot\! \vs H_{x,\mu}^{g\,  *} \right)\right\rangle
\left. \sin(s\psi_x/2)\sin(s\psi_{x+\hat{\mu}}/2)
\str \right\}\strr \ .
\eea
All expressions for $h_1$ that are derived
for $SU(2)$ gauge fields 
in section~\ref{SECSU22acc}
and section~\ref{SECSU24acc} can be obtained from this formula by 
specifying the number of dimensions $d$,
the choice  of the gauge condition that determines the
$g$-matrices, and the form of the interpolation kernel $\psi$.

As an example, we derive the expression (\ref{2dgauge_deltah}) 
for time slice blocking in two dimensions:
We assume that the $g$-matrices are chosen according to the
block axial gauge (\ref{block_axial_gauge}) 
and that $\psi$ vanishes outside the
block $x_o'$.
Then 
\be
\half
\Tr \left(i\vn \!\cdot\! \vs \,
\Um{x}^g \, i\vn \!\cdot\! \vs H_{x,\mu}^{g\,  *} \right)
\ = \
- \half
\Tr \left(\Um{x}^g H_{x,\mu}^{g\,  *} \right) \ ,
\ee
since $\Um{x}^g = 1$ within the bottom of the block,
and we obtain eq.\ (\ref{2dgauge_deltah}):
$$
h_1 = \beta P \,
\sum_{x \in \Lamt}\sum_{\mu \neq \tau} 
\left[ 1- \cos \left( s(\psi_x-\psi_{x+\hat{\mu}})/2 \right) \right] \ .
$$
Here $P = \EW{\half \Tr(\Um{x} H_{x,\mu})} =
\EW{\half \Tr(U_{\cal P})}$ denotes the plaquette expectation value.

\clearpage

\section{Exact results for $SU(2)$ lattice gauge theory 
in two dimensions}

\setcounter{equation}{0}

\label{APPSU22x}

Two dimensional $SU(2)$ lattice gauge theory is defined 
by the partition function
\be
Z = \int \prod_{x,\mu} d U_{x,\mu} \,
\exp\bigl( - \calH(U) \bigr) \, .
\ee

The link variables $U_{x,\mu}$ take values in the gauge
group  $SU(2)$, and $dU$ denotes the corresponding
invariant Haar measure.
The standard Wilson action $\calH(U)$ is given by
\be 
{\cal H}(U)\,=\,\beta\sum_{\cal P} \bigl[ 1 -
\half \, \Tr \, U_{\cal P} \bigr] \ \ .
\ee
The sum is over all plaquettes in the lattice.
Wilson loops are defined by
\be
W(C) = \langle \half \Tr (U(C)) \rangle \ ,
\ee
where $U(C)$ is the parallel transporter 
around a Wilson loop $C$ of area $A$.
The exact result for $W(C)$ on the infinite lattice is \cite{bailandrouffe}
\be\label{tdwilson}
W(C) \ = \ \left(\frac{I_2(\beta)}{I_1(\beta)}\right)^A
\ = \ 1 - \frac{3A}{2\beta} + O\left(\frac{1}{\beta^2}\right) \ .
\ee
Here, $I_{\nu}(\beta)$ are the modified Bessel functions. 
The string tension $\kappa$ is defined by the asymptotic 
behavior of the Wilson loop
for large $A$
\be
W(C) \ \rellow{\sim}{A \rightarrow \infty} \ \mbox{e}^{-\kappa A} \ ,
\ee
 and has the dimension of a mass squared. Therefore 
\be
\kappa \ = \ - \log \left(\frac{I_2(\beta)}{I_1(\beta)}\right)
\ = \ \frac{3}{2\beta} + O\left(\frac{1}{\beta^2}\right) \ .
\ee
The string tension correlation length $\xi$ is given by 
\be
\xi \ = \ \frac{1}{\sqrt{\kappa}} \ = \
\left[ - \log \left(\frac{I_2(\beta)}{I_1(\beta)}\right)\right]^{-1/2}
\ = \ \sqrt{\frac{2\beta}{3}}\left[1 + O\left(\frac{1}{\beta}\right) \right] \ .
\ee

\clearpage

\section{Gauge conditions on the lattice}

\setcounter{equation}{0}

\label{APPgauge}

In this appendix we discuss the relation of gauge conditions on the lattice
that are formulated in terms of parallel transporters $\Um{x}$
with the corresponding gauge conditions given in terms
of a gauge potential $A_{x,\mu}$.
We follow the presentation of ref.\ \cite{zwanziger}.

Our conventions are as follows: A basis of the $SU(N)$ Lie-algebra is
given by traceless, anti-hermitean $N \times N$ matrices $t^a$,
normalized to $\Tr(t^a t^b) = - \delta^{ab}/2$, which satisfy
$[t^a,t^b] = f^{abc}t^c$, where $f^{abc}$ are the structure 
constants of the group.
In particular for $SU(2)$ we have $t^a = i/2 \sigma_a$,
where $\sigma_a$ are the Pauli matrices, and $f^{abc} = \varepsilon^{abc}$.

A gauge potential $A_{x,\mu}$ on the lattice can be defined by
\be \label{defpot}
A_{x,\mu} = A_{x,\mu}^a t^a = \frac{1}2[\Um{x} - \Um{x}^*]
- \frac{1}{2N}\Tr[\Um{x} - \Um{x}^*] \ .
\ee
This definition is motivated by the fact that 
$
\Um{x} = \exp(A_{x,\mu}) 
$
for infinitesimal $A_{x,\mu}$.

\subsection{The Landau gauge}

The Landau gauge on the lattice is defined by \cite{wilson}
\be\label{landaug}
G_L(U,g) = \sum_{x \in \Lambda_0} \sum_{\mu=1}^d
\mbox{Re}\,\Tr\left(g_x \Um{x} g_{x+\mu}^*\right) 
 \,\stackrel{\mbox{!}}{=} \, \mbox{maximal}\ .
\ee
By performing the gauge transformation
$
\Um{x} \ \rightarrow \ \Um{x}^g = g_x \Um{x} g_{x+\mu}^* \ ,
$
where $g_x$ is obtained by the absolute maximum of $G_L(U,g)$,
the absolute maximum is shifted to $g_x = 1$.
The gauge configuration $U$ such that $G_L(U,g)$ for $g=1$ is in an 
absolute maximum has the property that the 
corresponding gauge potential
is transverse:
\be \label{divergence}
\nabla_{\mu} A_{x,\mu}^a = 
\sum_{\mu=1}^d \left[ A_{x+\mu,\mu}^a -  A_{x,\mu}^a
\right]  = 0 \ .
\ee
$\nabla_{\mu} A_{x,\mu}^a$ denotes the lattice divergence of
$A_{x,\mu}^a$.
This can be shown as follows:
Consider the one parameter subgroup $g(\tau)$ of the local gauge group
defined by 
\be
g(\tau)  = \{g_x(\tau)\} = \{ \exp(\tau \omega_x)\}\ , \; \; \; 
\omega^* = - \omega \ .
\ee
Here, $\omega_x$ is an abitrary element of the local Lie algebra
of the form $\omega_x = t^a \omega_x^a$.
For fixed $\omega$ and $U$, let $U(\tau)$ be the one parameter curve on the 
gauge orbit through $U$ defined by
\be
\Um{x}(\tau) = g_x(\tau) \Um{x} g_{x+\mu}^*(\tau) \ .
\ee
Let $G(\tau)$ be defined by
\be
G(\tau) = G_L\left(U,g(\tau)\right) = 
\sum_{x \in \Lambda_0} \sum_{\mu=1}^d
\mbox{Re}\,\Tr\left(\exp(\tau\omega_x)\Um{x}\exp(-\tau\omega_{x+\mu})\right) 
\ .
\ee
We have
$$
\frac{dG(\tau)}{d\tau}  = G'(\tau) = 
\sum_{x \in \Lambda_0} \sum_{\mu=1}^d
\mbox{Re}\,\Tr\left[(\omega_x-\omega_{x+\mu})\Um{x}(\tau)\right] 
$$
$$
 =\sum_{x \in \Lambda_0} \sum_{\mu=1}^d
\mbox{Re}\,\Tr\left[\omega_x(\Um{x}(\tau)-\Um{x-\mu}(\tau))\right] \ .
$$
Now we use the fact that 
$
\mbox{Re}\,\Tr(\omega_x\Um{x}(\tau)) \ =
 \ - \mbox{Re}\,\Tr(\omega_x\Um{x}^*(\tau))
$
and get
$$
G'(\tau) = \half \sum_{x \in \Lambda_0} \sum_{\mu=1}^d
\mbox{Re}\,\Tr\left[\omega_x(
\Um{x}(\tau)-\Um{x}^*(\tau)
-\Um{x-\mu}(\tau)+\Um{x-\mu}^*(\tau))\right] \ .
$$
With the definition (\ref{defpot}) of the gauge potential
and taking into account that $\omega_x$ is traceless, we obtain
$$
G'(\tau) = \sum_{x \in \Lambda_0} \sum_{\mu=1}^d
\mbox{Re}\,\Tr\left[\omega_x(
A_{x,\mu}(\tau)-A_{x-\mu,\mu}(\tau))\right] \ 
=-\half \sum_{x \in \Lambda_0} 
\omega_{x-\mu}^a\sum_{\mu=1}^d(A_{x+\mu,\mu}^a(\tau) - A_{x,\mu}^a(\tau)) \ .
$$
The $\tau$-dependence of $A$ is induced by the $\tau$-dependence of $U$.
Finally, $G'(\tau)$ can be expressed by the lattice 
divergence (\ref{divergence}) of the gauge potential:
\be
G'(\tau) =-\half \sum_{x \in \Lambda_0} 
\omega_{x-\mu}^a\nabla_{\mu}A_{x,\mu}^a(\tau) \ .
\ee
If $U$ is a stationary point of the gauge functional $G_L(U,g)$ 
at $g=1$, we have $G'(0) = 0$ for all $\omega_x^a$, which implies
\be
\nabla_{\mu}A_{x,\mu}^a \  = \ 0 \ .
\ee
Therefore, the transversality of the gauge potential $A$ is 
the condition for the functional $G_L(U,g)$ to be stationary at
$g=1$.

\subsection{The Coulomb gauge}

The Coulomb gauge condition is 
\be
G_C(U,g) = \sum_{x \in \Lambda_0} \sum_{\mu=1}^{d-1}
\mbox{Re}\,\Tr\left(g_x \Um{x} g_{x+\mu}^*\right) 
 \,\stackrel{\mbox{!}}{=} \, \mbox{maximal}\ .
\ee
Note that now only spatial link variables enter in the gauge
functional.
A similar argumentation as above shows that this gauge condition
leads to a transversality condition for the 
spatial components of the gauge potential
\be
\vec{\nabla}_{\mu} \vec{A}_{x}^a = 
\sum_{\mu=1}^{d-1} \left[ A_{x+\mu,\mu}^a -  A_{x,\mu}^a
\right]  = 0 \ .
\ee
Here, $\vec{\nabla}_{\mu} \vec{A}_{x}^a$ denotes the 
$d-1$ dimensional lattice divergence of the spatial components
$\vec{A}_{x}^a = (A_{x,1}^a,\dots,A_{x,d-1}^a)^T$
of the gauge potential.

\subsection{An alternative formulation of the Landau gauge condition}

The Landau gauge condition can also be formulated as a minimization
of a quadratic form
\be\label{qgauge}
Q(U,g) \ = \ (g^*,-\Delta g^*)
 \,\stackrel{\mbox{!}}{=} \, \mbox{minimal}\ ,
\ee
under the constraint that the 
$g_x$ are elements of the gauge group.
Here $\Delta$ denotes the gauge covariant Laplacian 
\be
\left(\Delta \phi\right)_x =
 \sum_{\mu=1}^d[\Um{x} \ph_{x+\mu} + \Um{x-\mu}^* \ph_{x-\mu} - 2\ph_x] \ ,
\ee
and the scalar product is defined by
\be
(\ph,\psi) = \sum_{x \in \Lambda_0}\foN \mbox{Re}\,\Tr(\ph_x^* \psi_x) \ .
\ee
The subtraction of a constant of the form 
\be
const = 2 \sum_{x \in \Lambda_0} \foN \mbox{Re}\,\Tr(g_x g_x^*)
\ee
from $Q(U,g)$ will not alter the solution of the minimization problem.
Using this, it is simple to check that the gauge conditions
(\ref{landaug}) and (\ref{qgauge}) are equivalent.

\clearpage

\addcontentsline{toc}{section}{\protect \numberline{} {References}}

{\,}

%%%%%%%%%%%%%%%%%%%%%%%%%%%%%%%%%%%%%%%%%%%%%%%%%%%%%%%%%%%%%%%%%%%%%%%%%

\begin{thebibliography}{99}
\bibitem{klausmartin}
The cluster algorithm for the Sine Gordon model was
developed by M.\ Hasenbusch and K.\ Pinn.
Their program was used for comparison.
\end{thebibliography}
\end{document}